%% file: bdc-arxiv.tex
\documentclass[10pt,a4paper]{article}
\usepackage{geometry}
\usepackage{graphicx}
\input{parts/TexHeader.tex}
\def\inclgraphics{\includegraphics[width=\textwidth]}

\author{Johannes Harth-Kitzerow\setcounter{footnote}{0}\footnote{jharthki@mpa-garching.mpg.de}
    \footnote{
        Max-Planck-Institut f{\"u}r Astrophysik,
        Karl-Schwarzschild-Str. 1,
        85748 Garching, Germany
    }
    \footnote{
        Ludwig-Maximilians-Universit{\"a}t M{\"u}nchen,
        Geschwister-Scholl-Platz 1,
        80539 Munich, Germany
    }
    \footnote{
        Technische Universit{\"a}t M{\"u}nchen,
        James-Franck-Str. 1,
        85748 Garching, Germany
    }
    \footnote{
        Exzellenzcluster ORIGINS,
        Boltzmannstr. 2,
        85748 Garching, Germany
    }
    \and
    Reimar~H.~Leike\footnotemark[2]
    \and
    Philipp~Arras\footnotemark[2] \footnotemark[3] \footnote{
        Technische Universit{\"a}t M{\"u}nchen,
        Boltzmannstr. 3,
        85748 Garching, Germany
    }
    \and
    Torsten~A.~En{\ss}lin\footnotemark[2] \footnotemark[3]\footnotemark[4]
}

\begin{document}

\title{Towards Bayesian Data Compression} 

\maketitle

\input{parts/abstract}

\newpage
\input{parts/main_part}

\textbf{Acknowledgements} \par 
    \emph{J.~Harth-Kitzerow acknowledges financial support by the
    Deutsche Forschungsgemeinschaft (DFG, German Research Foundation) under
    Germany's Excellence Strategy -- EXC-2094 -- 390783311.}

    \emph{P.~Arras acknowledges financial support by the German Federal
    Ministry of Education and Research (BMBF) under grant 05A17PB1
    (Verbundprojekt D-MeerKAT).}
    
    \emph{We thank Philipp Frank for help with the coding.}

\medskip

%
\bibliographystyle{plain}
\bibliography{bdc_bib}

\newpage
\input{parts/appendix}


\end{document}

%% file: parts/TexHeader.tex
\usepackage{algorithm}
\usepackage[noend]{algpseudocode}
\usepackage{mathrsfs}
\usepackage{amsmath}
\usepackage{amsfonts}
\usepackage{amssymb}
\usepackage{dsfont}

\usepackage{hyperref}
\usepackage{csquotes}

\def\nwpage{}

\def\linktoGit{\url{https://gitlab.mpcdf.mpg.de/jharthki/bdc}}
\def\P{\mathcal{P}}
\def\tr{\mathrm{tr}}
\newcommand{\norm}[1]{\left\lVert#1\right\rVert}
\def\compr{\mathrm{c}}
\def\orig{\mathrm{o}}
\def\kl{\mathrm{KL}_{\orig, \compr}}
\def\ie{\emph{i.e.}\ }
\def\eg{\emph{e.g.}\ }

%% file: parts/abstract.tex
\section*{Abstract}


In order to handle large data sets omnipresent in modern science, efficient
compression algorithms are necessary.
Here, a Bayesian data compression (BDC) algorithm that adapts to the specific
measurement situation is derived in the context of signal reconstruction.
BDC compresses a data set under conservation of its posterior
structure with minimal information loss given the prior knowledge on the signal, the quantity of interest.
Its basic form is valid for Gaussian priors and likelihoods.
For constant noise standard deviation, basic BDC becomes equivalent to a Bayesian
analog of principal component analysis.
Using Metric Gaussian Variational Inference, BDC generalizes to non-linear settings .
In its current form, BDC requires the storage of effective instrument response
functions for the compressed data and corresponding noise encoding the
posterior covariance structure.
Their memory demand counteract the compression gain.
In order to improve this, sparsity of the compressed responses can be obtained
by separating the data into patches and compressing them separately.
The applicability of BDC is demonstrated by applying it to synthetic
data and radio astronomical data.
Still the algorithm needs further improvement as the computation time of the
compression and subsequent inference exceeds the time of the inference with the original data.

%% file: parts/main_part.tex
\section*{Introduction}

One of the challenges in contemporary signal processing is dealing with large data sets.
Those data sets need to be stored, processed, and analysed.
They often reach the limit of the available computational power and storage.
Examples include urban technology \cite{Kong2020UESS}, internet searches
\cite{Google2020HowSearchWorks}, bio-informatics
\cite{Marx2013BigData} and radio astronomy \cite{Grainge2017SKA}.
In this paper we discuss, how such huge data sets can be handled efficiently by compression.

In general, there are two categories of data compression methods:
Lossless compression and lossy compression.
From lossless compressed data one can regain the full uncompressed data.
This limits the amount of compression as only redundant information
can be taken away by a lossless scheme.
Lossy compression is more effective in terms of reducing the storage needed by
the compressed data.
This is possible at the cost of loss of information.

In this work, we focus on lossy compression methods.
The considered scenario is: compressing data which carries information about some quantity of interest, which we call the \emph{signal}.
Only the relevant information for this signal needs to be conserved.
Therefore, there is no need to regain the full original data in such
applications.

Many lossy compression schemes have been developed:
Rate distortion theory~\cite[p.\,301-307]{CoverElementsOfInformationTheory} gives a general approach
stating the need of a loss function, which shall be minimized in order to find the
best compressed representation of some original data $d$.
As a consequence of the Karhunen-Lo{\'e}ve
theorem~\cite{karhunen1947lineare,loeve1948functions,kosambi1943statistics},
principal component analysis~(PCA)~\cite{Pearson1901LinesAndPlanes,Jolliffe2010PCA}
can also be used for data compression.
Its aim is to compress some data, such that the compressed data carries the
same statistic properties as the original.
It was shown that PCA minimizes an upper bound
of the mutual information of the original and the compressed data about some
relevant signal \cite{Geiger2012SignalEnhancement}.
Both methods aim to reproduce the original data from the compressed data,
but are not specifically optimized to recover information about the actual
quantity of interest.

Before compressing data, one should be clear about the signal 
on which one wants to keep as much information as possible.
In a Bayesian setting, this means that the posterior probability of the signal
conditional to the compressed data should be as close
as possible to the original posterior that was conditioned on the original
data.

The natural distance measure between the original and compressed posterior to be
used as the action principle is the \emph{Kullback Leibler} (KL) divergence~\cite{LeikeOBA}.
From this we derive \emph{Bayesian Data Compression} (BDC).
Using the KL divergence as the loss function reduces the problem of finding the compressed
data representation to an eigenvalue problem equivalent to the generalized
eigenvalue problem found by~\cite{Giraldi2018OptimalProjection}.
In this work we give a didactic derivation and show how this
approach can be extended to nonlinear and non-Gaussian measurement situations
as well as to large inference problems.
This is verified using synthetic data with linear and nonlinear signal models
and in a nonlinear astronomical measurement setup.

This publication is structured as follows:
In sect.\,\ref{sec:KL}, we assume the setting of the \emph{generalized Wiener
Filter}: a linear measurement equation, for a Gaussian signal sensed under
Gaussian noise~\cite{Wiener1949Extrapolation}:
There, the optimal compression for prior mean reduces to an eigenvalue problem.
In sect.\,\ref{sec:Generalizations} this is generalized to nonlinear and
non-Gaussian measurements.
Furthermore, we show how a sparse structure can be utilized on the compression
algorithm making it possible to handle large data sets in a reasonable amount
of time.
In sect.\,\ref{sec:Application}, BDC is applied to
synthetic data resulting from a linear measurement in one dimension,
a nonlinear measurement in two dimensions, and
real data from the Giant Metrewave Radio Telescope.

\section{Linear Compression Algorithm}\label{sec:KL}

We approach the problem of compression from a probabilistic perspective.
To this end we juxtapose the posterior probability distribution of the full
inference problem with a posterior coming from a virtual likelihood together
with the same prior.
The goal is to derive an algorithm which takes the original likelihood and the
prior as input and returns a new, virtual likelihood that is computationally
less expensive than the original likelihood.
This shall happen such that the resulting posterior probability distribution
differs as little as possible from the original posterior.

The natural measure to compare the information content of a probability
distribution and an approximation to it in the absence of other clearly defined
loss functions is the KL divergence, as shown in~\cite{LeikeOBA}.
Minimizing the KL divergence completely leads to the criteria for the most
informative likelihood.

\subsection{Assumptions and General Problem}

The new likelihood needs to be parameterized such that the KL divergence can be
minimized.
Initially, we make the following assumptions:

\begin{enumerate}
\item The signal $s$, which is a priori Gaussian distributed with
    known covariance $S$, has been measured with a linear response function
        $R_\orig$.
  The resulting original data $d_\orig$ is subject to additive Gaussian noise
        with known covariance $N_\orig$.
  In summary,
    \begin{align}\label{eq:do=Ros+no}
      d_\orig := R_\orig s + n_\orig,
    \end{align}
        where we denote definitions by \enquote{$:=$}, with \enquote{$:$} standing at the side of the new
        defined variable,
        and $s\hookleftarrow\mathscr G (s, S)$
        and $n_\orig\hookleftarrow\mathscr G (n_\orig, N_\orig)$
        are drawn from zero centered Gaussian distributions.
        The notation $s\hookleftarrow\mathscr G (s-s_0, S)$ indicates that
        $s$ is drawn from a Gaussian distribution with mean $s_0$ and
        covariance $S$. For the signal prior, $s_0$ is zero.

\item The compressed data $d_\compr$, which is going to be lower dimensional
    compared to the original data $d_\orig$, is related to the signal $s$ linearly through a measurement
        process with additive Gaussian noise with covariance
        $N_\compr$ and response $R_\compr$,
        which need to be determined,
    \begin{align}
        d_\compr = R_\compr s + n_\compr.
    \end{align}
\end{enumerate}
In this setup, the likelihoods $\P(d_i|s)$ as well as the \emph{posteriors of} both
\emph{the original and the compressed inference problem} $\P_i(s) :=  \P(s|d_i)$ are
Gaussian again~\cite{Wiener1949Extrapolation}
(once $N_\compr$ and $R_\compr$ are specified):
\begin{align}
    \P(d_i|s) &= \mathscr G (d_i-Rs,N_i) \\
    \P_i(s) &= \mathscr{G}(s-m_i, D_i), \quad i \in \{\orig, \compr
    \}\label{eq:DefPosterior}.
\end{align}
The mean $m_i$ and covariance $D_i$ are
\begin{align}
        m_i &:= D_i R_i^\dagger N_i^{-1} d_i \label{eq:Defmi}
\end{align}
and
\begin{align}
    D_i &:= (S^{-1}+ M_i)^{-1}, \quad \text{with} \label{eq:DefDi}\\
    M_i &:= R_i^\dagger N_i^{-1} R_i. \label{eq:DefMi}
\end{align}

We call $M_i$ the \emph{measurement precision matrix}.
Our goal is to find the \emph{compressed measurement parameters} $(d_\compr, R_\compr, N_\compr)$ such that the
least amount of information on the signal $s$ is lost as compared to $(d_\orig, R_\orig, N_\orig)$.

This means we want to adjust $d_\compr, R_\compr$ and $N_\compr$, such that the
difference of the two posteriors of the signal, given the compressed data
$d_\compr$ and given the original data $d_\orig$, is minimal.
To this end, we minimize the KL divergence under the constraint that the
compressed data vector shall not exceed a certain number of dimensions
$k_\compr$:
\begin{align}
    \kl &:= D_\mathrm{KL}( \P_\orig ||  \P_ c) \nonumber \\
    &:= \int \mathrm{d}s \, \P(s|d_\orig) \ln \frac{ \P(s|d_\orig)}{\P(s|d_\compr)} \nonumber \\
    &=: \big \langle \ln  \P_\orig \big\rangle_{ \P_\orig} - \big\langle \ln  \P_\compr \big\rangle_{ \P_\orig}.
\end{align}
In this notation, $k_{\compr}$ is suppressed. It will become explicit in the next section.
A detailed derivation showing that the KL divergence is indeed the appropriate
measure to decide on the optimality of the compression and a discussion about the order of its arguments can be found in~\cite{LeikeOBA}.

For Gaussian posteriors, the KL divergence becomes
\begin{align}
  \kl(d_\compr, R_\compr, N_\compr) = \frac{1}{2} \tr \bigg[ D_\compr^{-1}
    D_\orig - \mathds{1} - \ln(D_\compr^{-1} D_\orig)
  + D_\compr^{-1} (m_\compr-m_\orig)(m_\compr-m_\orig)^\dagger  \bigg], \label{eq:KLoc}
\end{align}
where the compressed posterior mean $m_\compr$ and covariance $D_\compr$ depend on the
compressed measurement parameters $d_\compr, R_\compr$ and $N_\compr$
through \eqref{eq:Defmi} and \eqref{eq:DefDi}.
The superscript $\dagger$ denotes the adjoint, \ie the transposed
complex conjugate of a vector or linear operator.
Thus, we have formulated the original compression problem as a minimization problem:
\begin{equation}
  d_\compr, R_\compr, N_\compr = \mathrm{argmin}_{d_\compr, R_\compr, N_\compr} \kl(d_\compr, R_\compr, N_\compr).
\end{equation}

In order to arrive at the optimal choice of compressed measurement parameters, we minimize
$\kl$ sequentially with respect to its arguments $d_\compr, R_\compr$ and $N_\compr$.
In that procedure we keep not yet optimized parameters as given and express already optimized parameters as functions of their given parameters during their optimization.
Minimization of \eqref{eq:KLoc} with respect to the compressed data $d_\compr$
for given response $R_\compr$ and compressed noise covariance $N_\compr$ yields:
\begin{align}
    d_\compr(R_\compr, N_\compr)
    &= N_\compr \left( R_\compr D_\compr R_\compr^\dagger \right)^{-1} R_\compr
    m_\orig \nonumber \\
    &= (R_\compr S R_\compr^\dagger + N_\compr)(R_\compr S R_\compr^\dagger)^{-1} R_\compr m_\orig, \label{eq:dc}
\end{align}
using the identity
\begin{align}\label{eq:Identity}
  D R^\dagger &= S R^\dagger \left(R S R^\dagger + N\right)^{-1} N.
\end{align}
Defining the \emph{compressed} and \emph{original Wiener filter operator}
\begin{align}
        W_i &:= S R_i^\dagger (R_i S R_i^\dagger + N_i)^{-1} \nonumber \\
        &= D_i R_i^\dagger N_i^{-1}, \quad i \in \{\orig, \compr\},
\end{align}
we see that \eqref{eq:dc} is equivalent to $R_\compr W_\compr d_\compr =
R_\compr W_\orig d_\orig$ or $R_\compr m_\compr = R_\compr m_\orig$,
meaning that the original and compressed posterior means are indistinguishable
for the compressed response.

Inserting Equation~\eqref{eq:dc} back into \eqref{eq:KLoc}, we can define the already
$d_\compr$-minimized $\kl$, which only depends on the still to be optimized
parameters $R_\compr$ and $N_\compr$:
\begin{align}
    \kl(R_\compr,N_\compr) :=& \kl(d_\compr(R_\compr,N_\compr),R_\compr,N_\compr)
    \nonumber \\
    \widehat{=}& \frac12 \bigg( \tr \left[ M_\compr D_\orig - \ln(S^{-1}+M_\compr) \right]
    - m_\orig^\dagger R_\compr^\dagger (R_\compr S R_\compr^\dagger)^{-1} R_\compr m_\orig \bigg), \label{eq:KL(RcNc)}
\end{align}
where equalities up to constants that are irrelevant for the minimization, since they do not change the position of the minimum of $\kl$, are denoted by \enquote{$\widehat{=}$}.
However, note that the specific values of $\kl$ do not equal the divergence of the two posteriors anymore.

Only the last term of this new $\kl$ in Equation~\eqref{eq:KL(RcNc)} depends on the original data.
The more the compressed response $R_\compr$ preserves the information of the original posterior
mean $m_\orig$, the smaller this term becomes which reduces $\kl$.
Thus, a response $R_\compr$ that is sensitive to $m_\orig$ is favoured.
The original posterior mean $m_\orig$ will typically exhibit large absolute values in signal space
where the original response was of largest absolute values.
This gives an incentive for the compressed response towards the original one.

In this section we looked at the posterior distributions in the context of
Gaussian prior and Gaussian likelihood in a linear measurement setup.
By minimizing the Kullback-Leibler divergence with respect to the compressed
data, we found an expression for the latter.
Plugging in this expression, $\kl$ only depends on the compressed response and
noise covariance.

\subsection{Information Gain from Compressed Data}\label{sec:InformationGain}

The remaining task is to minimize $\kl(R_\compr,N_\compr)$ with respect to the
compressed response $R_\compr$ and noise $N_\compr$.
Both appear in the loss function of our choice only combined in the measurement precision
matrix $M_\compr=R_\compr^\dagger N_\compr^{-1} R_\compr$ and as $R_\compr$
only in the last term.
Let $U$ be a unitary transformation that diagonalizes the compressed noise
covariance $N_\compr$.
One can show that the transformation
\begin{align}
    \begin{split}
        R_\compr &\to R'_\compr=U R_\compr,\\
        N_\compr &\to N'_\compr = U N_\compr U^\dagger
    \end{split}
\end{align}
leaves \eqref{eq:KL(RcNc)} invariant:
\begin{align}
  m_\orig^\dagger {R'}_\compr^\dagger ({R'}_\compr S {R'}_\compr^\dagger)^{-1} {R'}_\compr m_\orig
  &= m_\orig^\dagger R_\compr^\dagger U^\dagger U (R_\compr S R_\compr^\dagger)^{-1} U^\dagger U R_\compr m_\orig \nonumber\\
  &= m_\orig^\dagger R_\compr^\dagger (R_\compr S R_\compr^\dagger)^{-1} R_\compr m_\orig
\end{align}
and analogous for the terms containing $M_\compr' = R_\compr^\dagger U^\dagger U
N_\compr^{-1} U^\dagger U R_\compr = R_\compr^\dagger N_\compr^{-1} R_\compr = M_\compr$.
We use this to find a parametrization of the compressed measurement precision matrix $M_\compr$ with a set of vectors $r := (r_i)_{i=1}^{k_\compr}$, such that
\begin{align}\label{eq:SetMc}
  M_\compr = \sum_{i=1}^{k_\compr} r_i r_i^\dagger
\end{align}
with $k_\compr$ being the number of entries of the compressed data vector $d_\compr$.
In its eigenbasis, the inverse compressed noise covariance reads
\begin{align}\label{eq:NcDecomposition}
  N_\compr^{-1} &= \sum_{i=1}^{k_\compr} \mu_i^2 e_i e_i^\dagger,
\end{align}
with $e_i$ the normalized eigenvectors of $N_\compr^{-1}$ and $\mu_i^2$ the corresponding positive eigenvalues.
For reasons that become clear later on, let us choose
\begin{align}
    R_\compr &:= \sum_{i=1}^{k_\compr} e_i \hat{r}_i^\dagger, \label{eq:DefRc}\\
    \hat{r}_i &:= \frac{r_i}{\| r_i \|_S},\\
    \mu_i &= \| r_i \|_S , \label{eq:mu}
\end{align}
where $\norm{\cdot}_S$ is the norm induced by the prior covariance:
$\norm{x}_S := \sqrt{ x^\dagger S x }$.
With these definitions we can interpret $\hat{r}_i$ as the \emph{(normalized)
compressed measurement direction}, \ie the direction in which the
signal $s$ is measured leading to the $i$th compressed data point.
$\hat{r} := (\hat{r}_i)_{i=1}^{k_\compr}$ shall then be orthonormal with
respect to the scalar product induced by $S$.
The noise of this measurement is given by the corresponding \emph{compressed noise
variance} $\mu_i$.
With these definitions \eqref{eq:SetMc} can be verified:
\begin{align}
    M_\compr &= R_\compr^\dagger N_\compr^{-1} R_\compr \nonumber \\
    &= \sum_{i,j,k=1}^{k_\compr} \hat{r}_i \underbrace{e_i^\dagger
    e_j}_{\delta_{ij}} \underbrace{e_j^\dagger e_k}_{\delta_{jk}}
    \hat{r}_k^\dagger \mu_j^2 \nonumber \\
    &= \sum_{i=1}^{k_\compr} r_i r_i^\dagger.
\end{align}
Thus, the relevant degrees of freedom of the compressed response $R_\compr$ and
noise covariance $N_\compr$ are encoded in the compressed measurement
directions $r=(r_i)_{i=1}^{k_\compr}$ and we can write
\begin{align}
  \kl(r) :=& \kl(R_\compr(r),N_\compr(r)) \nonumber \\
  \widehat =& \frac{1}{2} \left( \tr\left[\big(\sum_i r_i r_i^\dagger\big) D_\orig - \ln \left( S^{-1}+ \sum_i r_i r_i^\dagger \right) \right]
  - m_\orig^\dagger R_\compr^\dagger (R_\compr S R_\compr^\dagger)^{-1} R_\compr m_\orig\right). \label{eq:KLoc(r)}
\end{align}
We are left to evaluate the trace in \eqref{eq:KL(RcNc)}.
The main issue is the logarithmic term.
By decomposing $S = \sqrt S \sqrt S^\dagger $ it can be simplified:
\begin{align}
  \tr \ln \left( S^{-1}+ \sum_i r_i r_i^\dagger \right) \widehat{=} \tr \ln
    \left( \mathds{1} + \sum_i \sqrt{S}^\dagger r_i r_i^\dagger \sqrt{S} \right). \label{eq:trlog}
\end{align}
We call $\sqrt{S}$ \emph{prior dispersion}.
This nomenclature deviates from the common convention of using the names dispersion and covariance synonymously.
An \emph{excitation} $\xi$ following a
standard normal distribution $\mathcal{G}(s,\mathds{1})$ can be amplified by
$\sqrt{S}$, \ie $\sqrt{S} \xi$.
This amplified field follows the same statistics as $s$.
We thus distinguish between variables defined in \emph{signal space} and
\emph{excitation space}, with the dispersion $\sqrt{S}$ (resp.\
$\sqrt{S}^\dagger$) and its inverse serving as transformations between both.

The compressed measurement directions in excitation space are then
\begin{align}
    \begin{split}
      w_i &:= \sqrt S^\dagger r_i, \\
        \hat{w}_i &:= \frac{w_i }{ \norm{w_i} }, \quad \forall i \in \{1,\dots, k_\compr\}.
    \end{split}\label{eq:Defwi}
\end{align}
The orthonormal basis $\hat{w} := (\hat{w}_i)_{i=1}^{k_\compr}$ diagonalizes both summands of \eqref{eq:trlog} simultaneously:
\begin{align}
  \tr \ln \left( \mathds{1} + \sum_i \sqrt S^\dagger r_i r_i^\dagger \sqrt S \right) &= \tr \ln \left( \mathds{1} + \sum_i \|w_i\|^2 \hat w_i \hat w_i^\dagger \right) \nonumber \\
  &= \tr \left[ \sum_i \ln \left( 1+\|w_i\|^2 \right) \hat w_i \hat w_i^\dagger \right] \nonumber \\
  &= \sum_i \ln \left( 1+ w_i^\dagger w_i \right) \nonumber \\
    &= \sum_i \ln \left( 1 + r_i^\dagger S r_i \right)
\end{align}
In addition, the last term of \eqref{eq:KLoc(r)} reduces to
\begin{align}
  m_\orig^\dagger R_\compr^\dagger (R_\compr S R_\compr^\dagger)^{-1} R_\compr m_\orig
  &= m_\orig^\dagger \sum_i \hat{r}_i e_i^\dagger \bigg(\sum_{j,k} e_j \underbrace{r_j^\dagger S r_k}_{\delta_{jk}} e_k^\dagger\bigg)^{-1} \sum_l e_l \hat{r}_l^\dagger m_\orig \nonumber \\
  &= m_\orig^\dagger \sum_{i,l} \hat{r}_i \underbrace{e_i^\dagger
    e_l}_{\delta_{il}} \hat{r}_l^\dagger m_\orig \nonumber \\
    &= m_\orig^\dagger \sum_i \hat{r}_i \hat{r}_i^\dagger m_\orig ,
\end{align}
such that we get
\begin{align}
  \kl(r) &\widehat = \frac{1}{2} \bigg( \tr \big[S^{-1} m_\orig m_\orig^\dagger  \big]
  + \sum_{i=1}^{k_\compr} \big[ r_i^\dagger D_\orig r_i - \ln(1+ r_i^\dagger S r_i)- m_\orig^\dagger \hat r_i \hat r_i^\dagger m_\orig\big]\bigg) \nonumber \\
  &\widehat{=} - \sum_{i=1}^{k_\compr} \Delta I(r_i), \label{eq:KLoc(wkc)}
\end{align}
with \emph{information gain} $\Delta I(r_i)$ of a single compressed measurement
direction $r_i$.
Since the compressed measurement directions in excitation space are orthogonal
with respect to the standard scalar product, we proceed with the calculations in
excitation space.
There, the information gain becomes
\begin{align}
    \Delta I (w_i) &:= \frac{1}{2} \left[- w_i^\dagger \mathcal{D}_\orig w_i
    + \ln(1+ w_i^\dagger w_i) + \hat w_i^\dagger \tilde m_\orig \tilde m_\orig^\dagger \hat w_i \right], \label{eq:DeltaKLDef}
\end{align}
with posterior mean and covariance in excitation space
\begin{align}
        \tilde m_\orig &:= \sqrt{S}^{-1} m_\orig, \\
        \mathcal{D}_\orig &:= \sqrt{S}^{-1} D_\orig \sqrt{S}^{-\dagger}, \\
        \sqrt{S}^{-\dagger} &:= {\sqrt{S}^\dagger}^{-1} \nonumber.
\end{align}

We see that the relevant part of $\kl$ splits up into a sum over independent
contributions $- \Delta I(w_i)$ associated to the individual compressed
measurement direction $w_i$, each of which belongs to a specific compressed data point $(d_\compr)_i$.
Since $\kl$ expressed this way is additive with respect to the inclusion of additional data points, the sum in \eqref{eq:KLoc(wkc)} can easily be extended.
To minimize $\kl$ with respect to $w$ (or $r$ respectively), the contributions
$(-\Delta I(w_i))$ can be minimized individually with respect to their
respective compressed measurement direction $w_i$.

The information gain $\Delta I$ depends on the normalized measurement direction $\hat{w}_n$ and its magnitude $\| w_n \|$,
\begin{align}\label{eq:DeltaI(ww)}
    \Delta I \left(\hat{w}_n, \|w_n\|\right) := \frac{1}{2} \left[- \|w_n\|^2 \hat w_n^\dagger \mathcal{D}_\orig \hat w_n
  + \ln(1+ \| w_n \|^2) + \hat w_n^\dagger \widetilde{m}_\orig
    \widetilde{m}_\orig^\dagger \hat w_n \right] .
\end{align}
We maximize this with respect to the magnitude, get
\begin{align}\label{eq:|ri|Ssqd}
 \norm{w_n}^2 = \frac{1}{\hat w_n \mathcal{D}_\orig \hat w_n} - 1,
\end{align}
and insert the result into \eqref{eq:DeltaI(ww)}.
This leaves us with our final expression of the information gain of a single compressed data point:
\begin{align}\label{eq:CalDeltaKL}
  \boxed{ 2 \Delta I (\hat{w}_n) = \hat w_n^\dagger \mathcal{D}_\orig \hat w_n - 1 - \ln(\hat w_n \mathcal{D}_\orig \hat w_n) + \hat w_n^\dagger \widetilde{m}_\orig \widetilde{m}_\orig^\dagger \hat w_n. }
\end{align}
Summarizing, the sets of vectors $\{e_i\}$ and $\{r_i\}$ are representations of the compressed noise covariance and
compressed response, such that the trace in $\kl$ splits up into
independent summands.
Each summand is the negative information gain when considering the corresponding
compressed measurement direction $\hat{w}_n$ and can be treated individually.
As a next step we need to maximize $\Delta I(\hat{w}_n)$.

\subsection{Optimal Expected Information Gain}\label{sec:ExpectedOptimum}

In order to find the compressed data point which adds most information to the
compressed likelihood, \eqref{eq:CalDeltaKL} needs to be maximized with respect
to the normalized vector $\hat{w}_n$.
For zero posterior mean $m_\orig=0$, this problem reduces to an eigenvalue problem as shown in Appendix~\ref{Sec:ZeroMeanOptimization}.
We proceed by treating the general case of non-vanishing $m_\orig$.

There is only one normalized vector $\hat{w}_i$ (respectively $\hat{r}_i$) left
to be determined for each $i \in \{1 , \dots , k_\compr \}$.
However, in the current form, \eqref{eq:CalDeltaKL} cannot be maximized analytically.
The main issue is the last term that contains $\tilde m_\orig$ and which in
the following we treat stochastically using the prior signal and noise
knowledge on the signal $s$ and the noise $n_\orig$ only.
Thereby, we can calculate the expected information gain.

Using the Gaussian priors of signal $s$ and noise $n_\orig$ with zero mean, as well as the measurement \eqref{eq:do=Ros+no} and the definition \eqref{eq:Defmi} of $m_\orig$, we get the expected posterior signal mean
\begin{align}
  \langle m_\orig \rangle_{\P(s,n_\orig)} &= D_\orig R_\orig^\dagger N_\orig^{-1} (R_\orig \underbrace{\langle s \rangle_{\P(s)}}_{=0} + \underbrace{\langle n_\orig \rangle_{\P(n_\orig)}}_{=0} )=0,
\end{align}
and variance
\begin{align}
  C &:= \langle m_\orig m_\orig^\dagger \rangle_{\P(s,n_\orig)}\\
    &= D_\orig R_\orig^\dagger N_\orig^{-1} (R_\orig SR_\orig^\dagger +
    N_\orig) N_\orig^{-1} R_\orig D_\orig \nonumber \\
    &=  (D_\orig R_\orig^\dagger N_\orig^{-1} (R_\orig S R_\orig^\dagger + N_\orig) N_\orig^{-1} R_\orig + \mathds{1}) D_\orig - D_\orig \nonumber \\
    &\stackrel{\eqref{eq:Identity}}{=} (S R_\orig^\dagger N_\orig^{-1} R_\orig
    + \mathds{1}) D_\orig - D_\orig \nonumber \\
    &\stackrel{\eqref{eq:DefDi}}{=} S - D_\orig.\label{eq:DefC}
\end{align}
Thus, as a sum of Gaussian distributed variables, $m_\orig$ again is Gaussian distributed with
\begin{align}
  \P(m_\orig)=\mathscr{G}(m_\orig, C).
\end{align}
Calculating the mean of the last term of \eqref{eq:CalDeltaKL} under this distribution then gives
\begin{align}
    \hat w_n^\dagger \langle \widetilde{m}_\orig
    \widetilde{m}_\orig^\dagger\rangle_{\P(m_\orig)} \hat w_n
    &= \hat w_n^\dagger \sqrt{S}^{-1} C \sqrt{S}^{-\dagger} \hat w_n \nonumber \\
    &\stackrel{\eqref{eq:DefC}}{=} \hat w_n^\dagger (\mathds{1}-
    \mathcal{D}_\orig)\hat w_n \nonumber \\
    &\stackrel{\eqref{eq:Defwi}}{=} 1 - \hat w_n^\dagger \mathcal{D}_\orig \hat w_n ,
\end{align}
which cancels the first two terms.
The expected information gain then is
\begin{align}\label{eq:DeltaIBDC}
  \boxed{ \langle \Delta I (\hat{w}_n) \rangle_{\P(m_\orig)} = - \frac{1}{2}
    \ln(\hat w_n^\dagger \mathcal{D}_\orig \hat w_n). }
\end{align}
This expected information gain is maximal, if and only if $\hat{w}_n$ is parallel to the eigenvector of $\mathcal{D}_\orig$ with smallest eigenvalue $\delta_n^2$.
This insight reduces the problem to the eigenvalue problem
\begin{align}\label{eq:CalDwIsdeltaw}
  \mathcal{D}_\orig w_n = \delta_n^{2} w_n.
\end{align}
In terms of the vectors $r$, which build the compressed measurement precision matrix $M_\compr$ in
\eqref{eq:SetMc}, and after inserting \eqref{eq:DefDi}, this states
\begin{align}\label{eq:DoinvSr=delta2invr}
    \begin{split}
        D_\orig^{-1} S r_n &=(\mathds{1} + M_\orig S) r_n \\
        &= \delta_n^{-2}  r_n.
    \end{split}
\end{align}
Combining \eqref{eq:mu}, \eqref{eq:|ri|Ssqd} and \eqref{eq:CalDwIsdeltaw} gives:
\begin{align}
    \mu_n^2 &= \| w_n \|^2 = \delta_n^{-2} -1.
\end{align}
Thus \eqref{eq:DoinvSr=delta2invr} is equivalent to
\begin{align} \label{eq:MoSx=mux}
  \boxed{M_\orig S \ r_n = \mu_n^2 \, r_n.}
\end{align}
We call $M_\orig S =: Q$ the \emph{fidelity matrix}.
For identity responses, $Q$ can be interpreted as signal-to-noise covariance ratio.
The largest eigenvalues of $Q$ give rise to the most informative compressed
measurement directions according to the minimization of the Kullback-Leibler
divergence.
At the same time $Q$ is the matrix product of the original measurement precision matrix
$M_\orig$ and the signal prior covariance $S$.
Thus, its largest eigenvalues and corresponding eigenvectors are those directions in
signal space, where the measurement is maximally precise while the prior is
maximally uncertain.
In other words, the directions at which the
original data update the prior the most, are exactly
those, which are the most informative.

For $r_n = R_\orig^\dagger v$, eigenvalue problem \eqref{eq:MoSx=mux} is equivalent to the generalized eigenvalue
problem of~\cite{Giraldi2018OptimalProjection},
\begin{align}
    R_\orig S R_\orig^\dagger v = \lambda N_\orig v,
\end{align}
multiplied with $R_\orig^\dagger N_\orig^{-1}$ from the left.

For constant noise $N_\orig = \sigma_{n_\orig}^{2} \mathds{1}$, eigenvalue
problem~\eqref{eq:MoSx=mux} becomes a Bayesian analog to principal
component analysis (PCA)~\cite{Pearson1901LinesAndPlanes,Jolliffe2010PCA}.
PCA takes the highest eigenvalues and corresponding eigenvectors of the data
covariance matrix.
In PCA this data covariance is built by the covariance of several 
measurements.
In a Bayesian setting we can determine the data covariance with prior
information using equation~\eqref{eq:do=Ros+no} instead:
\begin{align}
    \langle d_\orig d_\orig^\dagger \rangle_{\P(s,n)} &= R_\orig S R_\orig +
    N_\orig,
\end{align}
with corresponding eigenvalue problem
\begin{align}\label{eq:BayesPCA}
    ( R_\orig S R_\orig^\dagger + N_\orig ) v &= \lambda v.
\end{align}
Multiplying with $N_\orig^{-1}$ from the left gives;
\begin{align}
    (N_\orig^{-1}R_\orig S R_\orig^\dagger +\mathds{1})\,v &=
    \lambda N_\orig^{-1}\,v.
\end{align}
We can substract $v$ on both sides of the equation and for constant noise
$\sigma_{n_\orig}^{2}$ define $\mu_n^2 := \lambda \sigma_n^{-2}-1$, such that
\begin{align}
    N_\orig^{-1}R_\orig S R_\orig^\dagger v
    &=\underbrace{(\lambda \sigma_{n_\orig}^{-2} -1 )}_{\mu_n^2} v,
\end{align}
where we inserted the constant noise value only on the left hand side of the
equation to illustrate the similarity to equation~\eqref{eq:MoSx=mux} on the
right hand side.
Muliplying with $R_\orig^\dagger$ from the left and identifying $r_n =
R_\orig^\dagger v$ this gives equation~\eqref{eq:MoSx=mux}.
In contrast to original PCA, in a Bayesian setting we can already compress a
single measurement.
In contrast to this Baysian analog to PCA,
BDC is able to handle varying noise and compresses optimally with respect to
the information about the variable of interest $s$ as the result of our
derivation.

With Equation~\eqref{eq:MoSx=mux}, the expected information gain for including the compressed data point
${d_\compr}_n$ in BDC then is
\begin{align}
    \Delta I (\mu_n) &:= \langle \Delta I (\hat{r}_n)\rangle_{\P(m_\orig)} \nonumber \\
    &= - \frac{1}{2} \ln(\hat w_n \underbrace{\mathcal{D}_\orig \hat w_n}_{= \delta_n^2 \hat w_n} ) \nonumber \\
    &= - \ln(\delta_n) \nonumber \\
    &= \frac{1}{2} \ln \left( \mu_n^2+1 \right)
    \label{eq:DeltaKL=lndelta}.
\end{align}

Summarizing, we need to find the $k_\compr$ largest eigenvalues $\mu_n^2$ and
corresponding eigenvectors $r_n$ of \eqref{eq:MoSx=mux}.
With these, the compressed measurement parameters are
\begin{align}
    N_\compr^{-1} &= \sum_{n=1}^{k_\compr} \mu_n^2 e_n e_n^\dagger,
    \label{eq:Nc}\\
    R_\compr &= \sum_{n=1}^{k_\compr} e_n \hat{r}_n^\dagger, \label{eq:Rc}\\
  d_\compr &= (R_\compr S R_\compr^\dagger + N_\compr)(R_\compr S
    R_\compr^\dagger)^{-1} R_\compr m_\orig, \label{eq:dcFin}
\end{align}
with
\begin{align}
  \hat{r}_n &:= \frac{r_n}{\norm{r_n}_S} \quad \text{and}\quad \norm{r_n}_S = \mu_n.
\end{align}
These equations and \eqref{eq:Defmi} and \eqref{eq:DefDi} are all ingredients
needed to solve the compression problem.

\subsection{Algorithm}

Now, the previously derived method shall be turned into the actual BDC\@.
For that we need to solve eigenvalue problem~\eqref{eq:MoSx=mux}.
For compressing the data to $k_\compr$ data points, one needs to determine the
$k_\compr$ largest eigenvalues and corresponding eigenvectors
that belong to the most informative measurement directions.
First we derive an estimate for the fraction of information stored in the
compressed measurement parameters if we compute only a limited number of
\emph{eigenpairs}, \ie eigenvalues and corresponding eigenvectors.
Then, we discuss some details of how to compute the input parameters for
getting the compressed measurement parameters, \ie for the eigenvalue problem~\eqref{eq:MoSx=mux}
and how to solve it.

Due to computational limits, in general we cannot determine all $K$ eigenpairs
of \eqref{eq:MoSx=mux} carrying information.
We need to set the number $k_{\max}$ of most informative eigenpairs being determined
numerically.
For that limited number of eigenpairs, we derive a lower bound of the
information stored in the corresponding compressed measurement parameters in
the following.
If we are only interested in a certain amount of information we can use this
bound to find and neglect eigenpairs containing too little information, such that in the end,
we have $k_\compr \leq k_{\max}$ eigenpairs containing still enough
information.

The eigenpairs carrying information are those with non-zero eigenvalue $\mu_i$.
The number $K$ of non-zero eigenvectors is equal to the rank of $M_\orig S =
R_\orig^\dagger N_\orig^{-1} R_\orig S$.
As Gaussian covariances, $S$ and $N_\orig$ are positive definite and therefore have
full rank, the original response
$R_\orig$ has at most a rank equal to the smaller rank of both covariances.
Thus, with \eqref{eq:DefMi}, the rank of $M_\orig S$ and therefore the number
of informative eigenpairs $K$ is equal to the rank of $R_\orig$.
Altogether the compressed measurement parameters can maximally carry the \emph{total
information} $I := \sum_{i=1}^{K} \Delta I (\mu_i)$.
$I$ is the difference in information stored in the posterior, when considering
all informative eigendirections $k_\compr = K$ compared to having no compressed
data, $k_\compr =0$.
For no compressed data, the compressed posterior distribution becomes the prior
distribution with covariance $S$.
Thus, $I$ is the total information about the signal $s$ encoded in the original data
with respect to prior knowledge.

We can find an upper bound for the total information.
The eigenpairs are ordered such that the eigenvalues decrease with
growing index,
and therefore the contribution to the total information sum $I$.
Thus, the last determined eigenvalue is the smallest of all eigenvalues
and the least informative one.
Also it is larger than all eigenvalues that could not be computed.
We can use this to give an upper limit bound to the amount of information lost by
truncating the sum at $k_\compr$, and adding the number of eigenvalues ignored,
$K-k_{\max}$, times the amount of information provided by the last
eigenpair $I (\mu_{k_{\max}})$.
Thus,
\begin{equation}\label{eq:Ileq}
    I \leq \sum_{i=1}^{k_{\max}} \Delta I (\mu_i) +
    (K-k_{\max}) \Delta I (\mu_{k_{\max}}).
\end{equation}

We define the \emph{fraction of information} $\gamma$ of
total information $I$ \emph{stored in the compressed measurement parameters} which are
determined by the eigenpairs:
\begin{equation}\label{eq:DefGamma}
        \gamma := \frac{\sum_{i=1}^{k_\compr}\Delta I (\mu_i)}{I}
\end{equation}
With Equation~\eqref{eq:Ileq}, we can find a lower bound
\begin{equation}\label{eq:DefGammaMin}
    \begin{split}
        \gamma &\stackrel{\eqref{eq:Ileq}}{\geq} \frac{\sum_{i=1}^{k_\compr}\Delta I
            (\mu_i)}{\sum_{i=1}^{k_{\max}}\Delta I (\mu_i) + (K - k_{\max}) \Delta
            I (\mu_{k_{\max}})} \\
            &=: \gamma_{\min}\left(k_\compr,(\mu_i)_{i=1}^{k_{\max}},K\right).
    \end{split}
\end{equation}

Using $I \geq \sum_{i=1}^{k_{\max}} \Delta I(\mu_i)$, we analogously find an
upper bound
\begin{equation}\label{eq:DefGammaMax}
    \gamma \leq \gamma_{\max} :=
            \frac{\sum_{i=1}^{k_\compr}\Delta I
            (\mu_i)}{\sum_{i=1}^{k_{\max}}\Delta I (\mu_i)}
\end{equation}
for $\gamma$.
Those bounds can be used to narrow the fraction of information stored in the
compressed measurement parameters $\gamma$ by $\gamma \in
[\gamma_{\min},\gamma_{\max}]$.

Now we can find the minimum number of eigenpairs
containing at least $\gamma_{\min} I$ information by finding
the smallest number of eigenpairs $k_\compr$ such that \eqref{eq:DefGammaMin}
holds and then forget all eigenpairs with a larger index than $k_\compr$.

In case $k_{\max} > K$, some eigenpairs contain no additional information to
the prior knowledge and the information gain of the last eigenpair $\Delta I
(\mu_{k_{\max}})$ is zero.
Then we have stored all information ($\gamma = 1$) in the compressed data
with non-zero eigenvalue and Equation~\eqref{eq:DefGammaMin} is
automatically fulfilled for given $\gamma_{\min}$.

With the information fraction $\gamma$, we have found a quantification of how
much of the available information is stored in the compressed measurement
parameters.
For limited number of computed eigenpairs, one can still estimate $\gamma$ by
its upper and lower bounds.
Next we discuss some details of the computation of the input parameters of BDC
and its final implementation.

When evaluating \eqref{eq:Defmi} for the original posterior mean $m_\orig$ we have to avoid the inversion of the prior
covariance $S$ or the prior dispersion $\sqrt{S}$ as the explicit inversion of a $n\times n$ matrix is of $\mathcal O(n^3)$.
Such expensive operations can be partly avoided by getting $m_\orig$ via
\begin{align}
    m_\orig &= \sqrt S \widetilde{m}_\orig \nonumber \\
    &= \sqrt S \mathcal{D}_\orig \tilde{j}_\orig \nonumber \\
    &= \sqrt S (\mathds{1} + \sqrt{S}^\dagger R_\orig^\dagger N_\orig^{-1} R_\orig \sqrt{S} )^{-1} \sqrt{S}^\dagger R_\orig^\dagger N_\orig^{-1} d_\orig.
\end{align}
Compared to directly calculating $m_\orig$ with Equations~\eqref{eq:Defmi} and
\eqref{eq:DefDi}, this saves one inversion of the prior covariance $S^{-1}$.
Still the linear operator $(\mathds{1} + \sqrt{S}^\dagger R_\orig^\dagger N_\orig^{-1} R_\orig \sqrt{S} )$ needs to be inverted.
In this case we can make use of the conjugate gradient algorithm which computes the application of the inverse of a matrix to a vector in $\mathcal O (n)$.

The eigenvalue problem \eqref{eq:MoSx=mux} can then be solved by an Arnoldi
iteration~\cite{ARPACK}.

This leaves us with basic BDC as summarized in Algorithm~\ref{Alg:PatchComp}.
We first compute the original posterior mean and prior covariance using the
prior dispersion.
Given the original data $d_\orig$, response $R_\orig$ and inverse noise
covariance $N_\orig^{-1}$,
we can compute the fidelity matrix to solve eigenvalue
problem~\ref{eq:MoSx=mux} for the $k_{\max}$ largest eigenvalues.
If a minimal amount of information fraction that shall be encoded in the
compressed data is specified, we can determine the largest index $k_\compr$
so that Equation~\eqref{eq:DefGammaMin} holds and only save those $k_\compr$
eigenpairs that carry that much information.
Then we normalize the eigenvectors with respect to the norm induced by the
prior covariance $\norm{\cdot}_S$ to finally determine the compressed
measurement parameters $(d_\compr, R_\compr, N_\compr)$ according to
Equations~\eqref{eq:Nc}, \eqref{eq:Rc} and \eqref{eq:dcFin}.

\begin{algorithm} 
    \caption{Basic Bayesian Data Compression}\label{Alg:PatchComp}
  \begin{algorithmic}[1]
      \Procedure{compress}{$\sqrt S$, $R_\orig$, $N_\orig^{-1}$, $d_\orig$,
      $k_{\max}$, $\gamma_{\min}$}
    \State $m_\orig = \sqrt S (\mathds{1} + \sqrt{S}^\dagger R_\orig^\dagger N_\orig^{-1} R_\orig \sqrt{S} )^{-1} \sqrt{S}^\dagger R_\orig^\dagger N_\orig^{-1} d_\orig$
    \State $S = \sqrt{S} \sqrt{S}^\dagger$
      \State compute largest eigenpairs $(\mu_i^2, r_i)_{i=1}^{k_\mathrm{max}}$
      of $R_\orig^\dagger N_\orig^{-1} R_\orig S$
      \State find smallest $k_\compr$, such that \eqref{eq:DefGammaMin} holds.
      \For{every $i > k_\compr$}
    \State forget $(\mu_i^2, r_i)$
    \EndFor
      \For{every $i \leq k_\compr$}
    \State $r_i \leftarrow \frac{r_i}{\sqrt{r_i^\dagger S r_i}}$
    \EndFor
      \State $R_\compr = \sum_{i=1}^{k_\compr} e_i r_i^\dagger$, with unit
      vectors $\{e_i\}_{i=1}^{k_\compr}$
      \State $N_\compr^{-1} = \mathrm{diag}((\mu_i^2)_{i=1}^{k_\compr})$
    \State $d_\compr = (R_\compr S R_\compr^\dagger + N_\compr) (R_\compr S R_\compr^\dagger)^{-1} R_\compr m_\orig$
    \State \Return{$d_\compr$, $R_\compr$, $N_\compr^{-1}$}
    \EndProcedure
  \end{algorithmic}
\end{algorithm}

In the linear scenario the full Wiener filter needs to be solved.
Thus, the computational resources required to compute and store the compressed
measurement parameters exceed the resources saved by the compression.
It would be of benefit in a real world application if the eigenfunctions could
be re-used in repetitions of the same measurement and do not need to be
computed again.
BDC's main benefit lies in the nonlinear scenario
with a nonlinear response inside the measurement equation.
There, the inference appears to be more complicated,
but BDC enables us to exploit information stored in the data further
while calling the original data and response less often.

\section{Generalizations}\label{sec:Generalizations}
\subsection{Generalization to Nonlinear Case}\label{sec:NonlinearCase}

The derivation of BDC so far is based on a linear measurement equation.
In real world problems, however, often nonlinear measurement equations
\begin{align}
    d = R(s) + n
\end{align}
describe the relation of signal and data.
There, the response transforms the signal nonlinearly.
In addition, the signal parameters can be very non-Gaussian and inter-dependent through a deep
hierarchical model.
In those cases we need to adjust basic BDC\@.
Unlike basic BDC in the linear case, the adjusted BDC in nonlinear
scenarios can save computation time as the original data and response do not
need to be called as often as in the full reconstruction.

Let us assume that such complications are expressed via a deep hierarchical model.
Following~\cite{Knollmueller2018EncodingPriorKnowledge}, deep hierarchical models
can be transformed into independent standard normal distributed parameters
by encoding prior knowledge into the likelihood.
In this fashion the complexity of a deep hierarchical model is stored in a nonlinear function $f$.
This function relates the parameters $s$ of the hierarchical model -- the actual signal
-- to the parameters $\xi$ of a transformed, flattened, non-hierarchical model via
\begin{align}
  s = f(\xi).
\end{align}
The transformation $f$ has to be chosen such that the prior of $\xi$ becomes a standard normal distribution:
\begin{align}
  \P(d|s) \P(s) \mathrm{d}s= \P(d|f(\xi)) \mathscr{G}(\xi, \mathds{1}) \mathrm{d} \xi.
\end{align}
Thus, we call $\xi$ excitation field as defined in
Section~\ref{sec:InformationGain}.

For flattened models we can now deal with nonlinear measurement setups using Metric Gaussian
Variational Inference (MGVI)~\cite{Knollmueller2019MGVI}.
There, the posterior is approximated by a Gaussian $\mathscr G(\xi - \bar{\xi},
\Xi)$ with inverse Fisher information metric as uncertainty covariance $\Xi$
centered on some mean value $\bar{\xi}$.
The Fisher information metric is
\begin{align}
  M_{d|s} = \left\langle \frac{\partial \mathcal{H}(d|s)}{\partial
    s^\dagger} \frac{\partial \mathcal{H}(d|s)}{\partial s}
    \right\rangle_{P(d|s)}.
\end{align}
Here
\begin{align}
  \mathcal{H}(d|s) := - \ln \P(d|s)
\end{align}
is the information Hamiltonian of the likelihood.
In order to distinguish the approximate uncertainty $\Xi$ from the true posterior uncertainty covariance,
we call it \emph{variational uncertainty}.
With a standard Gaussian prior, the posterior covariance then states
\begin{align}
    \Xi^{-1} &= J_{\bar{\xi}}^\dagger M_{d|s} J_{\bar \xi} +
    \mathds{1},
\end{align}
with Jacobian
\begin{align}
    J_\xi &:= \frac{\partial f(\xi)}{\partial \xi}.
\end{align}

For many measurement situations, the response splits into a linear part $R_\mathrm{lin}$ and a
nonlinear part $R_\mathrm{nl}$
\begin{align}\label{eq:NonLinMeasurementEqn}
  d_\orig = R_\mathrm{lin} R_\mathrm{nl}(s) + n_\orig.
\end{align}
The linear part might describe a linear telescope response,
or just be an identity operator.
Then we redefine our signal $s' := R_\mathrm{nl}(s)$.
Before, $s$ were the parameters of the hierarchical model with a function $f$
transforming the standard Gaussian distributed excitation $\xi$ into $s$.
Now $s'$ are our parameters being related to $\xi$ via $R_\mathrm{nl} \circ f$.
Thus, the Jacobian states
\begin{align}
  J'_\xi = \frac{\partial R_\mathrm{nl}(f(\xi))}{\partial \xi}
\end{align}
and we define
\begin{align}\label{eq:SqrtS}
  \sqrt{S'} := \left. \frac{\partial R_\mathrm{nl}(f(\xi))}{\partial \xi} \right|_{\xi = \bar{\xi}}.
\end{align}
This is a linearization of the nonlinear part evaluated at $\bar{\xi}$, a reference value of $\xi$,
\eg the current mean location provided by the MGVI algorithm,
such that
\begin{align}
    {J'_{\bar{\xi}}}^\dagger M_{d|\bar{\xi}} J'_{\bar{\xi}} = \sqrt{S'}^\dagger
    M_{d|\bar{\xi}} \sqrt{S'}.
\end{align}
The compression is then applied to the linear measurement equation
\begin{align}
  d_\orig = R_\mathrm{lin} s' + n_\orig
\end{align}
with given noise covariance $N_\orig$.
This way, we have all the ingredients for BDC to work in the nonlinear case as well.

During the inference process, the approximated mean~$\bar{\xi}$,
at which the linearization is evaluated, changes.
With updated knowledge also the compression input will change.
This suggests the following strategy:
\begin{enumerate}
    \item Compress the original measurement parameters with prior knowledge and original
        measurement parameters as input.
    \item Infer the posterior mean given the compressed measurement parameters.
        This will only be an approximate solution.
    \item Approximate the original posterior around the inferred mean and use
        it as the new prior to start again with the first step.
\end{enumerate}
We will call the number of compressions,
\ie the number of total repetitions of those three steps,
$n_\mathrm{comp}$, the number of MGVI minimization steps to infer the mean with
compressed data in between $n_\mathrm{rep}$.
In total the original data and response only have to be used as often as in
$n_\mathrm{comp}$ minimization steps,
while in total we reach $n_\mathrm{comp}\times n_\mathrm{rep}$
minimization steps exploiting the information in the data.
A crucial step will be to find the optimal exploration ($n_\mathrm{comp}$)
versus exploitation ($n_\mathrm{rep}$)
ratio as in~\cite{Vergassola2007Infotaxis}.

\subsection{Utilization of Sparsity}\label{sec:Patching}

High dimensional data are difficult to handle simultaneously.
For the eigenvalue problem of BDC, it is more efficient to solve a larger
number of lower dimensional problems.
For the signal inference, it is beneficial to ensure that the response
$R_\compr$ and noise $N_\compr$ of the compressed system are sparse operators.
This can be achieved by dividing the data into patches to be compressed separately.
For that we use the fact that not every data point carries information about all
degrees of freedom of the signal at once.
Data points that inform about the same degrees of freedom of the signal can
then be compressed together exploiting sparsity of the compressed measurement
directions.
This also has the advantage of lower dimensional eigenvalue problems to be
solved, saving computation time.
The separately compressed data of the patches as well as corresponding
responses and noise covariances are finally concatenated.

An example would be data and signal that are connected via a linear mask hiding
parts of the signal from the data as discussed in Section~\ref{sec:NonLinMock}.
If then the signal is correlated in space,
we can divide the data into patches which carry
information about the same patch in signal space.

Alternatively this method can be used to compress data online,
\ie while data is measured one can collect and process it blockwise as suggested by
\cite{Cai2017OnlineImaging}
such that the full data never has to be stored completely.
After each compression, the reconstruction of the signal takes the
concatenated measurement parameters, where the compressed response is now
sparse, and solves the inference problem altogether.

Mathematically speaking, we divide the original data $d_\orig$ into separated
sets of data ${ d_\orig }_i$ with responses ${R_\orig}_i$ and noise covariance
${N_\orig}_i$ for every patch $i$.
The responses ${R_\orig}_i$ are already sparse not covering the whole signal
space as such.
Then we compress those data sets separately leading to ${ d_\compr }_i$, ${
    R_\compr }_i$ and ${ N_\compr }_i$.
Concatenating them back again leads to the final measurement equation
\begin{align}
  \underbrace{
  \begin{pmatrix} d_{c1} \\ \vdots\\ d_{cn} \end{pmatrix}
  }_{d_\compr}
  =
  \underbrace{
  \begin{pmatrix} R_{c1} \\ \vdots \\ R_{cn} \end{pmatrix}
  }_{R_\compr}
  s
  +
  \underbrace{
  \begin{pmatrix} n_{c1} \\\vdots\\ n_{cn} \end{pmatrix}
  }_{n_\compr}
\end{align}
with noise covariance
\begin{align}
  N_\compr = \begin{pmatrix} N_{c1} \\ &\ddots \\ &&N_{cn}\\ \end{pmatrix}.
\end{align}
We call this process \emph{patchwise compression}.
If the compression of all original data is done at once, we call it
\emph{joint compression}.
With patchwise compression, signal correlations between data points of
different patches cannot be exploited for the compression.
One should aim to assign strongly correlated data to the same patches
such that their correlation is considered in BDC\@.
For data correlated in space, correlations are strongest between data for
neighbouring signal locations.
It makes sense to choose the patches by vicinity.
The computational benefit due to sparsity of the response contrasts the
information loss due to patching.
One can increase the dimension of the compressed data to compensate that loss and
still requires less storage capacity.

The signal is not affected by the patching.
Signal correlations are still represented via the signal prior covariance
$S$, and therefore also present in the compressed signal posterior.
Since the reconstruction is running over the full problem,
its result is not biased due to patchwise compression.
In principle, any kind of compression could be specified via introduction of
arbitrary $R_\compr$ and $N_\compr$ into Equation~\ref{eq:dc}.
The resulting reconstruction would all be unbiased, but of course, less
accurate.

To summarize, we separate the data into patches.
The data of every patch is compressed separately
leading to compressed measurement parameters for every patch.
Prerequisite for treating the patches separately is that the noise of
the individual patches is uncorrelated between the patches.
By concatenating the compressed measurement parameters of all patches, we get
all operators needed for the compressed signal posterior.
This removes the need to store the compressed responses over the entire signal
domains.
Only their patch values have to be stored,
saving memory and computation time.

\section{Application}\label{sec:Application}
Now, the performance of BDC is discussed for applications of increasing
complexity,
first for a linear synthetic measurement setting and then for a nonlinear one.
For the latter, we demonstrate the advantage of dividing the data into patches and compressing them separately.
Finally, the compression of radio interferometric data from the Giant Metrewave Radio Telescope (GMRT) is discussed.

\subsection{Synthetic Data: Linear Case}\label{sec:AplWF}

First, the BDC is applied to synthetic data in the Wiener filter context.
This means all probability distributions such as prior, likelihood and posterior
are Gaussian and the data are connected to the signal via a linear measurement
equation $d=Rs+n$.
In this setup, we can test basic BDC in its actual, not approximated form, for 
changing noise and masked areas.
Also, we compare it with the Bayesian analog of PCA (BaPCA), reducing the expected
data covariance to its principal components.

The signal domain is a one dimensional regular grid with 256 pixels.
The synthetic signal and corresponding synthetic data are drawn from a zero
centered Gaussian prior.
The data is masked, such that only pixels 35 to 45 and 60 to 90 are measured
linearly, according to $d=Rs+n$.
Additionally white Gaussian noise is added with zero mean and standard
deviation of $\sigma_n = 2 \cdot 10^{-3}$ for measurements up to pixel 79,
and $4 \cdot 10^{-3}$ for pixels 80 to 90.
Those noisy data are then compressed to four data points, from which the signal is inferred in a last step.

\begin{figure}
  \centering
    \inclgraphics{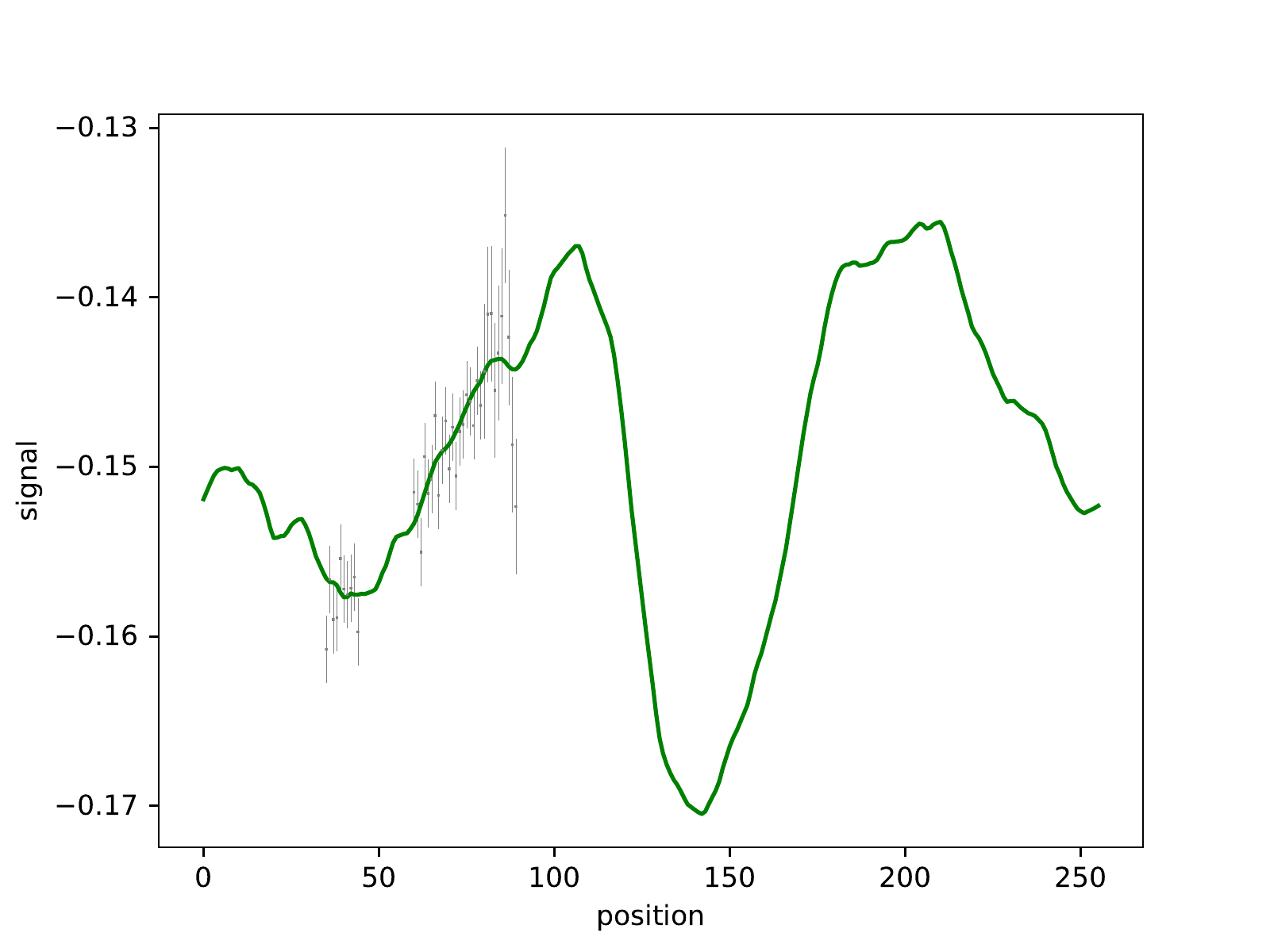}
  \caption{\label{fig:1DWienerFilterSetup}
    One dimensional synthetic data setup to test BDC\@.
    The synthetic signal is marked in green and the measured data in gray.
    Error bars at the data points mark the Gaussian noise standard deviation of
    $2 \cdot 10^{-3}$ for pixels before 79 and $4 \cdot 10^{-3}$ for pixels
    80 to 90.
    Those noisy data shall by compressed by BDC and used for reconstructing the
    synthetic signal back again.
  }
\end{figure}

The signal covariance is assumed to be diagonal in Fourier space,
with the power spectrum
\begin{align}
  P_s(k) := \frac{2 \cdot 10^4}{1 + \left(\frac{k}{20} \right)^4}.
\end{align}
The signal itself can be computed from the power spectrum via
\begin{align}
  s = \mathbb{F} \sqrt{P_s(k)} \xi_k
\end{align}
with a Fourier or Hartley transformation $\mathbb{F}$ and the Fourier modes
$\xi_k$ being drawn independently from a standard Gaussian $\mathscr{G}(\xi, \mathds{1})$.
The response is set to be a mask measuring pixels 35 to 45 and 60 to 90
directly, with a local and thereby unity response.
The measurement setup with signal mean, synthetic signal, and data are shown in Figure~\ref{fig:1DWienerFilterSetup}.

We apply basic BDC as described in Algorithm \ref{Alg:PatchComp}.
For the eigenvalue problem we use the implementation of the Arnoldi method in \texttt{scipy} (\texttt{scipy.sparse.linalg.eigs} \cite{Jones2001SciPy}).

\begin{figure}
  \centering
  \includegraphics[width=0.49\columnwidth]{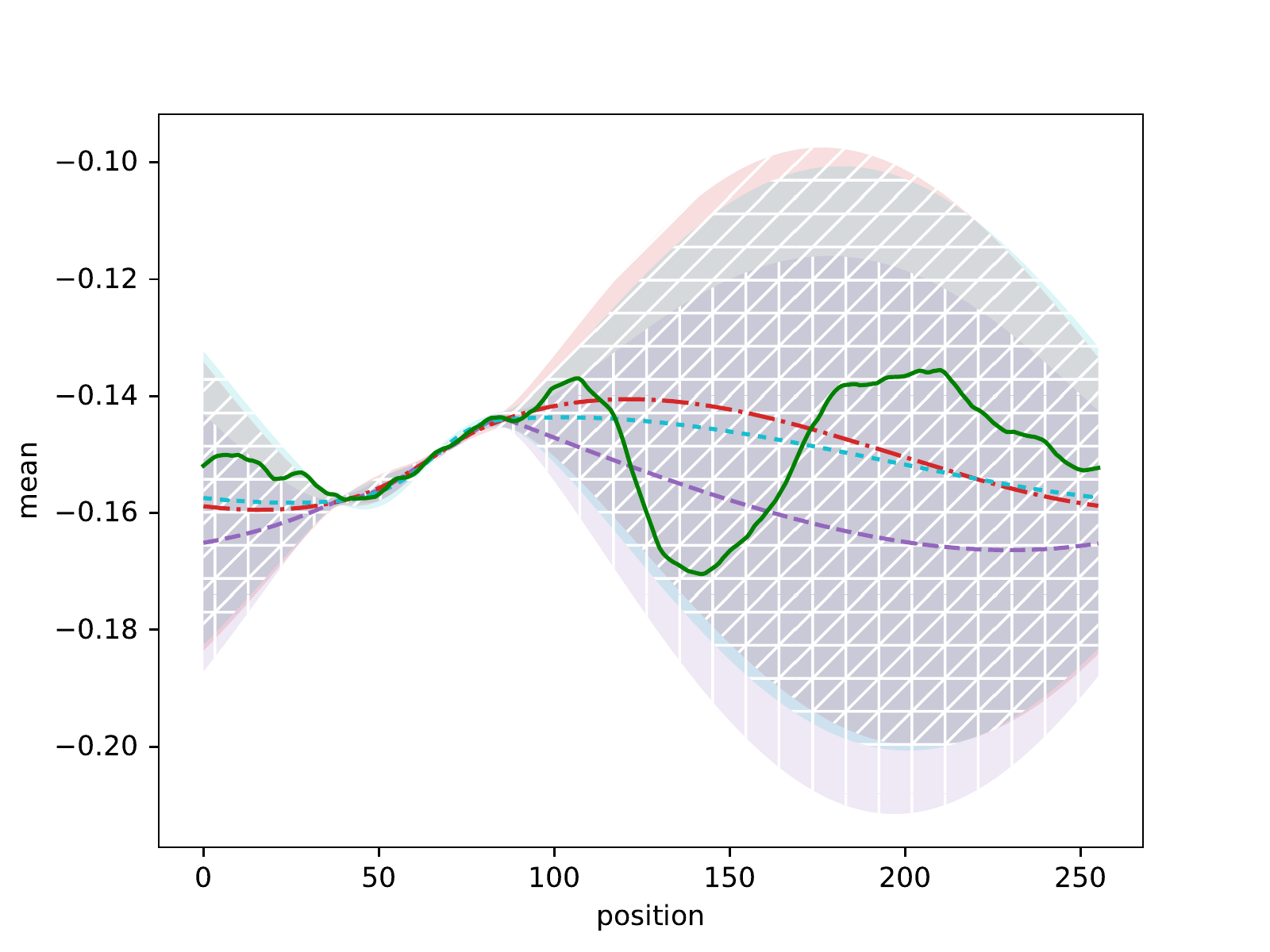}
  \includegraphics[width=0.49\columnwidth]{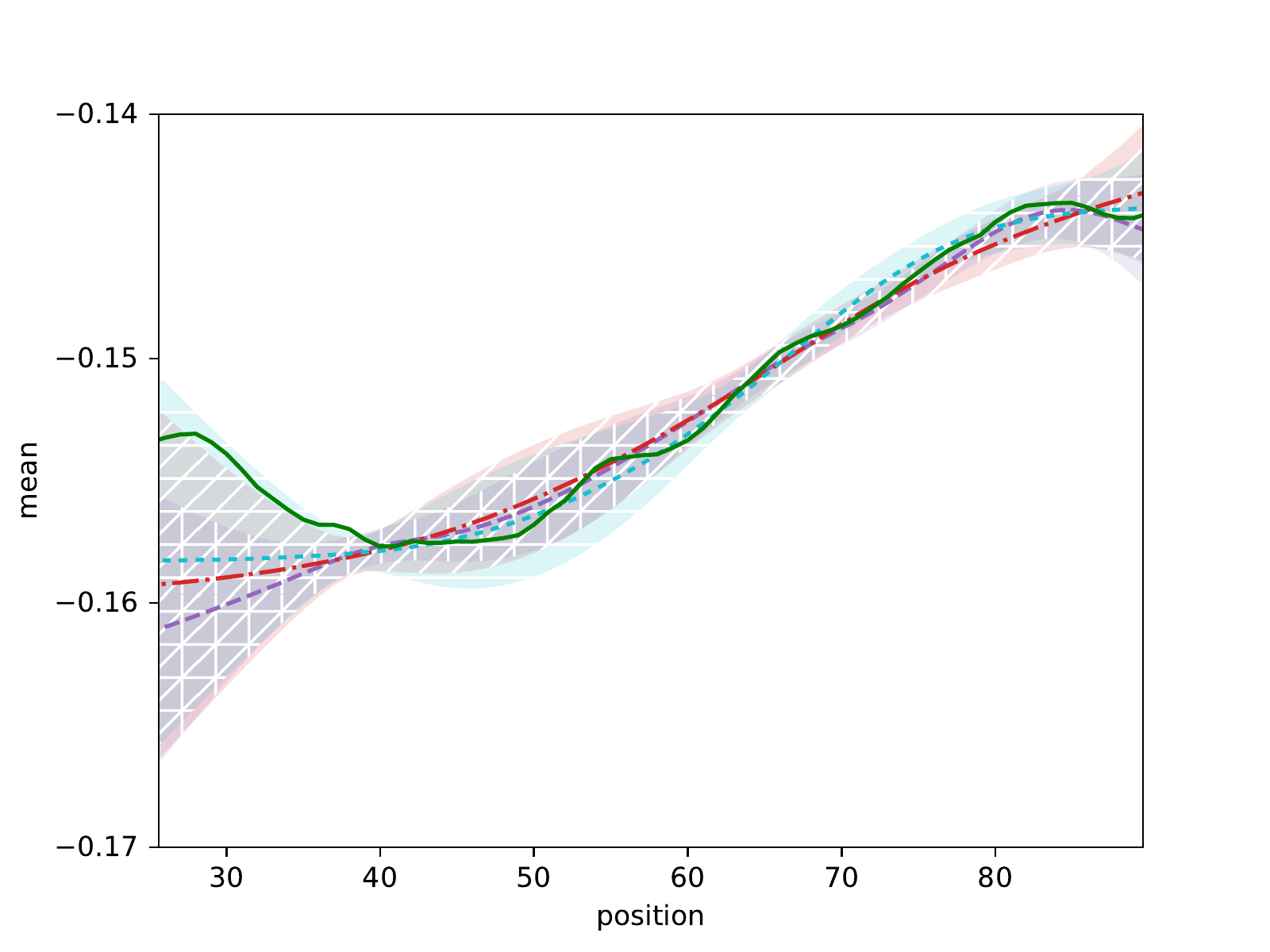}
  \caption{\label{fig:1DWienerFilterMeans}
    Inference of synthetic signal (green) in a linear Wiener Filter setup.
    The posterior mean given the original data shown in
    Figure~\ref{fig:1DWienerFilterSetup} is plotted as dashed
    purple line with vertically hatched shades marking its standard deviation.
    The same for the posterior mean given four compressed data
    points from BDC as dash-dotted red lines with diagonally hatched shades
    and for the posterior mean given four compressed data
    points from BaPCA as dotted cyan lines with horizontally hatched shades.
    The left graph shows the full domain, the right one a zoom in the measured
    area.
    The left and right subfigure show the full domain and the region constrained by data, respectively.
    For all methods the ground truth is consistent with the $1\sigma$ uncertainty.
  }
\end{figure}

After having compressed the data, we evaluate the reconstruction performance
using the compressed data.
With Equation~\eqref{eq:DefPosterior} the posterior can be calculated directly
from the compressed measurement parameters, and signal covariance.
The posterior mean and uncertainty for the original and the compressed data are
compared to the ground truth in Figure~\ref{fig:1DWienerFilterMeans}.
The original data has been compressed from 40 to 4 data points with a fraction
$\gamma$ of 83.7\% of the total information encoded in the compressed data. 
Especially at the measured areas, both the original and the compressed
reconstruction are close to the ground truth,
while the reconstructed means deviate from the ground truth at masked areas far
away from measured areas.
However, this deviation is still captured in the uncertainties.

\begin{figure}
  \centering
  \includegraphics[width=0.49\columnwidth]{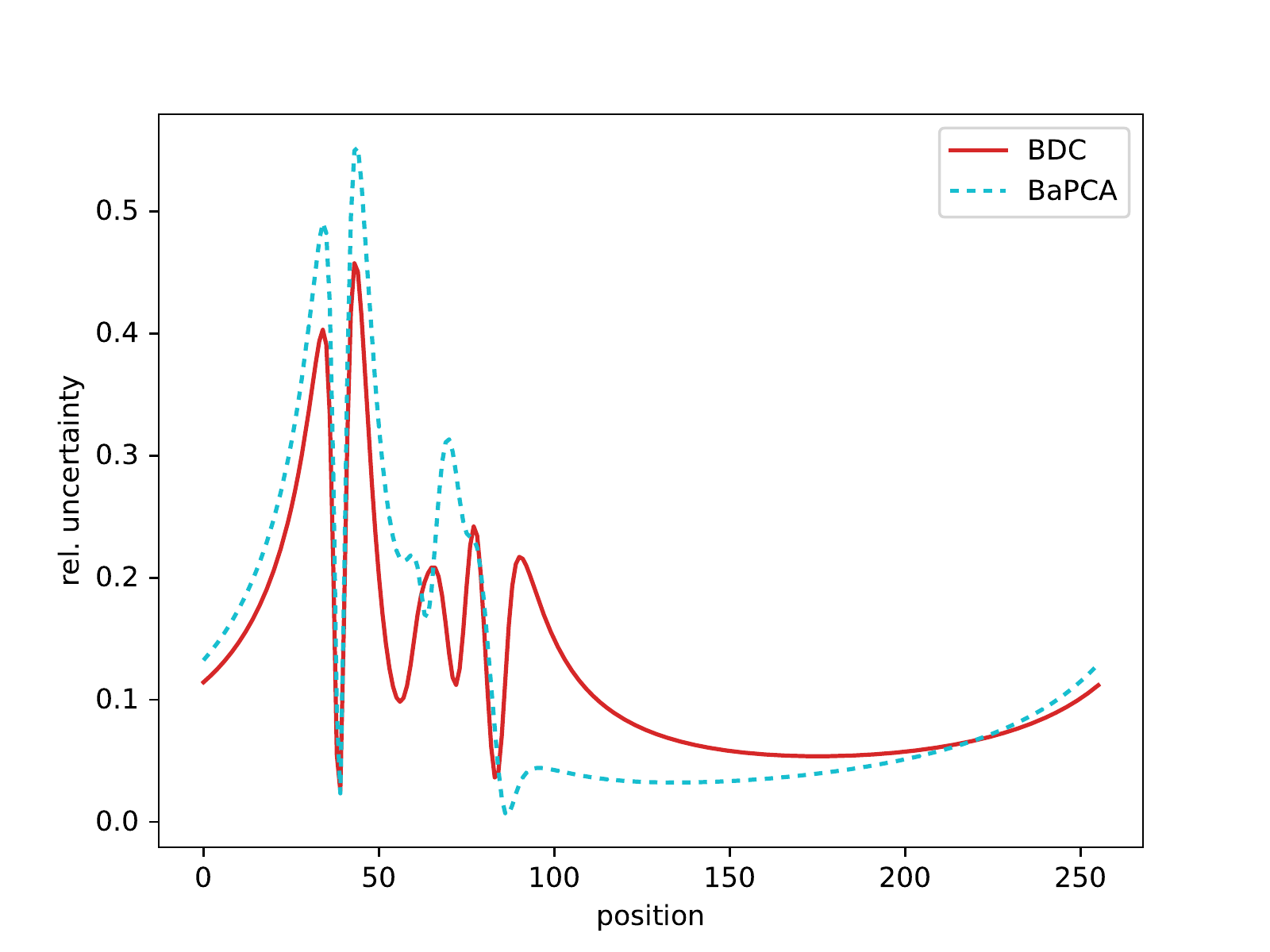}
  \includegraphics[width=0.49\columnwidth]{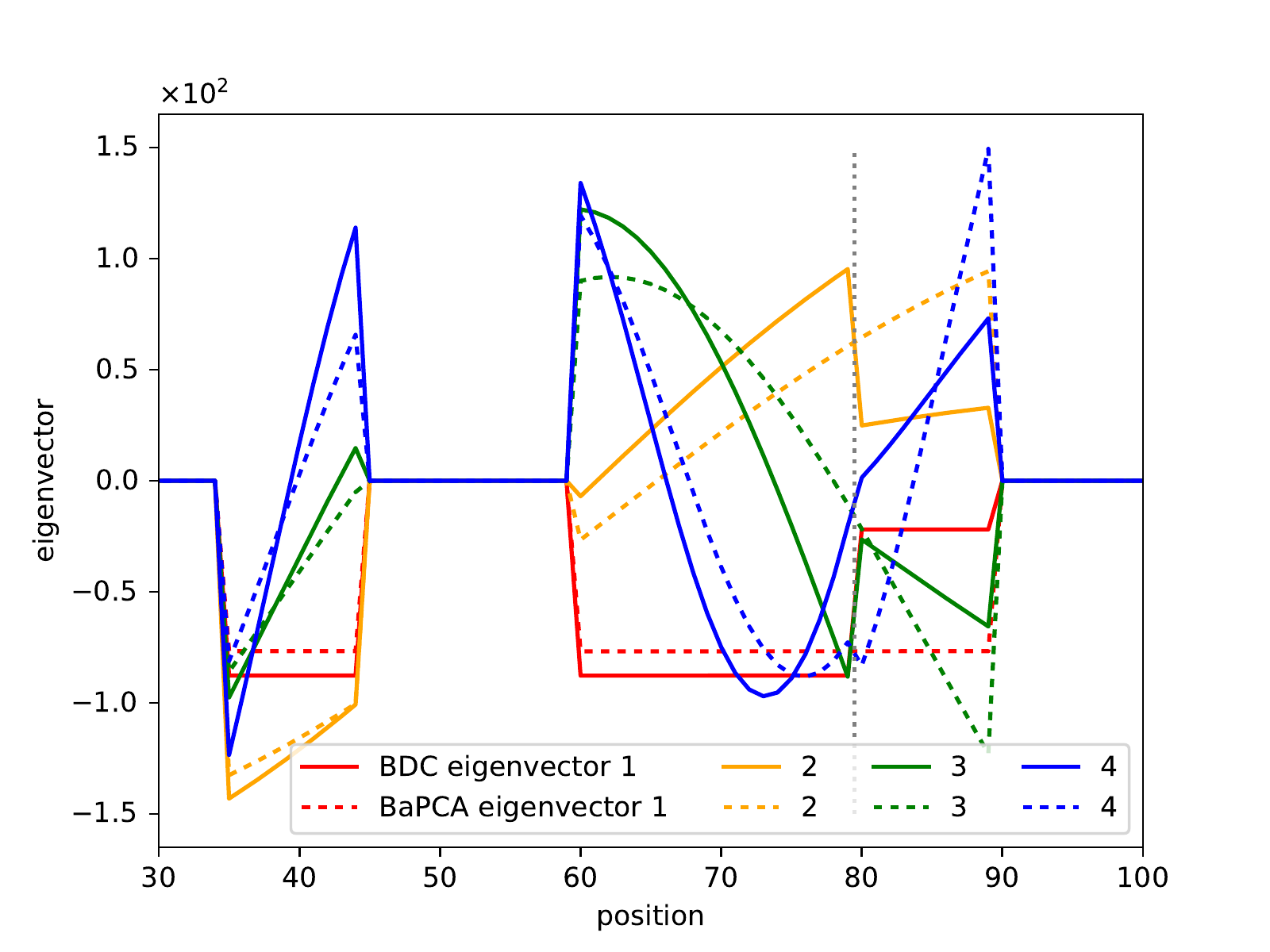}
  \caption{\label{fig:1DWienerFilterUncs}\label{fig:1DWienerFilterEigenVecs}
    Left: Relative uncertainty excess of the compressed reconstruction with
    respect to the one based on the original data in red.
    The same for the reconstruction from BaPCA compressed data with a dotted
    cyan line.
    Due to information loss the uncertainty of the BDC reconstruction is
    higher than the orignal one leading to positive values only.
    In the measured area, the relative uncertainty excess rises.
    The BaPCA reconstructed uncertainty has a similar slope with a higher
    uncertainty excess in regions with smaller noise standard deviation
    and smaller uncertainty excess in regions with higher noise standard deviation.\protect\linebreak
    Right: Relevant part of the eigenvectors of the eigenvalue problem~\eqref{eq:MoSx=mux}
    (BDC, solid lines)
    and of eigenvalue problem~\eqref{eq:BayesPCA} (BaPCA, dashed lines)
    for the synthetic data setup in
    Figure~\ref{fig:1DWienerFilterSetup} over the signal domain.
    The eigenvectors are colored in red, orange, green, and blue in descending
    order of their eigenvalues.
    The gray dotted line marks the noise level change.
    All eigenvectors are zero in the unmeasured area.
    In addition the ones from BDC decline in the area of higher noise while
    only the fourth eigenvector of BaPCA slightly changes its slope depending
    on the noise.
    A detailed discussion of the shape of BDC eigenvectors can be found in
    Appendix~\ref{sec:EigvecsWienerFilter}.
  }
\end{figure}

Figures~\ref{fig:1DWienerFilterMeans} and \ref{fig:1DWienerFilterUncs} show
that the compressed posterior has a higher variance than the original posterior.
Figure~\ref{fig:1DWienerFilterUncs} shows the relative uncertainty
difference of the compressed and original posterior,
\ie the compressed posterior uncertainty substracted from the
original posterior uncertainty divided by their mean at each pixel.
It is strictly positive in the measured areas as well as in the unmeasured
areas.
In the measured area compared to the unmeasured area,
the relative uncertainty excess of the compressed posterior is higher,
since there the absolute uncertainty is low.
Slight absolute increase of uncertainty there leads to a higher relative variation.
This proves the increase of uncertainty due to the compression.

The eigenvectors are plotted in Figure~\ref{fig:1DWienerFilterEigenVecs}.
At the masked pixels, the eigenvectors stay zero.
The changing noise covariance visibly impacts the shape of the eigenvectors.
Between pixels 79 and 80, where the noise increases, is a clear break in all the eigenvectors.
A higher noise standard deviation leads to abrupt drops in the eigenvectors.
A more detailed discussion of the eigenvectors can be found in
Appendix~\ref{sec:EigvecsWienerFilter}.
There we compare the shape of the eigenvectors in a simpler setup of a
continuous mask and constant noise to those of Chebyshev polynomials of the
first kind.
An analytical derivation of their form in this simple setting
is given in Appendix~\ref{sec:WFEigenVecDerivation}.

For comparison, we compressed the data of this setup with BaPCA defined by
eigenvalue problem~\eqref{eq:BayesPCA}
and performed a linear Wiener filter.
Note that in the original PCA, there is no noise defined.
We determined the largest four eigenvalues from Equation~\eqref{eq:BayesPCA}
and corresponing eigenvectors $v$.
Those eigenvectors were used as the row vectors of a transformaion $V$ which compresses the
original data.
The measurement equation for those compressed data then becomes
\begin{align}
    d_\text{pca} &= V d_\orig \nonumber \\
    &= V R_\orig s + V n_\orig.
\end{align}
Identifying $R_\text{pca} := V R_\orig$ as the response of this compressed
system and $n_\text{pca} := V n_\orig$ as its noise,
one can apply a linear Wiener Filter as in Equations~\eqref{eq:Defmi},
\eqref{eq:DefDi} and \eqref{eq:DefMi}.

We have plotted the corresponding reconstructions in
Figure~\ref{fig:1DWienerFilterMeans} with corresponding uncertainty as cyan
dotted line with horizontally hatched shades.
BaPCA reconstructs the mean and standard deviation similar to BDC.
For comparison we have also plotted the relative uncertainty in
Figure~\ref{fig:1DWienerFilterUncs} as we did for BDC as well as the eigenvectors
building the compressing transformation $V$.
The relative uncertainty from BaPCA clearly exceeds the one of BDC in areas of
low noise.
In the area of higher noise, the relative uncertainty excess from BaPCA is lower
than the one from BDC compared to the original posterior uncertainty.
The reason for this can be seen by comparing the eigenvectors of BaPCA and BDC:
BaPCA is more sensitive in high noise areas,
therefore having a lower posterior uncertainty there,
but also letting thereby more noise enter the compressed data.
Compared to BaPCA, BDC encodes more information in regions of lower noise, where
the data is more informative, 
and it keeps less information from regions of higher noise.

This can be seen when looking at the eigenvectors of both methods.
The amplitude of eigenvectors from BDC drop where the noise standard
deviation becomes higher.
For BaPCA, only the fourth eigenvector changes its amplitude in the region of
higher noise standard deviation.
The first three eigenvectors, however, do not vary their amplitude with 
noise change.
This coincides with the observation that BaPCA and BDC become equivalent for
constant noise standard deviation as discussed in
Section~\ref{sec:ExpectedOptimum}.

We found that the compression method reduces the dimension of the data
with minimal loss of information in the simple case of a linear 1D Wiener Filter inference.
Storing only four compressed data points still reconstructs the signal well,
compared to the reconstruction with original data.
Every compressed data point determines the amplitude of an eigenvector such that the signal is approximated appropriately.
The lossy compression leads to a slightly higher uncertainty, as information is lost.
In this application, BDC and BaPCA give similar results in terms of reconstruction.
Compared to a BaPCA, BDC focusses more on regions of lower noise standard
deviation, where the data are more informative.

\subsection{Synthetic Data: Nonlinear Case}\label{sec:NonLinMock}

Testing BDC on data from a nonlinear generated signal in two dimensions allows
us to verify the derivation of the nonlinear approximation in
Section~\ref{sec:NonlinearCase} and also test patchwise compression discussed
in Section~\ref{sec:Patching}.
Some nonlinear synthetic signal is generated and then inferred with the
original data, with compressed data, and with data, which has been first divided into patches and then compressed.
The results are discussed with respect to the quality of the inferred mean for the different methods, their standard deviation, their power spectrum, as well as the computation time.

\begin{figure*}
  \centering
    \includegraphics[width=0.49\columnwidth]{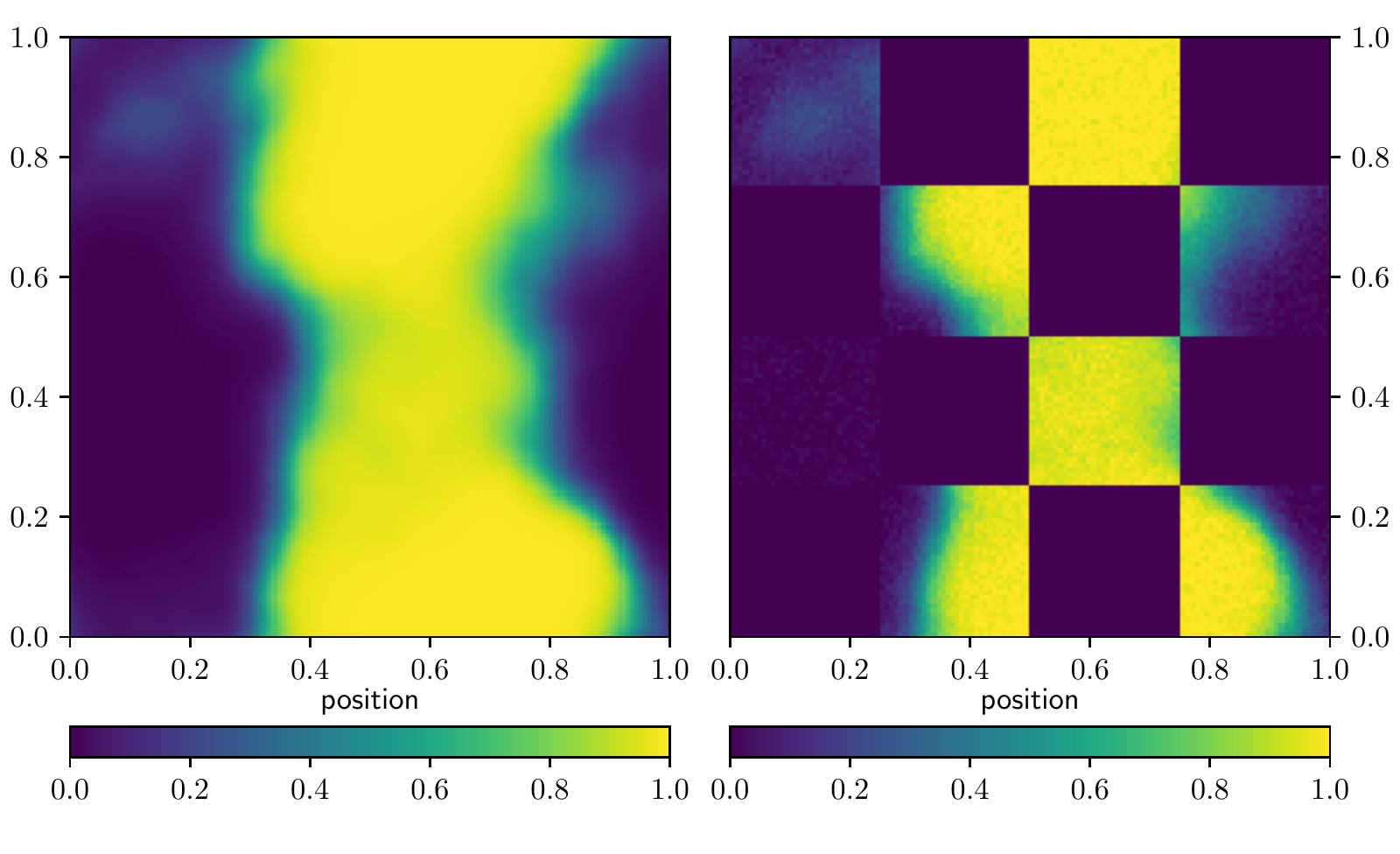}
    \includegraphics[width=0.49\columnwidth]{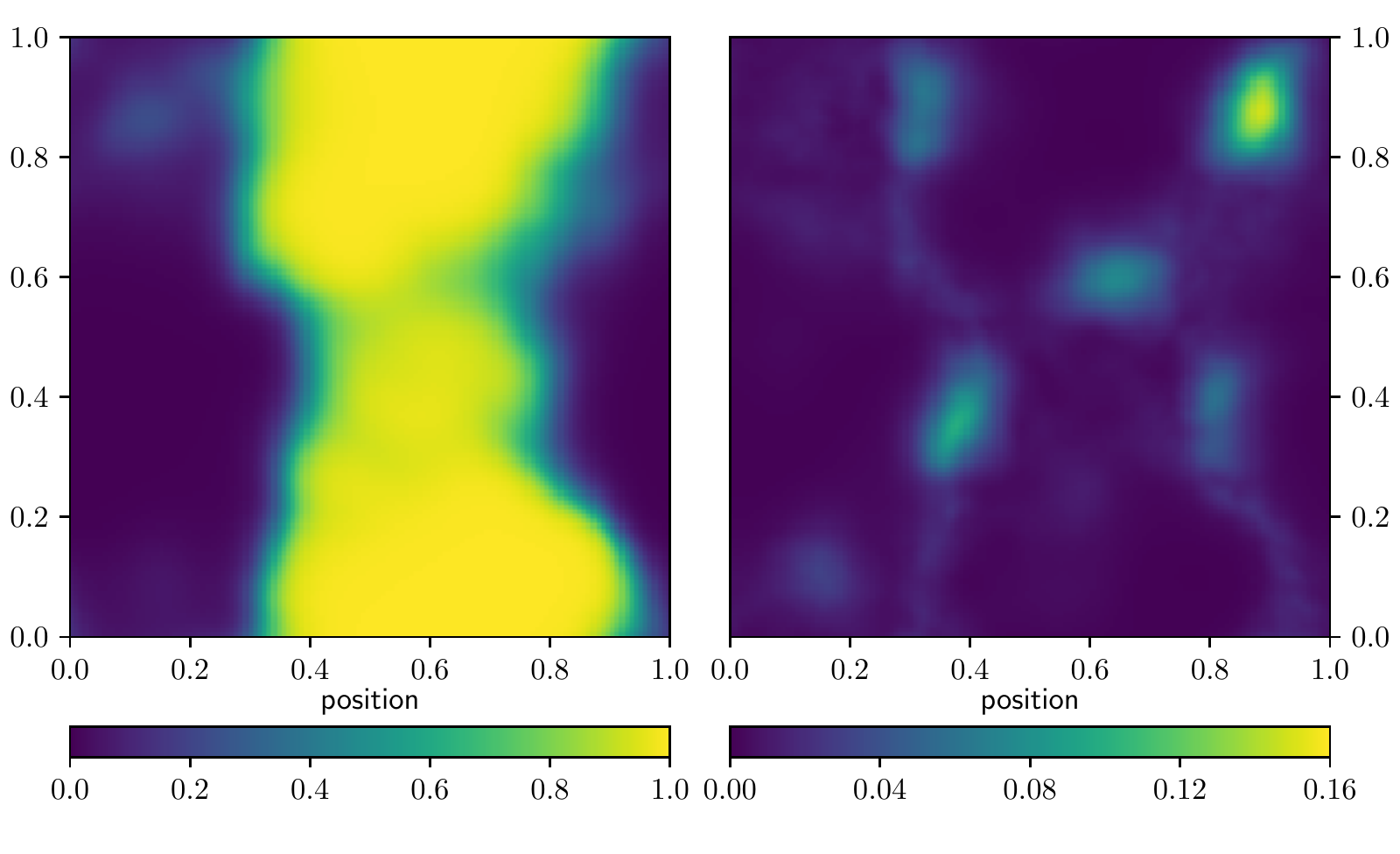}
    \includegraphics[width=0.49\columnwidth]{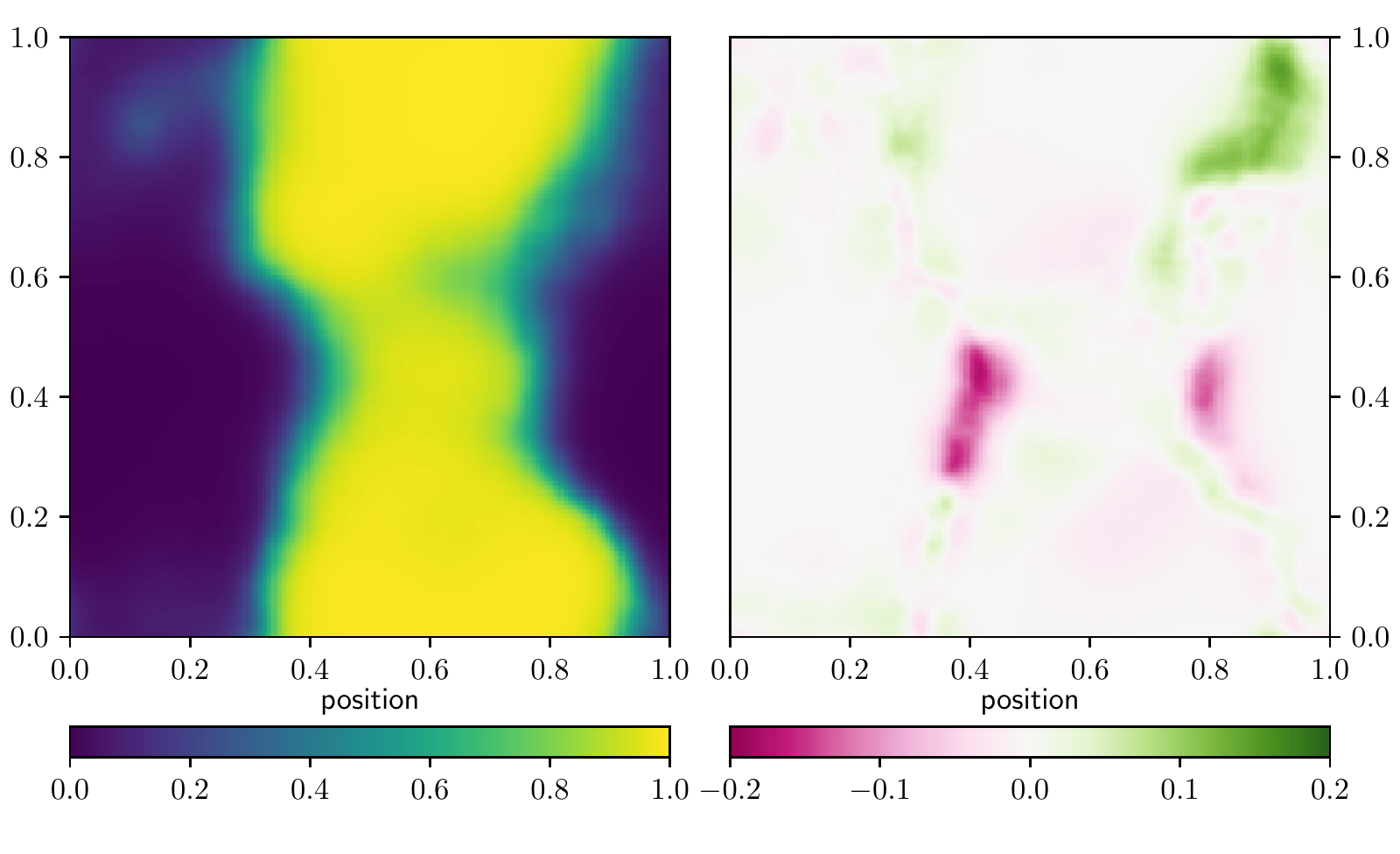}
    \includegraphics[width=0.49\columnwidth]{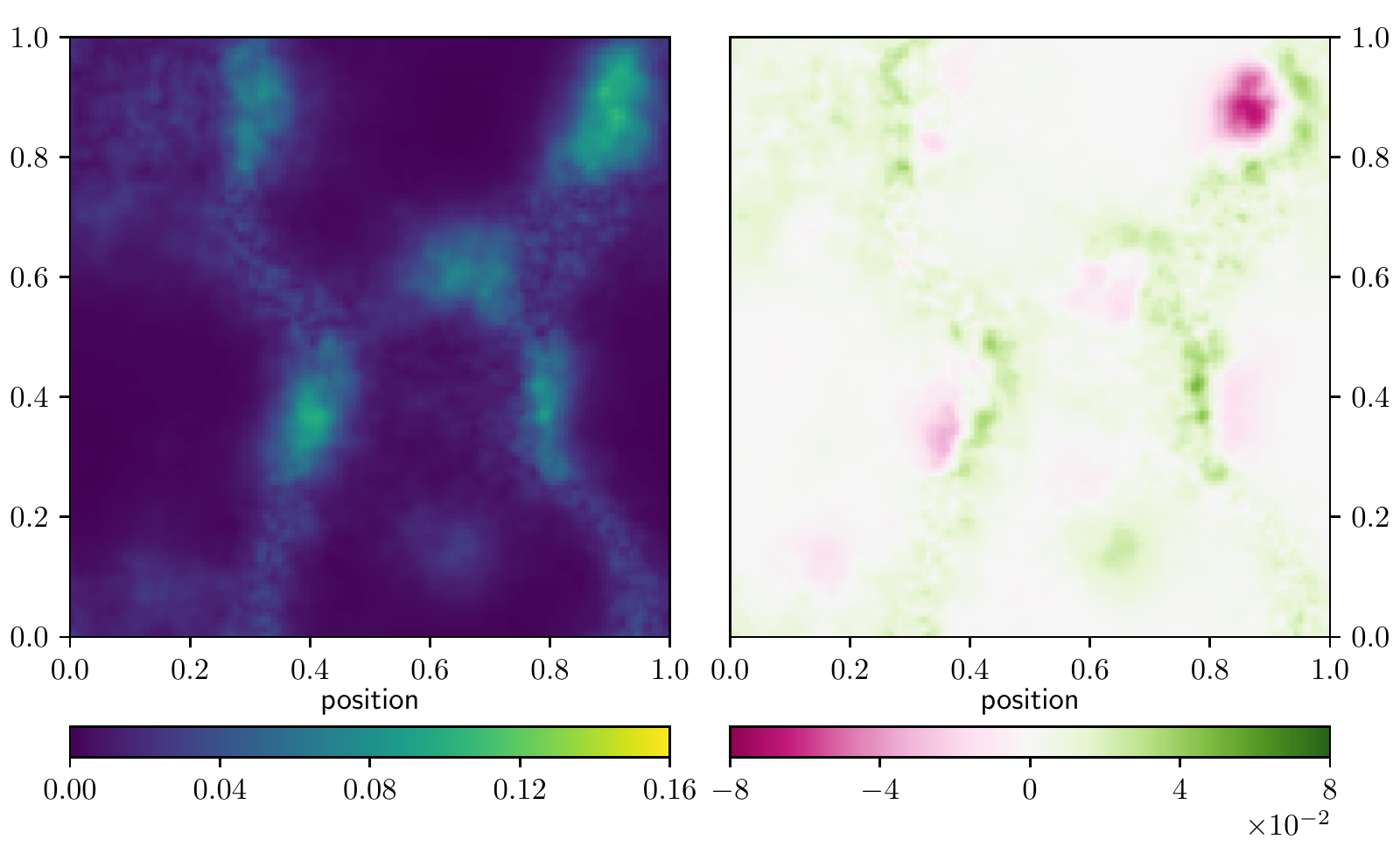}
    \includegraphics[width=0.49\columnwidth]{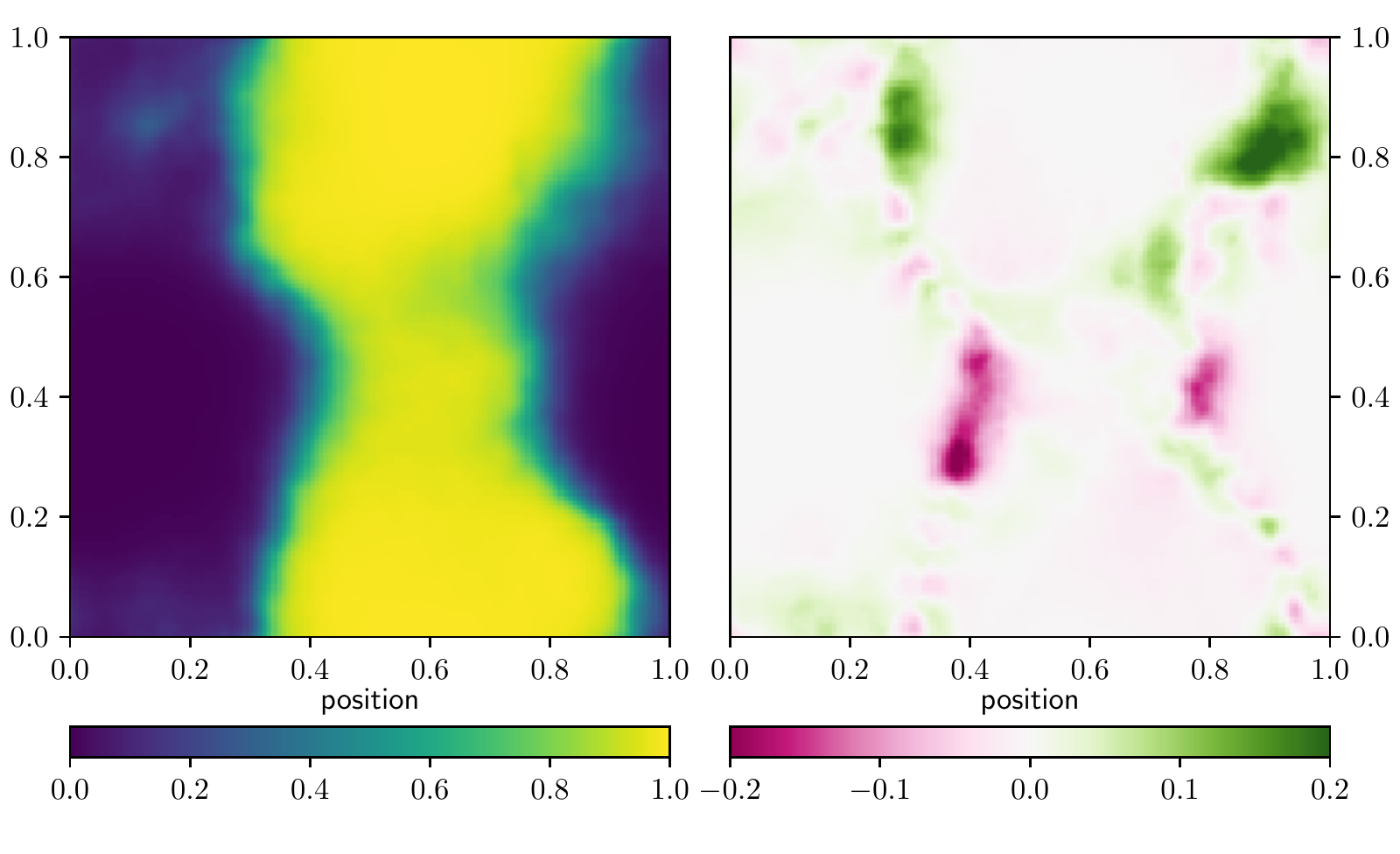}
    \includegraphics[width=0.49\columnwidth]{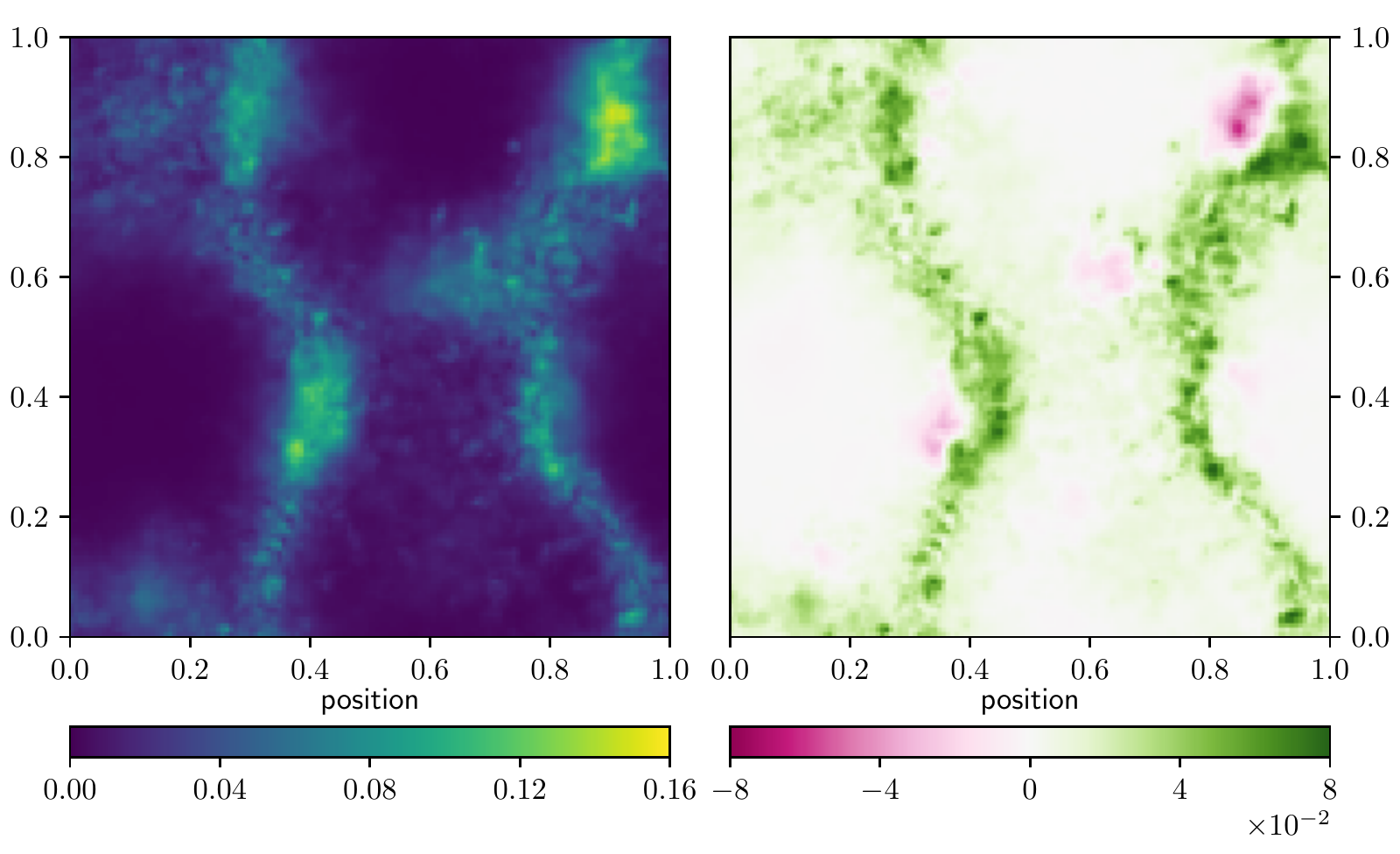}
  \caption{\label{fig:GroundTruthAndOriginalData}\label{fig:NLM-OriginalInference}\label{fig:NLM-CompressedMean}\label{fig:NLM-CompressedVariance}\label{fig:NLM-PatchCompMean}\label{fig:NLM-PatchCompVariance}
    \textit{Top row:}
    From left to right: First, the generated ground truth and second, the synthetic data for a synthetic signal created by the amplitude model and processed by a sigmoid function.
    A unity response connects the data and the signal with a four times four
    checkerboard mask hiding parts of the signal from the data.
    Third, using MGVI, the posterior mean of the signal and fourth, its
    uncertainty as inferred with the original data.
    This setup is used to test BDC\@.
    The reconstruction from original data serves as a reference for the reconstruction from compressed data.\protect\linebreak
    \textit{Middle row:}
    Results for jointly compressed data.
    Three compression and MGVI steps were performed.
    First, the posterior mean.
    Second, its difference to the mean of the inference without
    compression, \ie the compressed posterior mean subtracted from the
    original posterior mean.
    Third, the posterior variance of the inference with compressed data.
    And fourth, the same, subtracted from the variance of the posterior with the original data.
    The reconstructions from original and jointly compressed data deviate
    mainly at the edges of larger structures, at most at the order of $10^{-1}$.
    The compressed posterior uncertainty is higher than the one of the original
    posterior at most locations.
    \protect\linebreak
    \textit{Bottom row:}
    As in the middle, just with patchwise compressed data.
    As in the case of reconstruction from jointly compressed data, the
    posterior mean mainly deviates at the edges of large structures of the
    signal from the original posterior mean. Again, the standard deviation is
    mainly higher for the compressed posterior compared to the original one.
    }
\end{figure*}

The synthetic signal has been generated with a power spectrum created by a
nonlinear amplitude model as described in~\cite{Arras2019UnifiedRadioImaging}
deformed by a sigmoid function to create a nonlinear relation between signal
and data.
The code of the implementation can be found here:
\linktoGit.
The resulting ground truth lies on an $128 \times 128$ regular grid and is shown
on the top left most panel of Figure~\ref{fig:GroundTruthAndOriginalData}.
To test BDC on masked areas in a nonlinear setup as well,
this signal is covered by a four by four checkerboard mask with equally sized
$32 \times 32$ squares, as
displayed in the second top panel of Figure~\ref{fig:GroundTruthAndOriginalData}.
Additionally uncorrelated noise with zero mean and 0.02 standard deviation has been added.
From this incomplete and noisy data, the non-Gaussian signal as well as the
power spectrum of the underlying Gaussian process need to be inferred
simultaneously.
The results of the original inference are plotted in the third top panel of
Figure~\ref{fig:NLM-OriginalInference}.

After setting up the input parameters for BDC, the data was compressed from
8\,192 to $k_{\max} = 80$ data points altogether without sorting out less
informative data points.
Next Metric Gaussian Variational Inference~\cite{Knollmueller2019MGVI}
performed $n_{\mathrm{rep}}=2$ inference steps based on the compressed data,
each time finding a better approximation for the posterior mean and
approximating the posterior distribution again.
Then the original data was compressed another time using the current posterior mean as
the reference point $\overline{\xi}$.
This was done $n_{\mathrm{comp}}=3$ times in total.
To determine the amount of information contained in the resulting compressed data
another run has been started where 4096 eigenpairs have been computed.
This way the estimation of $\gamma$ is more exact using Equation~\eqref{eq:DefGammaMin} for
a lower bound $\gamma_{\min}$ and \eqref{eq:DefGammaMax} for an upper bound
$\gamma_{\max}$.
With this we estimate that the compressed data of size 80 contains $31.4\%$ to $32.2\%$ of the information.
The same way, one gets that 672 compressed data points contain $80\%$ of
the information.
It turns out that already 80 compressed data points contain enough information
to reconstruct the essential structures of the signal as one can see from
reconstruction and difference maps shown in Figure~\ref{fig:NLM-CompressedMean}.

The corresponding posterior mean is plotted in the center left of
Figure~\ref{fig:NLM-CompressedMean} together with the difference to the
originally inferred mean.
Overall the compression yields similar results.
Deviations appear at the edges of homogeneous structures,
while deviations inside homogeneous structures are neglectable.
The variance is plotted in the center right of Figure~\ref{fig:NLM-CompressedVariance}.
Again it differs mainly at the edges.
Since during the compression process information is lost, the results should have higher uncertainty in general.
This is almost everywhere the case, however, there are some parts, which report a better significance than the reconstruction without compression.
This either implies an inaccurateness of BDC or that BDC can partly compensate the approximation MGVI brings
into the inference
by providing it with measurement parameters that are better formatted for its
operation.

Let us discuss in more detail how such a high loss in information still
reproduces reasonable results.
The information loss is equivalent to a widening of the
posterior distribution.
Its quantitative value in terms of $(1-\gamma)$ does not take into account on
which scales the information is mainly lost.
As can be observed, a substantial fraction of the measurement information
constrains small scales.
Losing this part does not make such a large difference to the human eye,
in particular as the small scale structures are of smaller amplitude,
but the information loss is measured on relative changes.
Thus a loss of 70\% of  mainly small scale information is possible without increasing
the error budget significantly.

Now we investigate patchwise compression in the same setup.
The data in each of the eight measured squares were compressed to $k_{\max} =
10$ data points separately.
In total there are as many compressed data points for the patchwise compression as
for the joint one.
We can use that the response only masks the signal but does not transform it.
Thus, we can compress the data with prior information of the corresponding patch only.
This way we reduce the dimension of the eigenvalue problem \eqref{eq:MoSx=mux} to the size
of the patch.
Before the reconstruction, the resulting compressed responses are expanded to the full
signal space in a sparse form and concatenated as described in Section~\ref{sec:Patching}.
For the reconstruction, the whole signal is inferred altogether.
The resulting mean and variance of the inference for this method and their difference to the
original ones are shown in the lower part of Figure~\ref{fig:NLM-PatchCompMean}.
Both differ mainly at the edges of homogeneous structures from the original
posterior mean and variance.
Also for the case of patchwise compressed data, the variance at some points
becomes smaller than for the original inference.
Since patchwise BDC does not use knowledge about correlations between the
patches for the compression,
it compresses less optimal than joint BDC,
and thus mostly has a higher deviation of the mean and higher uncertainty in the
reconstruction.
Like in the case of compressed data, we improve the estimation of $\gamma$ by
computing 4096 eigenpairs, \ie 512 eigenpairs per patch.
It turns out that for every patch on average $34.8\%$ to $35.4\%$ of
information is kept about the signal inside this patch
when using the 10 most informative eigenpairs for the compressed data points,
where all patches but two contain $15\%$ to $25\%$.
Counting from left to right and top to bottom, 
the data compressed from the second and fifth patch with very homogeneous signal
contain more than $75\%$ of the information.
When comparing the $\gamma$ values of each patch,
one needs to consider that the patches are compressed individually.
Therefore the information of one and another patch might be partly redundant
and their individual $\gamma$s can not just be added or averaged in order to get the joint information
content.

\begin{figure}
  \centering
    \includegraphics[width=0.49\columnwidth]{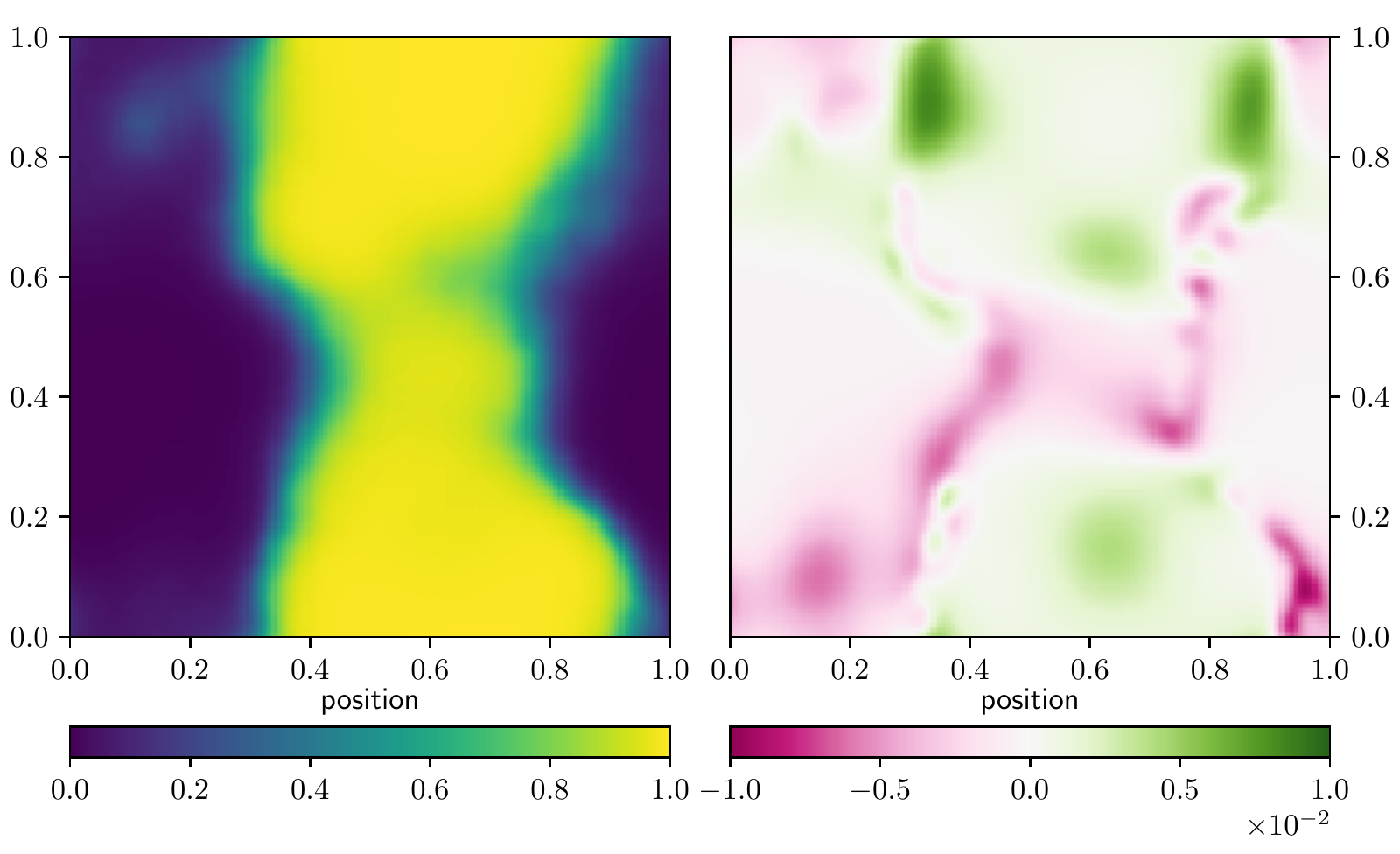}
    \includegraphics[width=0.49\columnwidth]{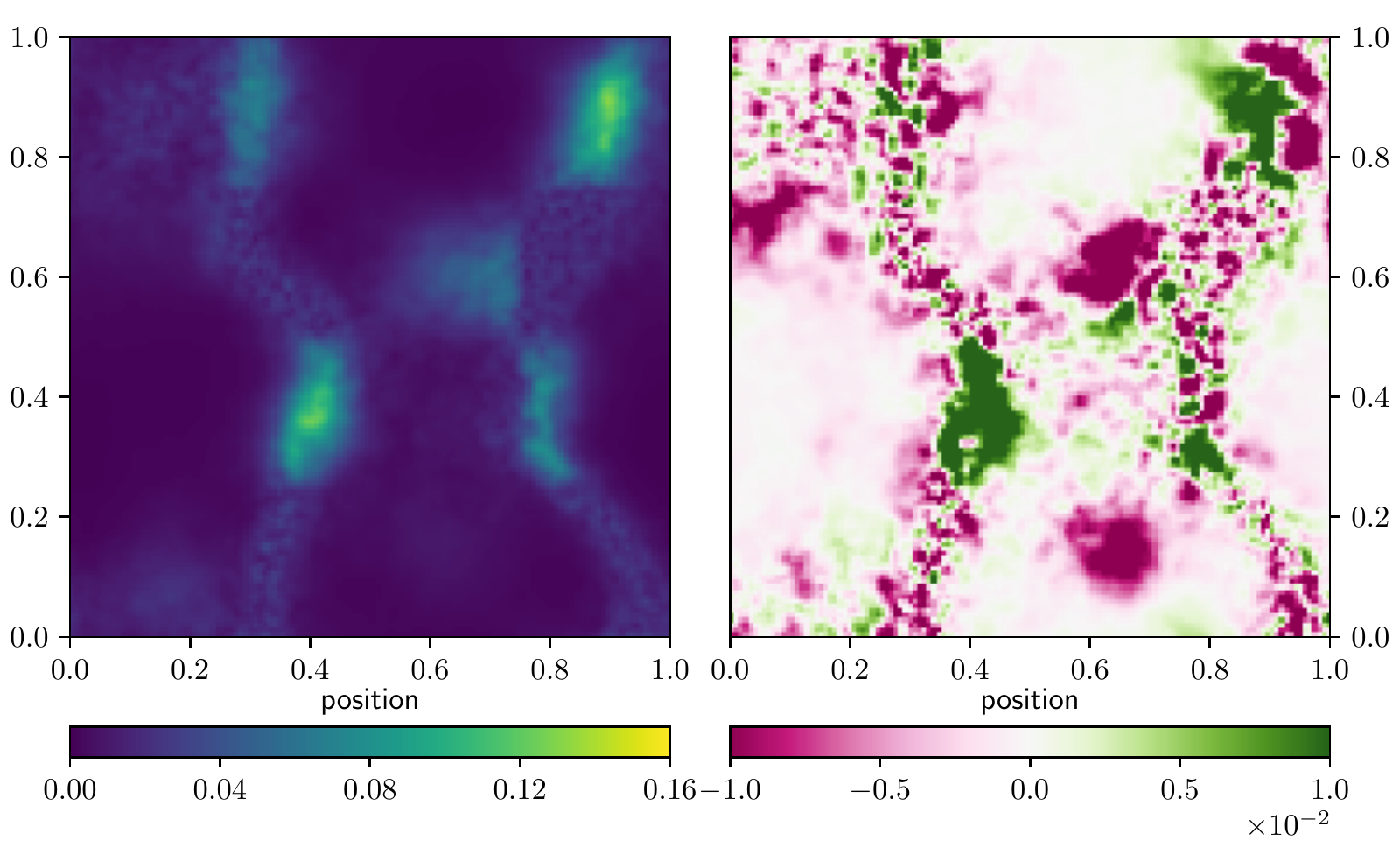}
  \caption{\label{fig:NLM-LinC}
    From left to right: Mean computed from the compressed measurement
    parameters directly from Equation~\eqref{eq:Defmi} without another
    minimization using MGVI\@.
    Second, its difference to the mean computed from the compressed data by MGVI as
    in Figure~\ref{fig:NLM-CompressedMean}.
    Third and forth, the same for the uncertainty computed from
    Equation~\eqref{eq:DefDi}.
    The structures are maintained well.
    Deviations are at the order of $10^{-2}$.
    This is one magnitude lower than the standard deviation.
  }
\end{figure}

After having computed the compressed measurement parameters, we can also
directly get the compressed posterior mean and covariance from
Equations~\eqref{eq:Defmi} and \eqref{eq:DefDi}.
The inference from the compressed data then reduces to a linear Wiener Filter
problem.
The resulting mean and variance are plotted in Figure~\ref{fig:NLM-LinC}
together with their difference to the mean and variance using one more MGVI inference.
Both deviate at the order of $10^{-2}$ which is one magnitude lower than the
uncertainty.
This illustrates that the compression helped to linearize the inference problem
around the posterior mean.

\begin{figure}
  \centering
    \inclgraphics{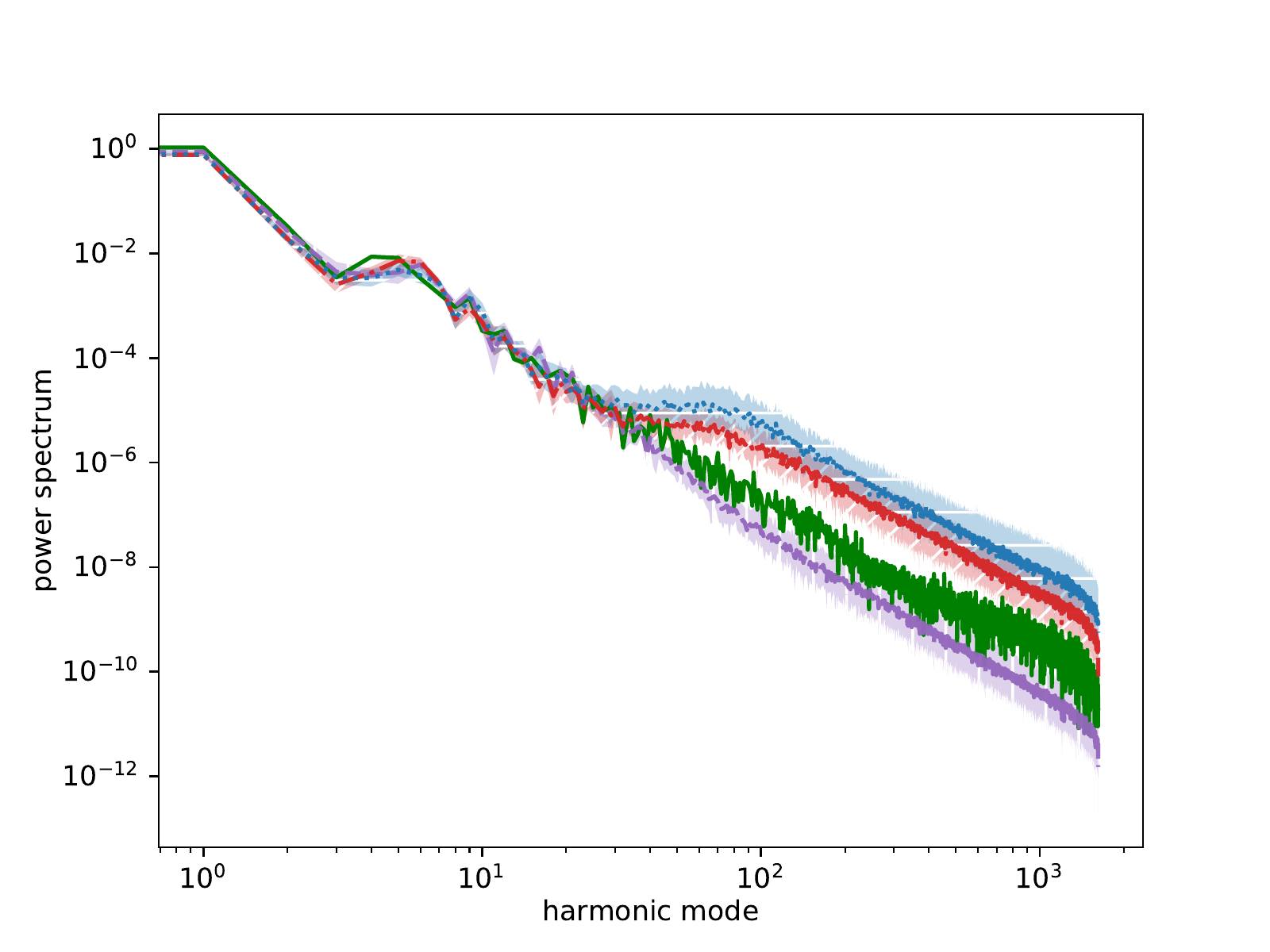}
  \caption{\label{fig:NLM-PowSpecs}
    The power spectrum of the underlying Gaussian process:
    ground truth (solid, green) and inferred from
    original (purple),
    jointly compressed (red)
    and patchwise compressed (blue) data.
    Qualitatively all power spectra reconstruct the ground truth well for small
    harmonic modes while they deviate at higher modes.
    There, the power spectrum reconstructed from original data tends to
    underestimate the amplitude of the original power spectrum while the
    power spectra reconstructed from jointly compressed data overestimate it.
    Compared to the joint compressed power spectrum, the patchwise compressed
    power spectrum deviates even more for higher harmonic modes.
    This inaccuracy contrasts the computational benefit of patchwise
    compression due to utilizing sparsity. 
  }
\end{figure}

Finally the results of the methods can be compared by looking at the inferred
power spectra of the underlying Gaussian process in Figure~\ref{fig:NLM-PowSpecs}.
All of the reconstructions recover the power spectrum well for high harmonic
modes up to the order of $10^1$.
For higher modes, the samples of the originally inferred power
spectrum tend to lie below the ground truth.
In contrast, the reconstructions of the two compression methods
overestimate the power spectrum for higher modes.
It is not completely clear why this is the case.
For higher modes, the signal to noise ratio is low.
In those regimes it is more difficult to reconstruct the power spectra.
This could be a reason for the deviation on high harmonic scales.
In addition variational inference methods tend to underestimate
uncertainties~\cite{Blei2017VariationalInference}.
Since we use MGVI for the reconstruction for all methods, this could cause the
inferred power spectra not to coinside within their uncertainties.

\begin{figure}
  \centering
    \includegraphics[width=0.49\columnwidth]{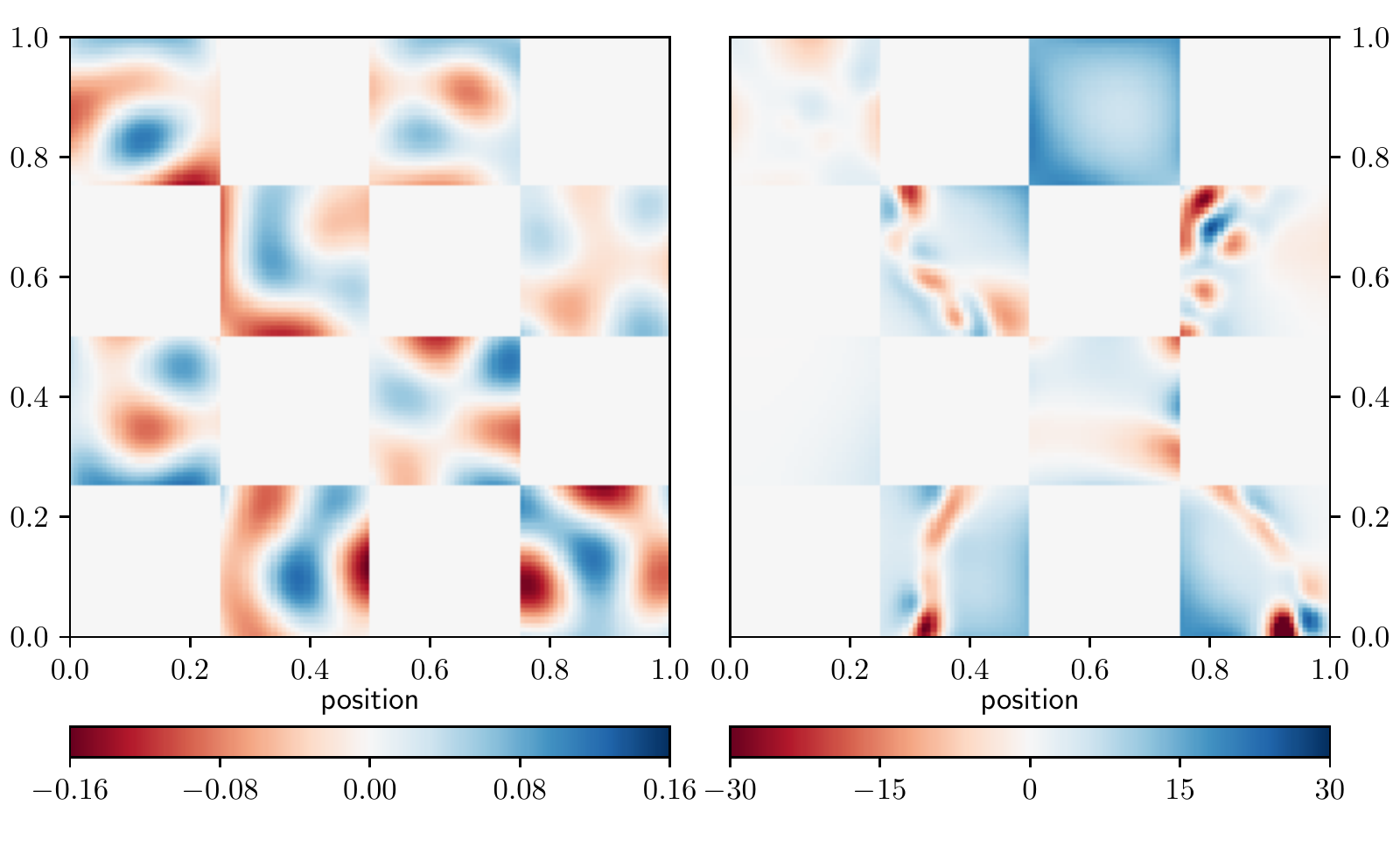}
  \caption{\label{fig:NLM-BackProjectionComp}
    The back projected jointly compressed data before (left) and after (right) the inference.
    Before, the structure is rather uniformly distributed while the
    back projection of the compressed data clearly focuses on the edges of the
    signal structures after the inference.
  }
\end{figure}

\begin{figure}
  \centering
    \includegraphics[width=0.49\columnwidth]{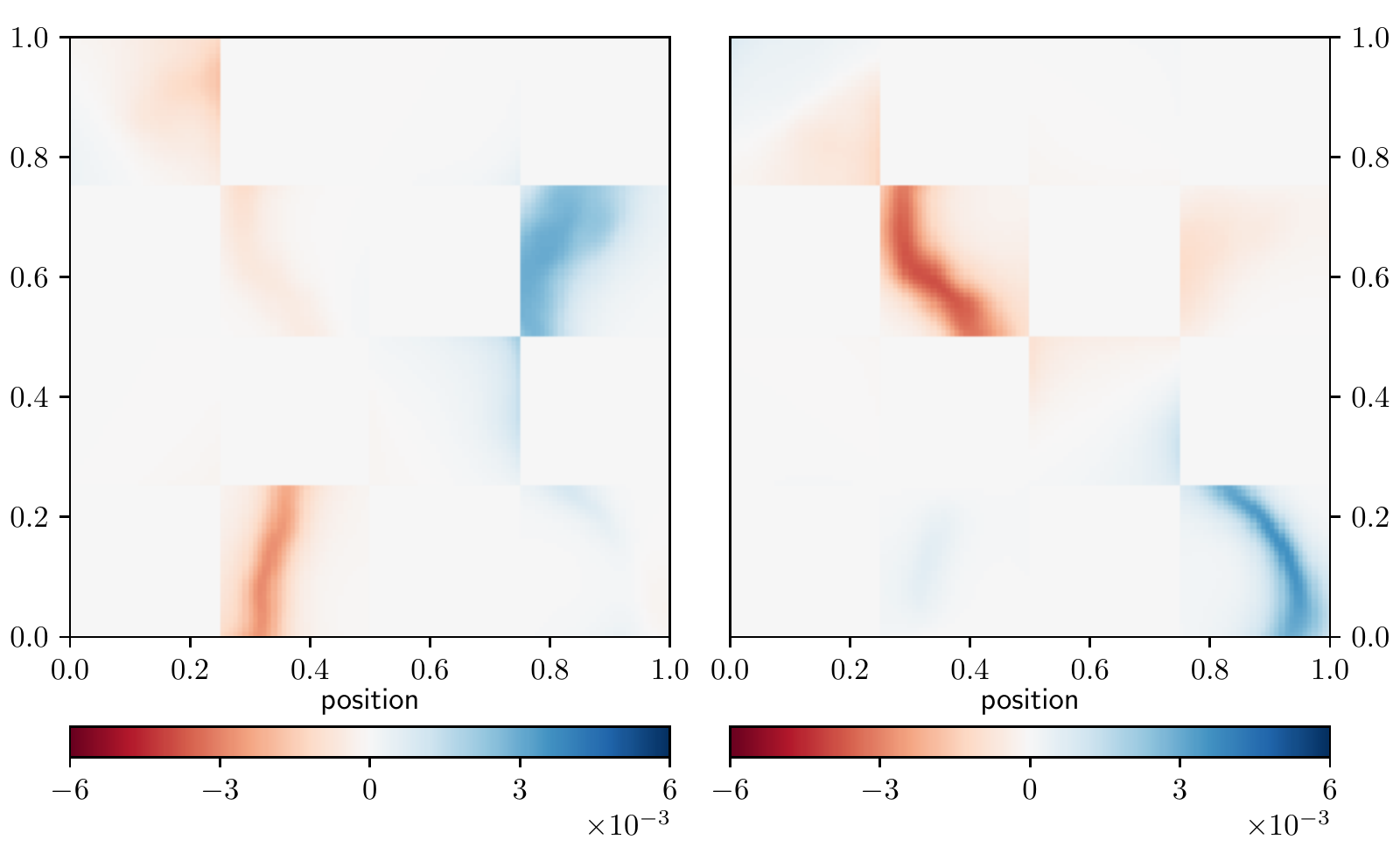}
    \includegraphics[width=0.49\columnwidth]{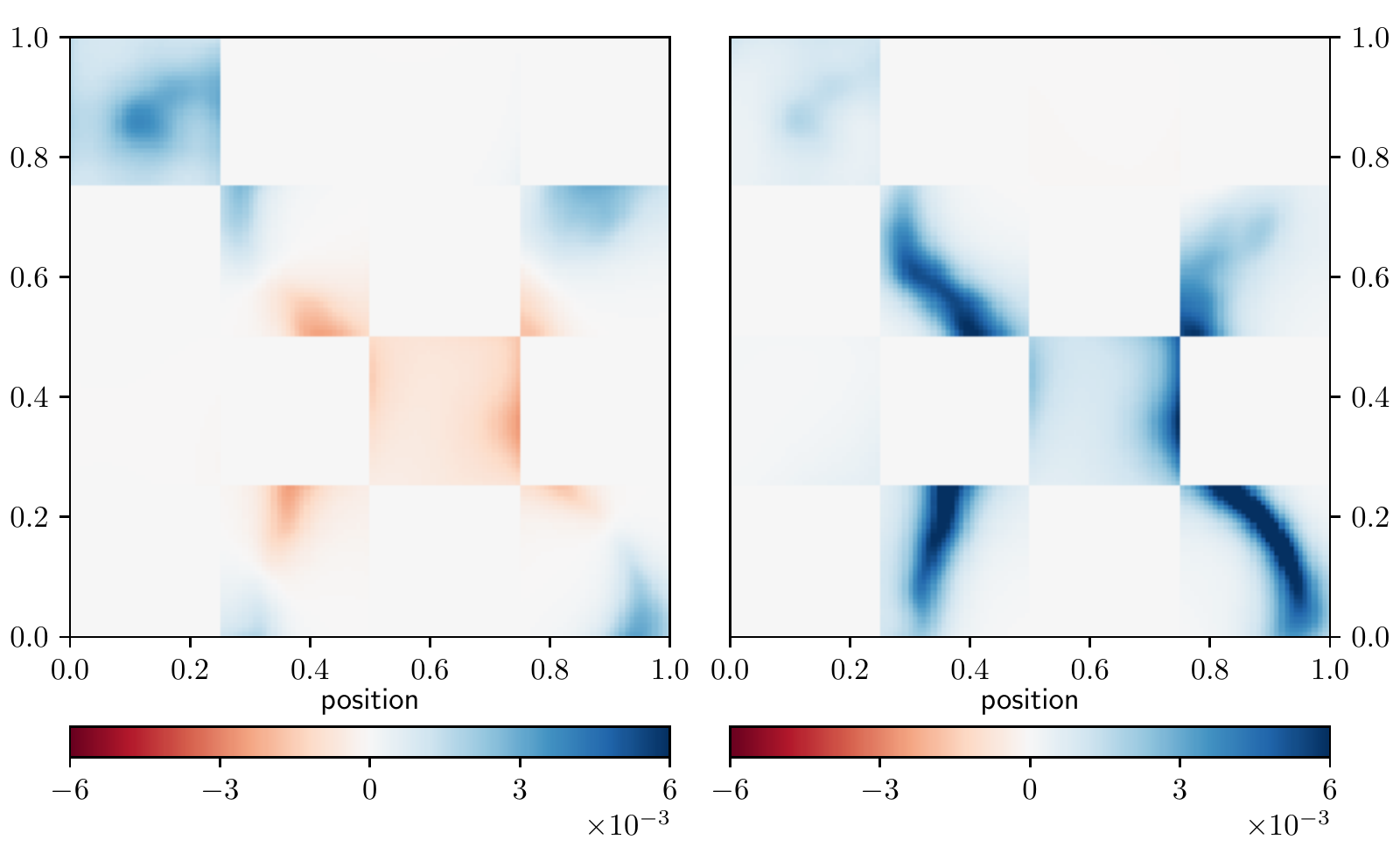}
  \caption{\label{fig:NLM-Rc}
    The eigenvectors, \ie the row vectors of the jointly compressed response
    after the inference corresponding to the highest eigenvalues in descending
    order from left to right. They clearly capture the shape of the signal with
    higher amplitude at the edges of larger structures.
  }
\end{figure}

It is interesting to have a closer look on the back projection of the
compressed data, \ie $R_\compr^\dagger d_\compr$ as well as on the projection of the eigenvectors building the compressed responses onto the space.
The back projection of the jointly compressed data before the first
inference,
\ie having looked at the original data only once, is shown in
Figure~\ref{fig:NLM-BackProjectionComp} on the left.
In contrast to the back projection after the minimization process in the right
plot, the data look quite uninformative, covering more or less uniformly the
whole probed signal domain.
After the inference, when the reference point around which the linearization
is made has changed, the jointly compressed data addresses mainly regions of
rapid changes in the signal.
Especially the contours at the edges are saved in the jointly compressed data.
This is even clearer visible in the projection of the eigenvectors $r_i$ building the
compressed response according to \eqref{eq:DefRc} in Figure~\ref{fig:NLM-Rc}.
The first two eigenvectors capture the frame of the large structure.
The third one mainly looks at the upper left corner, where also some structure
occurs, though it is less distinct than the large one.
None of the eigenvectors covers any structure in the second and fifth patches,
which was also the patches with largest $\gamma$, \ie least information loss
due to the compression.
Since the structure of the ground truth there is rather uniform,
it does not contain much information but the amplitude of the field,
which can easily be compressed to few data points.

\begin{figure}
  \centering
    \includegraphics[width=0.49\columnwidth]{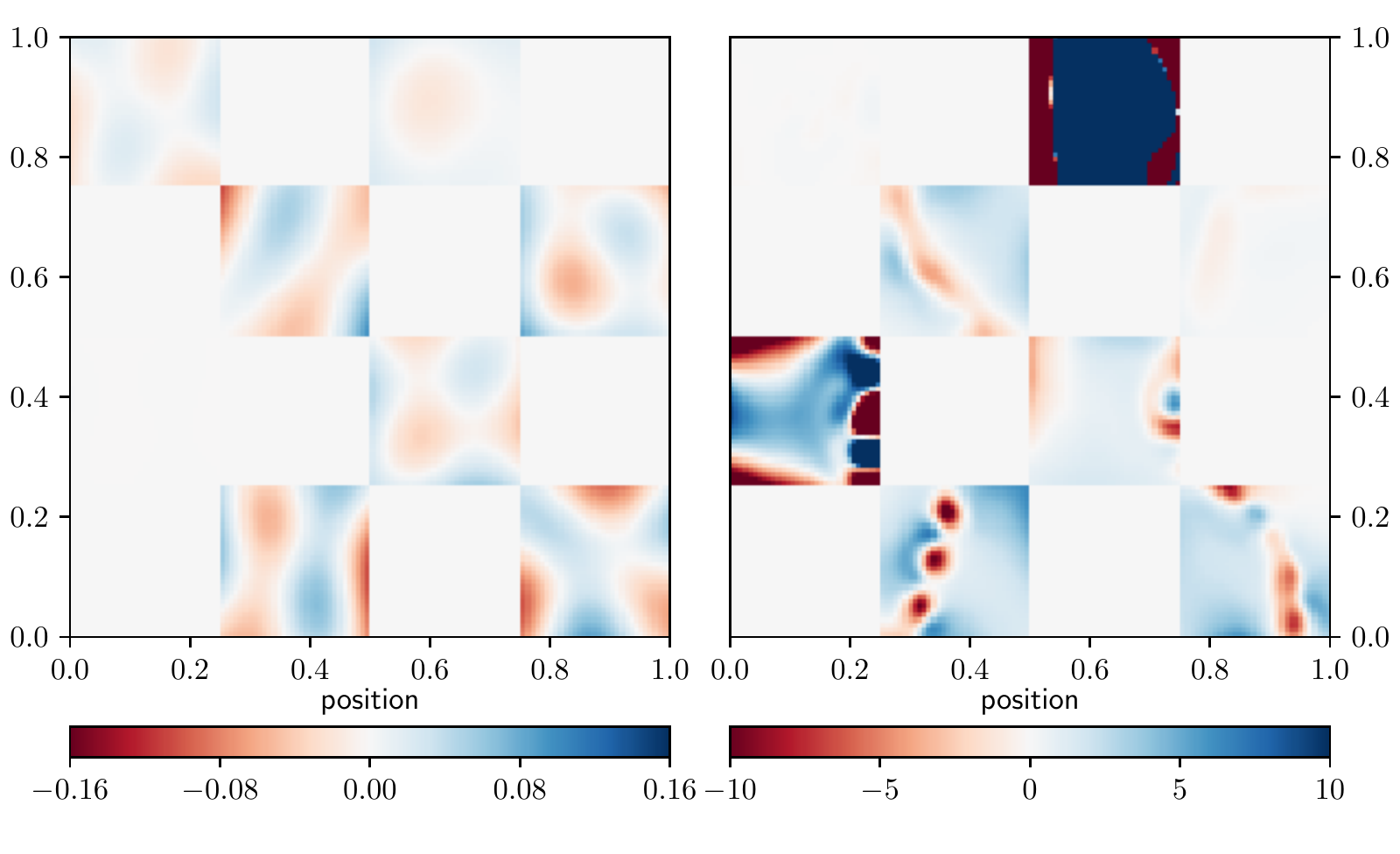}
  \caption{\label{fig:NLM-BackProjectionPatchComp}
    Back projection of the patchwise compressed data before (left) and after (right) the inference.
    As in the case of joint compression, structures in the signal can be
    recognized in the back projection of the compressed data after the
    inference.
    In addition the back projection is high in rather uniform areas of the
    signal.
  }
\end{figure}

Figure~\ref{fig:NLM-BackProjectionPatchComp} shows the back projection of the
patchwise compressed data before and after the inference.
Here, the change of the basis functions becomes apparent as well.

Table~\ref{tab:NLM-Time} shows the computation time of the different
compression methods and reconstructions.
As described above, it has been measured for
$(n_\mathrm{rep},n_\mathrm{comp})~=~(2,3)$, as well as
$(n_\mathrm{rep},n_\mathrm{comp})~=~(3,2)$.
In case of the original inference in total
$n_\mathrm{rep}\,\times\,n_\mathrm{comp}$ inference steps were performed such
that in there are
an identical number of inference steps for every method
The time has been measured for the inference only and for the total run of
separation of the data into patches, $n_\mathrm{comp}$ compressions with
$n_\mathrm{rep}$ inferences after each compression.
The average of all $n_\mathrm{rep}\,\times\,n_\mathrm{comp}$ inferences is given
in the first line of Table~\ref{tab:NLM-Time}.
The time for the total runs is given in the second line.
It has been measured on a single node of the FREYA computation facility of the
Max Planck Computing \& Data Facility restricted to 42 GB RAM.
In all categories the inference with patchwise compressed data is the fastest.
In contrast, joint compression takes the longest time.
There, one can clearly see the advantage of patching as discussed in
Section~\ref{sec:Patching}.
This leads to sparse responses which are more affordable in terms of
computation time and storage.
This is in agreement with the synthetic example discussed in this section.
It shows that such sparse representations are highly beneficial.

\input{parts/tables.tex}

The response is called, \ie applied, several times during the minimization.
In the application here, calling the response is inexpensive.
However, there are applications in which the response is expensive.
Then one aims to minimize the number of response calls,
as this determines the computation time.
The number of response calls were counted for
$(n_\mathrm{rep},n_\mathrm{comp})~=~(2,3)$.
In the inference with the original data, $R_\orig$ was called 686\,187 times.
In the process of compressing jointly it was called 4\,369 times.
During the patchwise compression, the patchwise original response has been
called 10\,776 times.
In the case of patchwise compression, the response only maps between the
single patches, \ie it is a factor 16 smaller than the full response.
Thus, effectively the full original response has been called only about 674 times in
the case of patchwise compression leading to a speed up factor of up to 1018
in case the response calculation is the dominant term.
In this application, patchwise compression lead to computation times consistent
with the computation time of inferring with original data.
Future steps to make BDC more rewarding could be to find representations for the
compressed response that are even more feasible.

\subsection{Real data: Radio interferometry}\label{sec:AplResolve}

Finally, we apply BDC to radio astronomical data from the supernova remnant Cassiopeia A observed by the GMRT~\cite{Ananthakrishnan1995GMRT, Raja2014FaradaySlicing}.
200\,000 data points from the measurement were selected randomly
and noise corrected according to \cite{Arras2020RESOLVE}.
Using those data, two images were constructed by the RESOLVE
algorithm~\cite{Junklewitz2016Resolve,Arras2020RESOLVE} that relies on MGVI,
where we make one image with compression and one
without compression for comparison.

The model used by RESOLVE as described in~\cite{Junklewitz2016Resolve} is
adopted for the inference.
To this end, the variables from the amplitude model
in~\cite{Junklewitz2016Resolve} are denoted as $\xi$ and transformed to the sky
$s$ -- the actual signal -- by $s = f(\xi)$ as described in
\cite{Knollmueller2018EncodingPriorKnowledge}.
The data are connected to the sky $s$ via a nonlinear measurement
equation of the form \eqref{eq:NonLinMeasurementEqn}.
The nonlinear part of the response $R_\mathrm{nl}$ contains a pointwise
exponentiation and a Fourier transformation onto a grid,
which leads to the variable of interest for the compression method $s'$.
The linear part of the response $R_\mathrm{lin}$ degrids the resulting points and
transforms them to the data $d_\orig$, \ie it projects the points lying on a
grid to continuous space.
The total response $R = R_\mathrm{lin} \circ R_\mathrm{nl}$ directly maps from
the sky $s$ to the data $d_\orig$, such that we have the measurement equations
\begin{align}
  d_\orig &= R(s) + n \nonumber \\
    &= R_\mathrm{lin}(R_\mathrm{nl}(s)) + n \nonumber \\
    &=: R_\mathrm{lin}(s') + n.
\end{align}
$R_\mathrm{lin}$ is computationally expensive and the data $d_\orig$ are large.
The aim is to compress the original data with the signal $s'$ as the variable of interest.
It turns out that the joint compression of all data is infeasible due to
its large computational costs.
So only the patchwise compression is tested.
Similar to the previous example, a non-Gaussian signal and the power spectrum
of the underlying Gaussian process need to be estimated simultaneously from
noisy and very incomplete data, only here the nonlinearity is an exponential
function and the sparse data live in Fourier space.

\begin{figure}
  \centering
  \inclgraphics{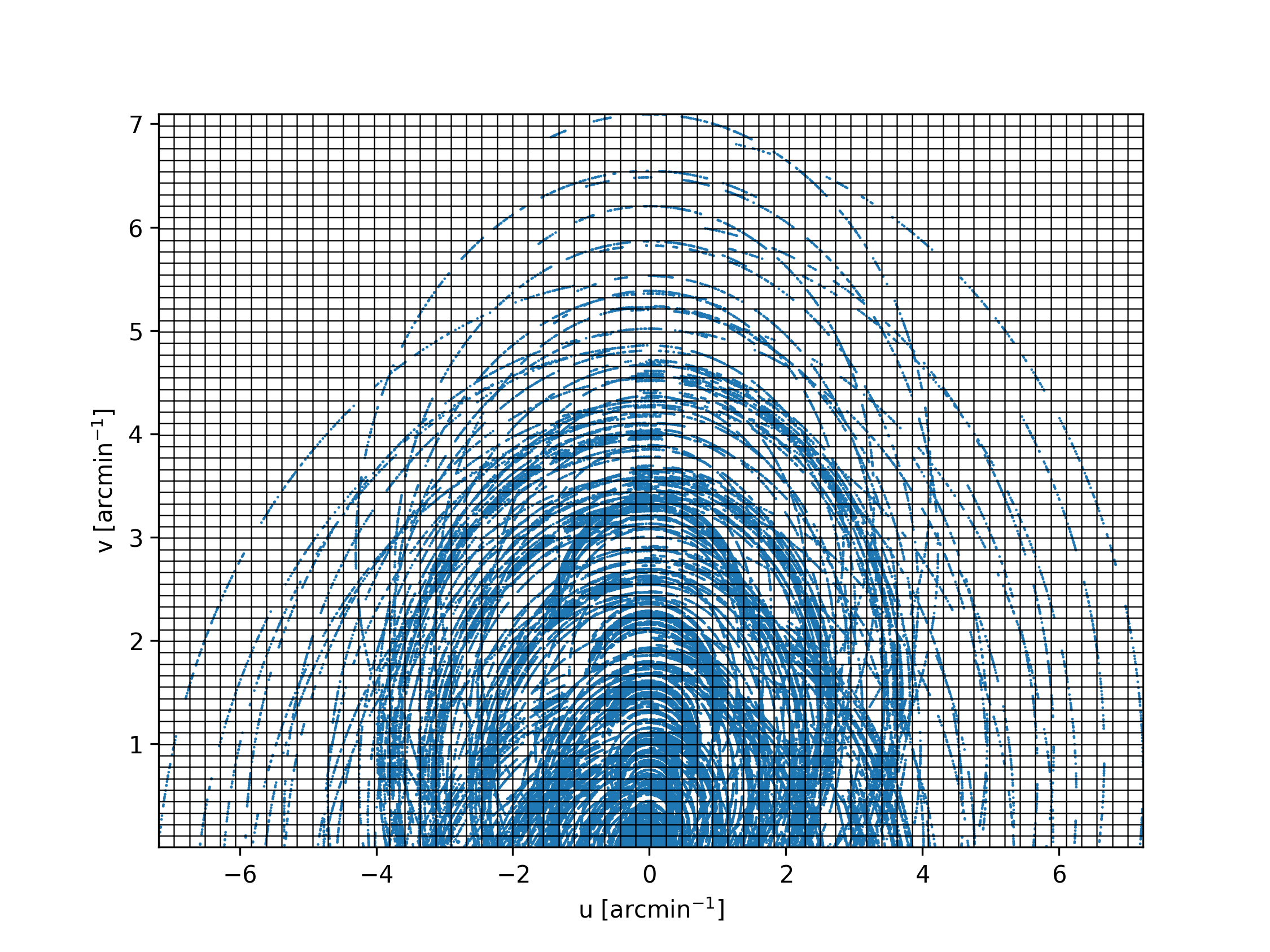}
  \caption{\label{fig:PatchedData}
    The position of all data points in Fourier space.
    The overlaid grid marks the data patches according to a Cartesian grid.
    Those patches are used for the patchwise compression.
    }
\end{figure}

We divide the Fourier plane into $64 \times 64$ squared patches, as shown in
Figure~\ref{fig:PatchedData}.
The location of the measurements in the Fourier plane are marked as well.
It is apparent that many patches are free of data, some contain some data, and
the highest density of data points occurs around the origin of the Fourier plane.

\begin{figure}
  \centering
    \includegraphics[width=0.5\textwidth]{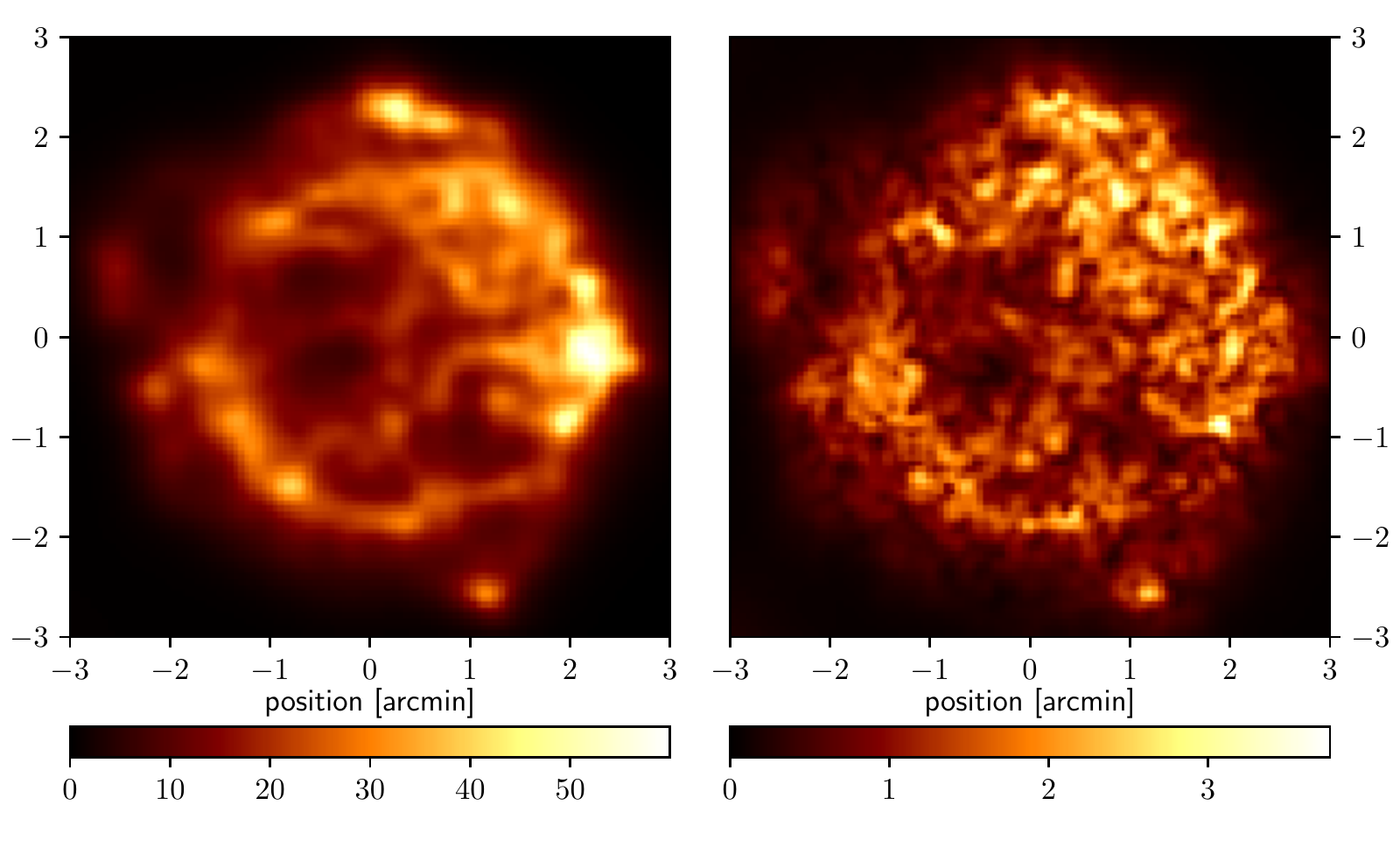}
    \includegraphics[width=0.5\textwidth]{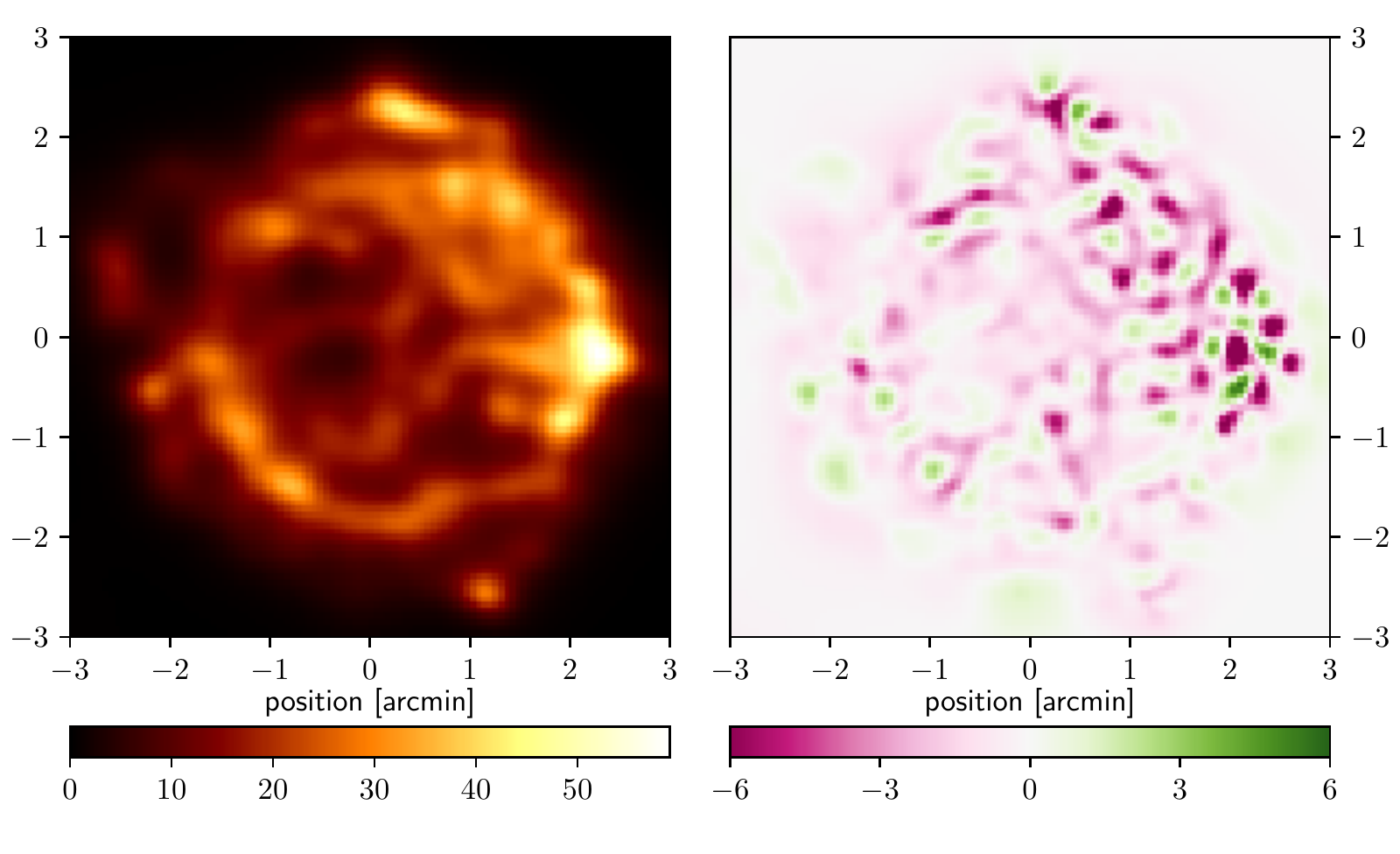}
    \includegraphics[width=0.5\textwidth]{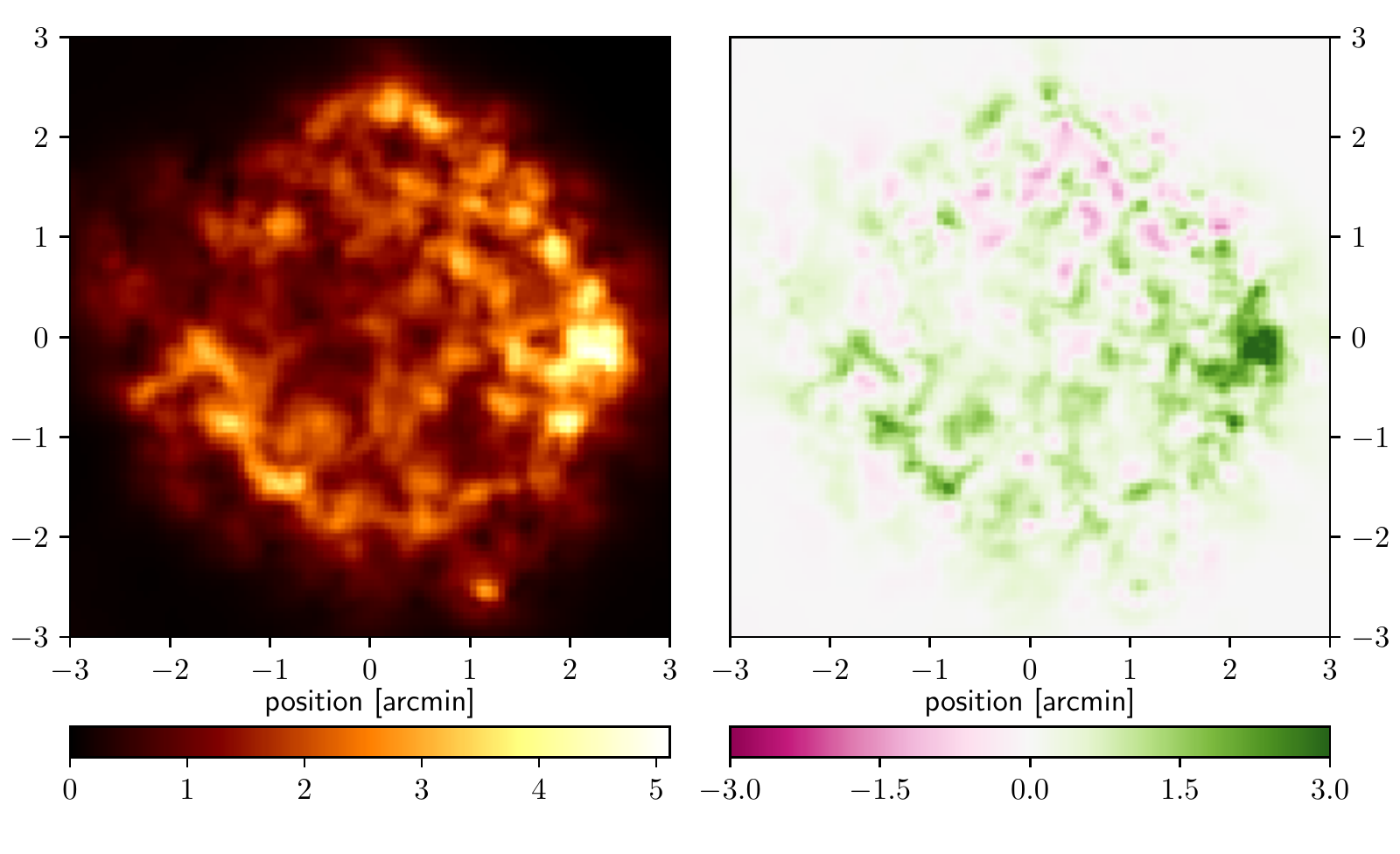}
  \caption{\label{fig:Rve:Rec}
    The reconstruction of supernova remnant Cassiopeia A from GMRT
    measurements. The colorbars have units $\text{Jy} \ \text{arcmin}^{-2}$.\protect\linebreak
    \textit{Top row:}
    To the left the resulting posterior mean, to the right its posterior
    uncertainty using original data.\protect\linebreak
    \textit{Middle row:}
    To the left reconstructed mean using patchwise compressed data.
    To the right its difference to the reconstructed mean from original data.
    The difference map shows that mainly small scale structures are lost due
    to the compression.
    \protect\linebreak
    \textit{Bottom row:}
    The same for the posterior uncertainty.
    Since information is lost due to the compression, the uncertainty is mainly
    higher than the one of the original posterior.
    }
\end{figure}

Figure~\ref{fig:Rve:Rec} shows the image obtained from the original
dataset and from the patchwise compressed dataset.
The mean is obtained after three MGVI iteration steps from the original data.
Using BDC, the data in each patch was compressed ideally under prior information.

The resulting data, noise covariance and responses were concatenated and used for inference in MGVI with three inference steps.
The received posterior distribution was used to compress the separated original data once more with updated knowledge, inferred with three minimization steps.
Doing this one more time resulted in the reconstruction shown here.
The bottom right plot in Figure~\ref{fig:Rve:Rec} shows that the uncertainty of the
reconstruction from patchwise compressed data is mostly higher than the
uncertainty from the original reconstruction.
This is expected due to the information loss of the compression.
The data points in every patch have been compressed to maximally $k_{\max} =
64$ data points per patch.
The minimal fraction of information stored in the compressed measurement
parameters $\gamma_{\min}$ has been set to 0.99.
In total this lead to 73\,239 compressed data points, which is a reduction of
the data size by a factor $2.73$.
Due to the large computation time, it is infeasible to determine more
eigenpairs.
Thus, estimates of the amount of information $\gamma$ contained in
the compressed data points are very vast.
On average, with Equations~\eqref{eq:DefGammaMin} and \eqref{eq:DefGammaMax}, the estimated range of $\gamma$ is $0.35\%$ to
$94.8\%$ for every patch with a standard deviation of $0.39\%$ and $17.3 \%$ respectively.
The dispersion of those values for different patches is high, caused by the
varying distribution of data points per patch.

\begin{figure}
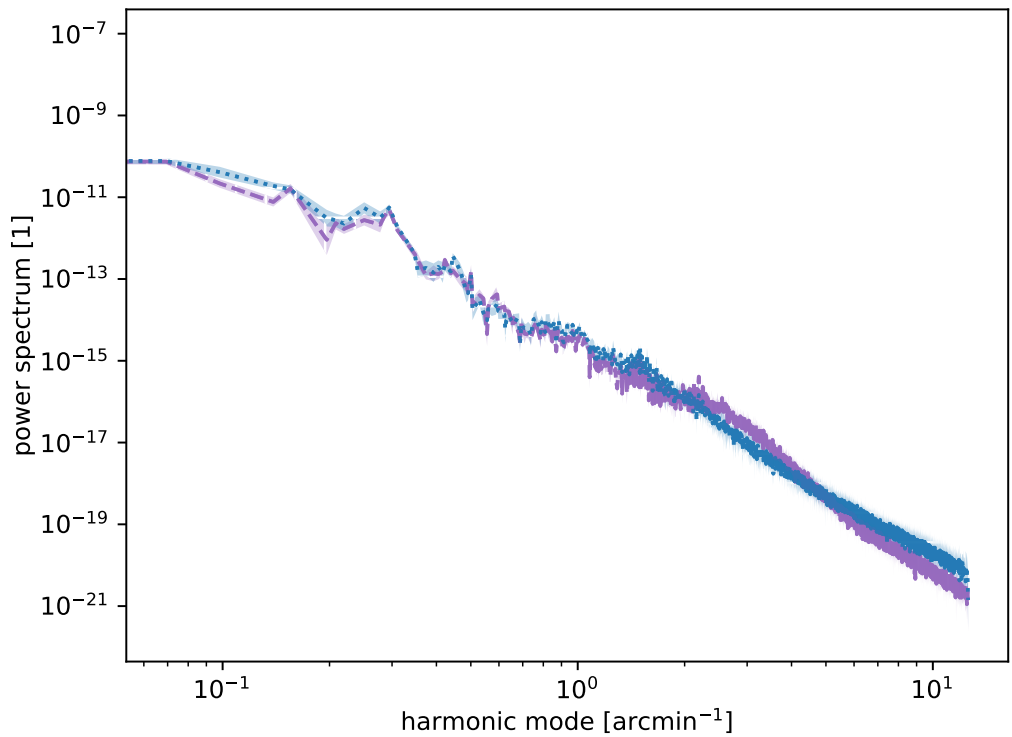

  \centering
    \inclgraphics{./figs/bdc_Resolve_0_99_CompTo_64_datapoints__4_9_o_PowSpecs}
  \caption{\label{fig:Rslv-PowSpecs}
    Similar to Figure~\ref{fig:NLM-PowSpecs}. The power spectra of the
    underlying Gaussian process inferred from original and patchwise compressed
    data are depicted in purple (dashed) and blue (dotted), respectively.
    They roughly agree in their slope.
  }
\end{figure}

The corresponding power spectra of the underlying Gaussian processes
are shown in Figure~\ref{fig:Rslv-PowSpecs}.
Their slopes qualitatively agree, but partly deviate outside their
uncertainties.
As discussed for the synthetic application in Section~\ref{sec:NonLinMock},
one need to take into account
that MGVI tends to underestimate the uncertainties.
In general, the power spectrum inferred from the original data is more distinct
in its slope than the one from compressed data in terms of deviations from a
straightly falling power spectrum.
The compressed spectrum is flatter especially in the higher harmonic regime.
This could be caused by a lower signal to noise ratio in this regime.
However, it is not completely clear, why BDC shows this behavior.

In addition to Euclidean gridded patches, we separated the data
into equiradial and equiangular patches.
This leads to a more even distribution of data points inside the patches, since
this way patches become larger further outside, where there are less data
points.
However, for this patch pattern small structures in the
reconstruction get lost.
The reason for this is that data points in the Fourier plane far away from the
origin store information about the small scale image structures.
Compressing them together therefore can be expected to lead to a loss of
information on small scale structures.
Thus, two criteria need to be considered for the choice of the patch geometry:
From a computational perspective patches with few data points are favoured.
From information theoretical perspective,
data points carrying similar information shall be compressed together.

This application shows that BDC is able to operate on real world data sets in the framework
of radioastronomical image reconstruction.
The run time still can be improved, though.
Since the compression for different patches works independently, this can perfectly be parallelized.
Another potential area for improvement would be the choice of the separation of the
patches.
One needs to aim for a patch geometry,
where the original data points are evenly distributed among the patches,
while also highly correlated data points are assigned to the same patch.

\section{Conclusion}\label{sec:Conclusion}

A generic Bayesian data compression algorithm has been derived, which compresses data in an
optimal way in the sense that as much information as possible about a signal
for which the correlation structure is assumed to be known a-priori.

Our derivation is based on the Kullback-Leiber divergence.
It reproduces the results of~\cite{Giraldi2018OptimalProjection} that optimizing the information
loss function leads to a generalized eigenvalue problem.
We generalized the method to the nonlinear case with the help of Metric Gaussian Variational Inference~\cite{Knollmueller2019MGVI}.
Also, we divided the data set into patches to limit the computational resources needed for the compression.
This leads to sparseness of the response, allowing to apply the method in high dimensional settings as well.

The method has been successfully applied to synthetic and real data problems.
In an illustrative one dimensional synthetic linear scenario, 40 data points
could be compressed to four data points with less than 20\% loss of information.
In a more complex, two dimensional and nonlinear synthetic measurement
scenario, 8192 measurements could be reduced to 80 data points with 70\% loss
of information that still capture the essential structures of the signal.
Dividing the data into patches resulted in a huge reduction of the required computation time for the compression itself, confirming the expected advantage.

Finally, the method has been applied to real astrophysical data.
The radio image of a supernova remnant has been reconstructed qualitatively
with a data reduction by a factor of almost 3.

For such scientific applications of BDC one needs to choose the variable of
interest, the signal $s$, such that $s$ represents the scientific interest best.
Then BDC can optimally adjust which information need to be stored in the
compressed data optimally.
In the chosen example, all degrees of freedom of the field had to be stored.
In principle, only certain (Fourier) scales or
certain areas of a field could be defined to be the quantity of interest.
This allows BDC to remove information on the other,
irrelevant scales or areas.

BDC compresses optimally with respect to the knowledge about this quantity.
It is a lossy compression method, \ie the compressed data contain
less information about the quantity of interest---information that is relevant
to answer the scientific question---than the original data.
This information loss consistently leads to higher uncertainty in the linear
case where the solution is exactly known and mostly higher uncertainty in the
nonlinear applications where only approximate solutions can be found.
To quantify the loss of information we have introduced the fraction $\gamma$ of
information about the quantity of interest $s$ stored in the
compressed data compared to the information in the full data.
This fraction can be used to reduce the dimension of the compressed data such
that they still contain the relevant amount of information.
In case the compression is too lossy, one needs to adjust the number
$k_{\max}$ of computationally determined eigenpairs that build the compressed
measurement parameters or increase the limit $\gamma_{\min}$ of information that
should be contained in the compressed data.

Still, the current BDC algorithm requires too much resources in terms of
storage needed for the responses and computation time.
In order to improve this even further, the choice of the data patches can still
be investigated and optimized
such that data points storing similar information are assigned to the same patch.
Up to now, data have been patched which are neighbouring in real or Fourier
space.
However also data points could be informationally connected non-locally.
One would need to look at the Kullback-Leibler divergence again to find those
connections and group the data accordingly.

Another problem is the computational cost of the response.
In the course of our derivation we represented it in a vector decomposition.
One could demand further restrictions to those vectors such as a certain
parametrization or find other representations to
find a computationally ideal basis for the responses.
This could lead to a higher reduction factor in applications such as the
astrophysical application in Section~\ref{sec:AplResolve}.

As a final step BDC needs to prove its advantage in real applications.
A promising application could be online compression such as
\cite{Cai2017OnlineImaging} suggest.
In a scenario, where data come in blockwise, those blocks can be treated as the
patches and compressed separately.
This is for example applicable in any experiment running over time.
There, time periods imply the measurement blocks.
This way the original data never needs to be stored at all, but compressed
immediately and optimally under the current knowledge.

%% file: parts/tables.tex
\begin{table}
    \caption{Computation times for the
            inference with original, compressed, and patchwise compressed data in
            the nonlinear synthetic application.
            Also the number of original response calls is stated.
            The original data has been inferred $n_\mathrm{rep} \times
            n_\mathrm{comp}$ times.
        }
    \begin{tabular}[\textwidth]{@{}cccc@{}}
        \hline 
        $(n_\mathrm{rep},n_\mathrm{comp}) = (3,2)$
        &original
        &comp 
        &patchcomp
        \\ 
        \hline 
        Inference time [sec]
        &317
        &984
        &289
        \\ 
        Totel run time [sec]
        &633
        &1972
        &586
        \\ 
        No. response calls
        &686187
        &2913
        &6062
        \\
        \hline 
        $(n_\mathrm{rep},n_\mathrm{comp}) = (2,3)$ & & & 
        \\ 
        \hline
        Inference time [sec]
        & 205
        & 637
        & 200
        \\
        Total run time [sec]
        &615
        &1917
        &613
        \\ 
        No. response calls
        &686187
        &4369
        &10776
        \\
        \hline
        
        \label{tab:NLM-Time} 
    \end{tabular}
\end{table}

%% file: parts/appendix.tex
\section{Optimality of BDC for Zero Posterior Mean}\label{Sec:ZeroMeanOptimization}

In this section of the appendix we prove that for zero original posterior mean $m_\orig
= 0$ the compression is optimal, if $\hat w_i$ is the eigenvector to the
smallest eigenvalue of $\mathcal{D}_\orig$.
Optimal means that the information gain $2 \Delta I(\hat{w}_i)$ in
\eqref{eq:CalDeltaKL} is maximal with respect to $\hat{w}_i$.

\paragraph{Proof:} For the proof, the $\hat w_i$ dependence in \eqref{eq:CalDeltaKL} needs to be shown explicitly:
\begin{align}
  2 \Delta I(\hat w_i) := \hat w_i^\dagger \mathcal{D}_\orig \hat w_i - 1 - \ln(\hat w_i \mathcal{D}_\orig \hat w_i)
\end{align}

Let $\{ (v_i, \delta_i) \}$ be the eigenpairs of $\mathcal{D}_\orig$.
The eigenvectors $\{v_i \}_i$ form a complete orthonormal basis.
Then $\hat w_i$ can be written as $\hat w_i = \sum_j \omega_{ij} v_j$ with $\sum_j \omega_{ij}^2 = 1$, such that $\hat w_i$ is normalized, and
\begin{align}
  2 \Delta I(\hat w_i) &= 2 \Delta I(\sum_j \omega_{ij} v_j) \nonumber \\
                       &= \sum_j \omega_{ij}^2 \delta_j - 1 - \ln(\sum_j \omega_{ij}^2 \delta_j).
\end{align}

We will use that $f(x) = x - 1 - \ln x$ is a convex function.
This can be easily verified by calculating the first derivative
\begin{align}
  \partial_x f (x) = 1 - \frac{1}{x}
\end{align}
and the second one
\begin{align}
  \partial^2_x f (x) = \frac{1}{x^2} > 0.
\end{align}
We observe that $2 \Delta I(\hat{w}_i) = f(\sum_j \omega_{ij}^2 \delta_j)$.
By Jensen's inequality we get
\begin{align}
  2 f(\sum_j \omega_{ij} \delta_j) &\leq \sum_j \omega_{ij}^2 2 f(\delta_j) \nonumber \\
                                   &\leq \sum_j \omega_{ij}^2 2 f(\min_k(\delta_k)) \nonumber \\
                                   &= 2 f(\min_k(\delta_k)) \underbrace{\sum_j \omega_{ij}^2}_{=1}.
\end{align}

Note, we used here that $f(x)$ has its minimum at $x=1$ and that the
eigenvalues of $\mathcal{D}_\orig$ are between 0 and 1, \ie the smaller eigenvalues maximize $f(\delta_i)$.
This way we got an upper bound reached for $\omega_i = \delta_{0i}$, where $v_0$ is the eigenvector corresponding to the smallest eigenvalue $\delta_0 := \min_k(\delta_k)$.
$\Box$

Doing data compression by just considering the smallest eigenvalues of
$D_\orig$ will be found to be the right choice when considering $\langle
\Delta \kl (\hat{r}_i)\rangle_{P(m_\orig)}$ as a loss
function in the section~\ref{sec:EigvecsWienerFilter}.
This gives the expected loss for the expected mean of $m_\orig=0$ under $\mathcal{P}(m_\orig) = \mathcal{P}(m_\orig|S,R_\orig,N_\orig)$.

\nwpage
\section{1D Wiener Filter Data Compression}\label{sec:EigvecsWienerFilter}

In Section~\ref{sec:AplWF} we applied our data compression method to synthetic data in the  context of the generalized Wiener Filter with a linear measurement equation.
In this section, we are going to investigate the shape of the eigenfunctions
corresponding to the eigenproblem of Equation~\eqref{eq:MoSx=mux} in this setting.

Therefore consider an easy set up without varying noise nor a complex mask.
To ensure a certain definition, we choose the signal space to be a one dimensional regular grid with 2048 lattice points in one dimension.
The synthetic signal and corresponding synthetic data are drawn from the prior
specified in section~\ref{sec:AplWF}.
The data is masked, such that only the central 256 pixels are measured.
Those data are then compressed to four data points, from which the signal is inferred in a last step.

\begin{figure}
  \centering
  \inclgraphics{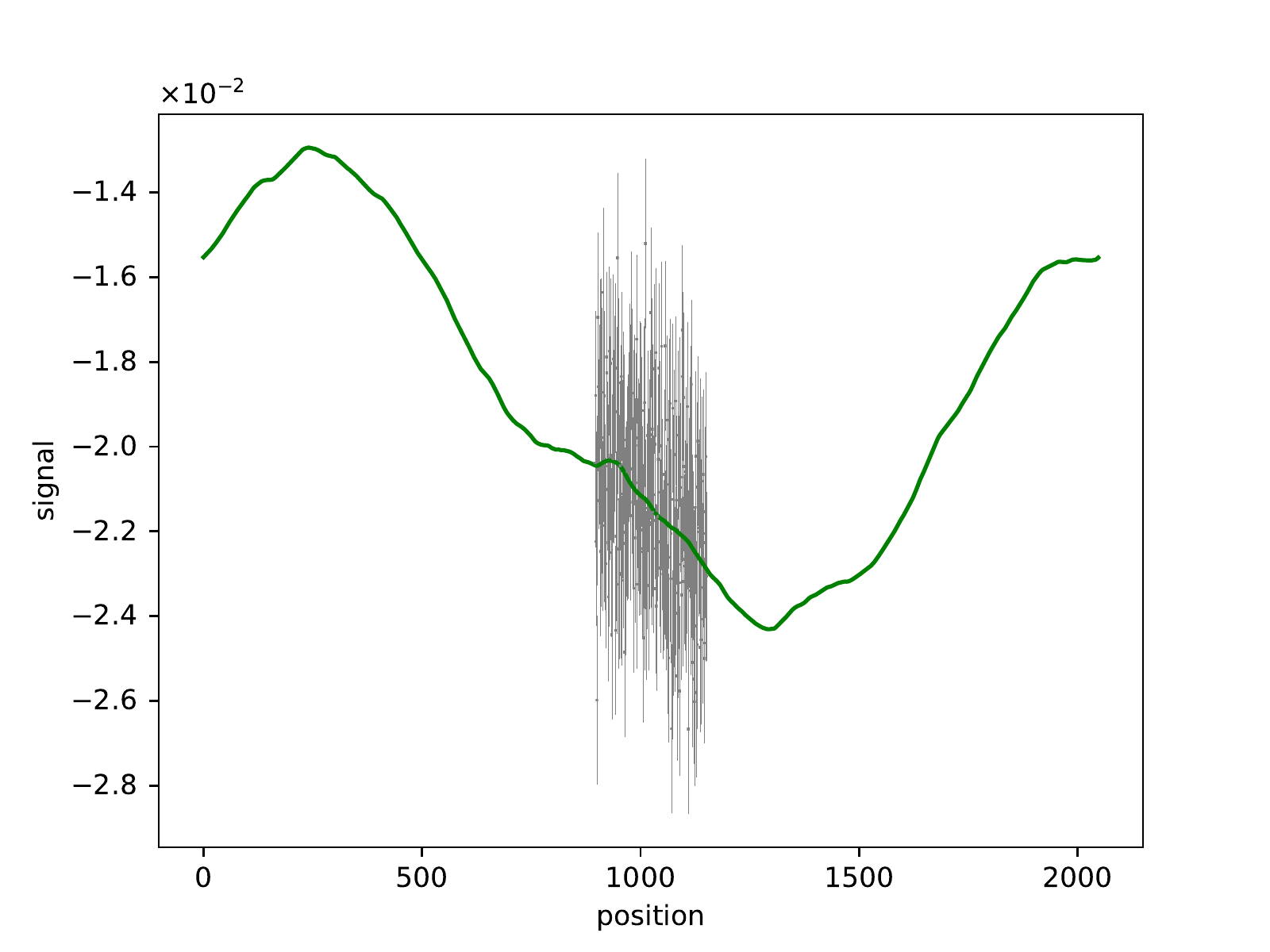}
  \caption{\label{fig:1DWFM_easy_Setup}
    One dimensional synthetic data setup to investigate BDC.
    The synthetic signal is marked in green and the measured data in gray.
  }
\end{figure}

The synthetic signal and data are computed as before.
However, the noise standard deviation now is constantly $0.2 \cdot 10^{-2}$ and the response is set to be a mask measuring pixels 896 to 1152 leading to a transparent window of 256 pixels in the center of the grid.
The measurement setup with signal mean, synthetic signal and data can be seen in Figure~\ref{fig:1DWFM_easy_Setup}.

\begin{figure}
  \centering
    \includegraphics[width=.49\columnwidth]{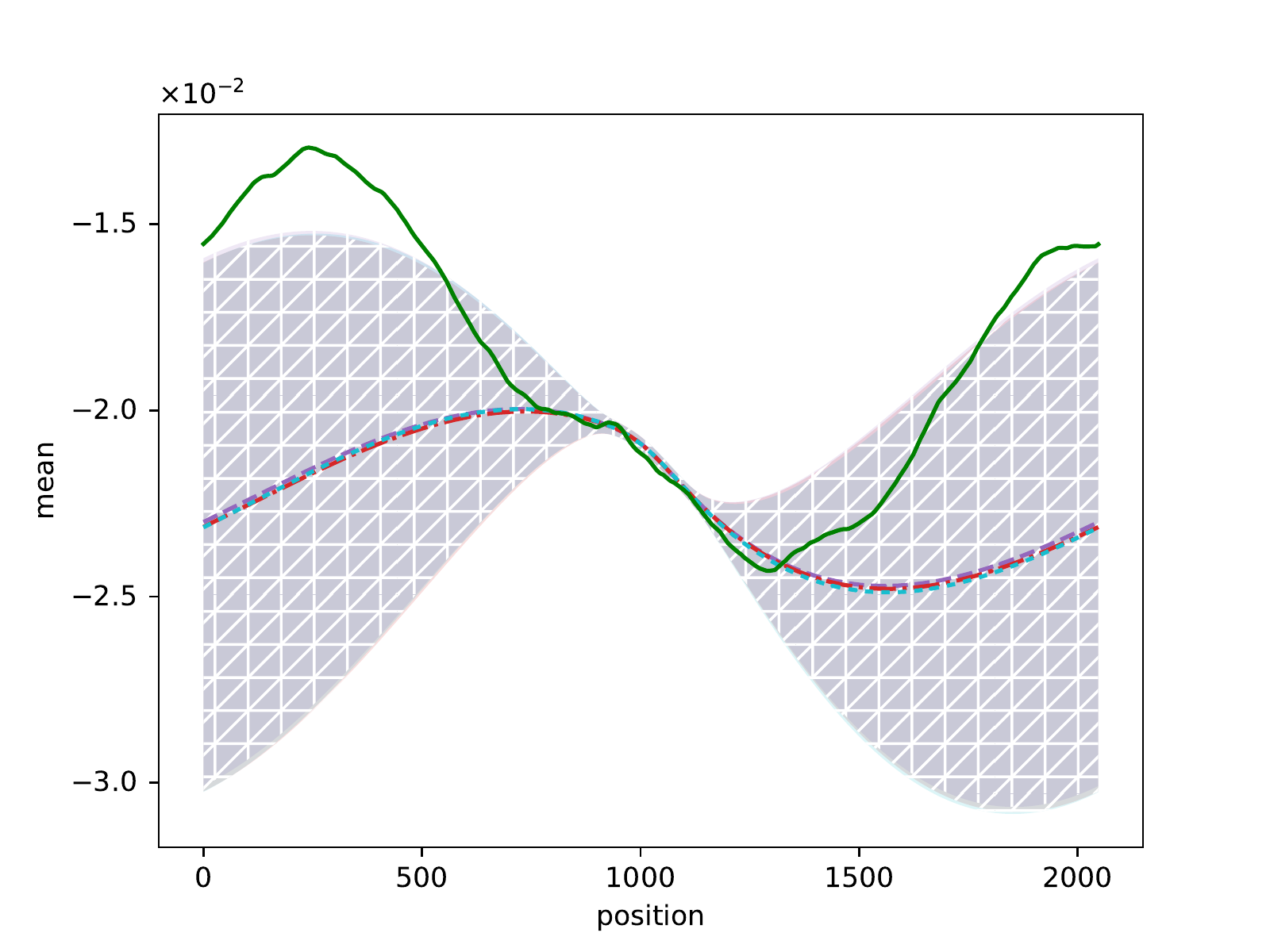}
  \includegraphics[width=.49\columnwidth]{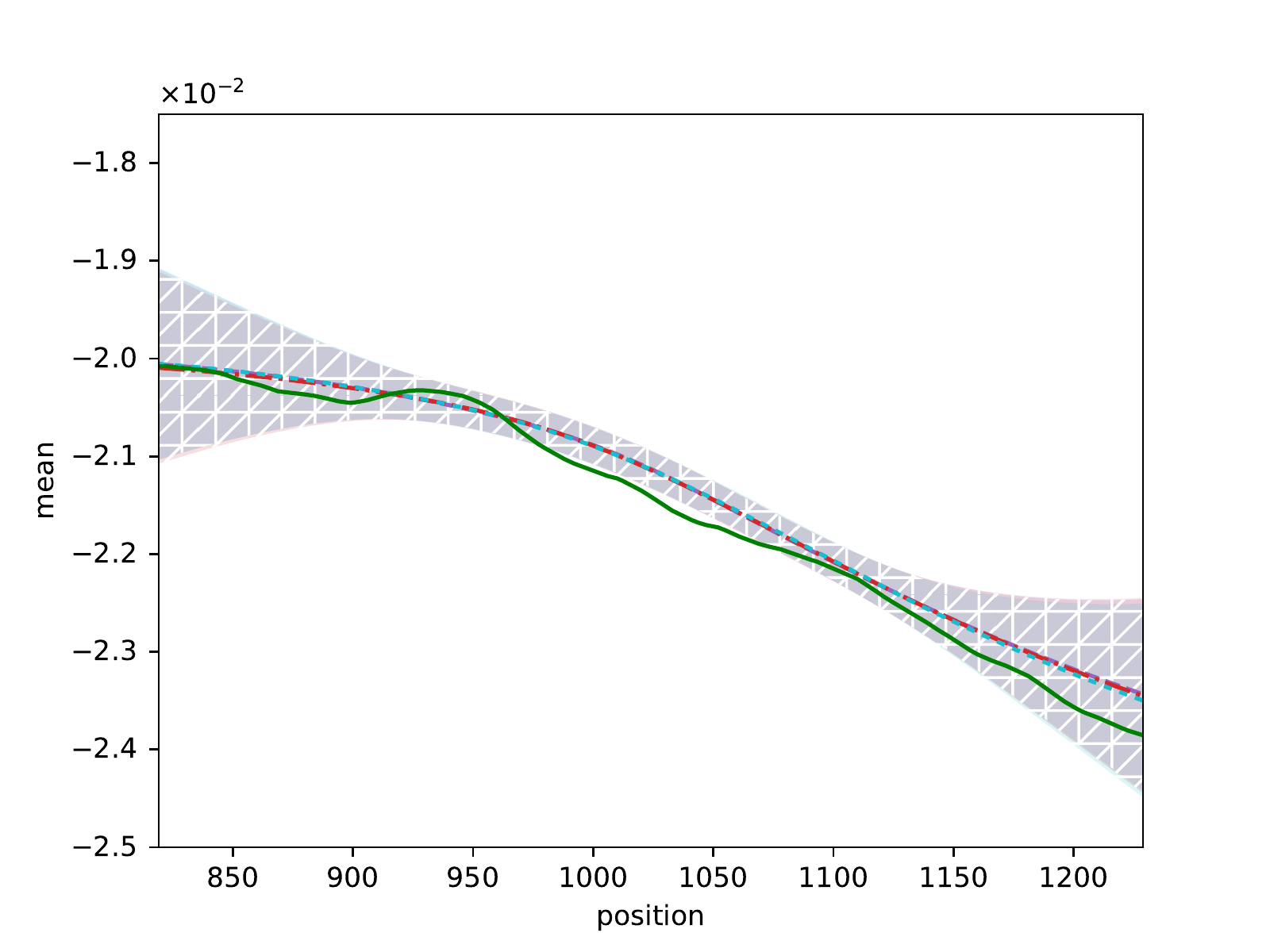}
  \caption{\label{fig:1DWFM_easy_Means}
    The same as Figure~\ref{fig:1DWienerFilterMeans} but for the setup shown in
    Figure~\ref{fig:1DWFM_easy_Setup}.
    In this simple set up original and compressed reconstruction almost
    perfectly coincide.
  }
\end{figure}

The consequential mean and uncertainty for the inference with the original data
and the ones with the compressed data are plotted together with the ground
truth in figure \ref{fig:1DWFM_easy_Means}.
The original data has been compressed from 256 to 4 data points.

\begin{figure}
  \centering
  \inclgraphics{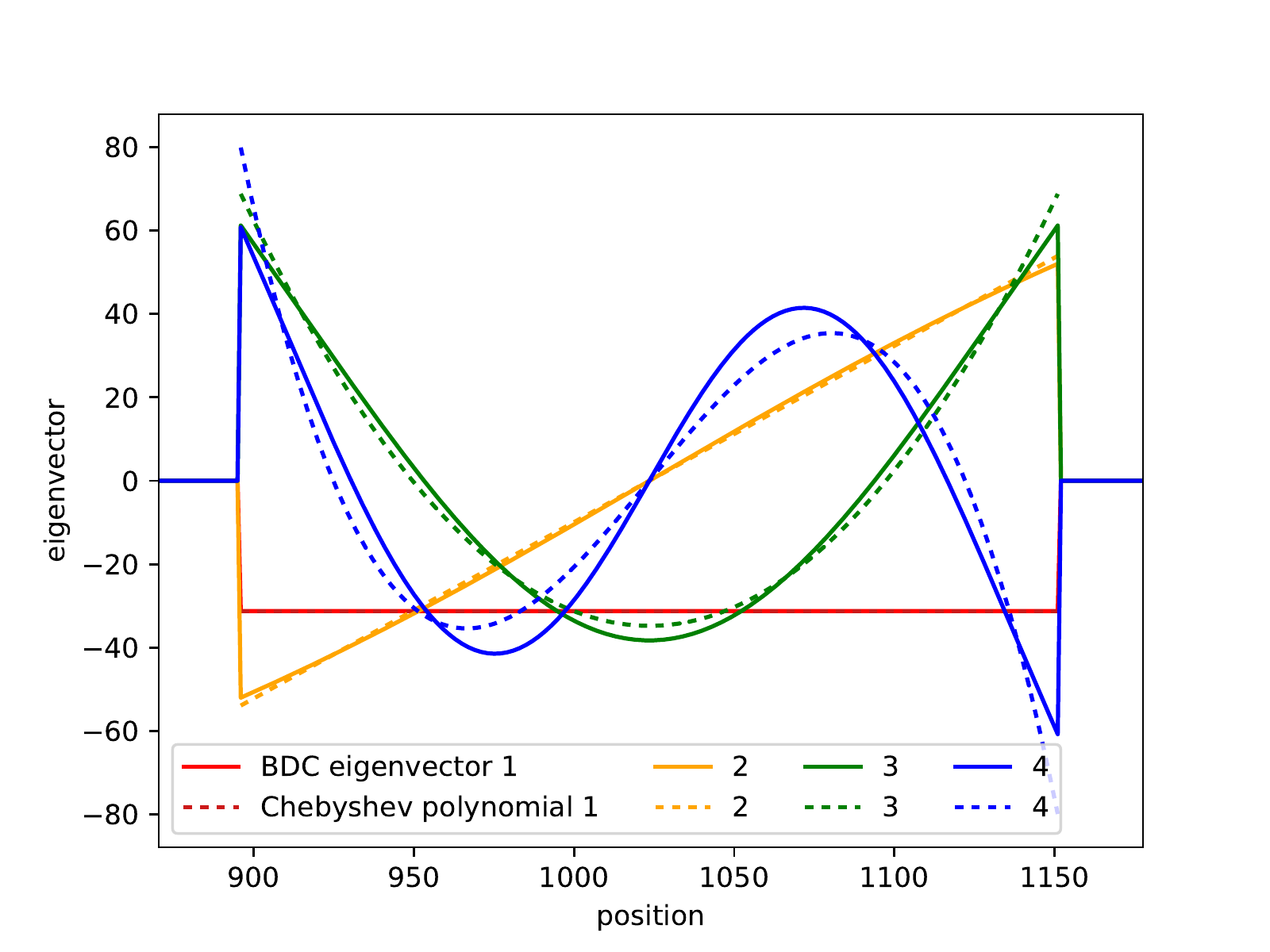}
  \caption{\label{fig:1DWFM_easy_Vgl}
    The first four eigenvectors (solid lines) as in
    Figure~\ref{fig:1DWienerFilterEigenVecs} but for the setup shown in 
    Figure~\ref{fig:1DWFM_easy_Setup} together with the first four
    Chebyshev polynomials (dashed lines).
    Their coincidence reveals the operating mode of BDC using polynomials fits
    to compress the information about the signal.
    Then the compressed data represent the amplitude of those
    polynomials and the compressed response stores their shape.
  }
\end{figure}

Now let us have a closer look at the eigenvectors plotted in Figure~\ref{fig:1DWFM_easy_Vgl}, which correspond to a back projection with $R_\compr$ of the corresponding single data point being one and all the others being zero.
These functions remind of Chebyshev polynomials of the first kind.
The Chebyshev polynomials were fitted to the eigenvectors minimizing the mean squared error and are plotted in the same figure as the eigenvectors.
One can clearly see their similarity.
The lower order polynomials fit the best, while higher order polynomials deviate especially at the edges.
This hints at BDC transferring the compression problem to a polynomial fit.
Then the compressed data points are the amplitudes of the polynomials while the
compressed response stores their individual shapes.

An analytical analysis is done in the next section.

\section{Analytical solution of the 1D Wiener Filter Data Compression}\label{sec:WFEigenVecDerivation}

In this section we will derive the eigenfunctions of \eqref{eq:MoSx=mux} analytically for some signal on a one dimensional line covered by a mask of length $L$ starting at $x=0$.
The response of the linear measurement equation is
\begin{align}
    \begin{split}
        {R_\orig}_{x x'} &= \delta(x-x') \chi_{[0,L]}(x'),\\
        &\quad \text{with} \quad \chi_{[0,L]}(x) := \begin{cases} 1 & \text{for } x \in [0,L] \\ 0 &
        \text{else} \end{cases}.
    \end{split}
\end{align}
The Gaussian noise $n_\orig$ in the measured area has the covariance
\begin{align}
  {N_\orig}_{x x'} &= n_{\mathrm{const}} \delta(x-x').
\end{align}
Having a look on the eigenvalue problem \eqref{eq:MoSx=mux}
\begin{align}
  M_\orig S r_i &= \mu_i^2 r_i
\end{align}
with
\begin{align}
  M_\orig &= R_\orig^\dagger N_\orig R_\orig,
\end{align}
we see, that $r_i(x) = 0$ for $x \notin [0,L]$ and $\mu_i \neq 0$. For $x \in [0,L]$
\begin{align}
    S r_i &= \frac{\mu_i^2}{n_{\mathrm{const}}} r_i \nonumber \\
    &=: \lambda_i r_i.
\end{align}
We specified $S$ by a falling power spectrum following a power law with
spectral index of $-2\alpha$ in Hartley space.
\begin{align}
    S &= \mathbb{H} P(|k|) \mathbb{H}^\dagger \nonumber \\
    &= \left( \mathbb{H} \frac{1}{k^\alpha} \right)^{1+\dagger} \nonumber \\
    &= \left( \Delta^{-\frac{\alpha}{2}} \right)^\dagger \Delta^{-\frac{\alpha}{2}}
\end{align}
with Hartley transform
\begin{align}
  \mathbb{H}_{xk} &= \frac{1}{\sqrt{2 \pi}} \int \mathrm{d}k \left[ \cos(kx) + \sin(kx) \right]
\end{align}
and Laplace operator
\begin{align}
  \Delta := \nabla^2 = \sum_i \partial_{x_i}^2.
\end{align}
Then
\begin{align}
    \lambda_i r_i &= S r_i \nonumber \\
    &= \Delta^{-\alpha} r_i \nonumber \\
    &= {\lambda_\Delta}_i^{-\alpha} r_i
\end{align}
with eigenfunction $r_i$ and eigenvalue ${\lambda_\Delta}_i$ of the Laplace operator.
This is equivalent to the Helmholtz equation with opposite sign.
Its eigenfunctions in 2D are the Bessel functions.
In one dimension with $\alpha=2$ as in section \ref{sec:AplWF}, the covariance
operator becomes $S = \partial_x^{-4}$, thus
\begin{align}
    \partial_x^{-4} r_i = {\lambda^{-2}_\Delta}_i r_i.
\end{align}
The square of the eigenvalue ensures the eigenvalues of the prior covariance to be positive.
Since $\lambda_i = {\lambda_\Delta}_i^{-\alpha}$, $\lambda_i$ does not
become zero.

The solution to this problem are super positions of exponential functions of
the same eigenvalue
\begin{align}
  r_i(x) = a e^{+ \sqrt{\lambda_i} x} + b e^{- \sqrt{\lambda_i} x},
\end{align}
such as $\cosh(\sqrt{\lambda_i} x)$ and $\sinh( \sqrt{\lambda_i}x)$ for
positive $\lambda_i$, as well as $\cos(\sqrt{-\lambda_i} x)$ and
$\sin(\sqrt{-\lambda_i} x)$ for negative $\lambda_i$.
For eigenvalue ${\lambda_\Delta}_i = 0$, also polynomials up to third order
are eigenfunctions to the Laplace operator.
Those belong to the largest eigenvalues of $S$, which are also the most
informative ones according to Equation~\eqref{eq:DeltaIBDC}.
This explains the proximity of the eigenvectors to Chebyshev polynomials as
observed in Figure \ref{fig:1DWFM_easy_Vgl} for a signal power spectrum that
asymptotically follows $k^{-4}$.

%% file: bdc-arxiv.bbl
\begin{thebibliography}{10}

\bibitem{Google2020HowSearchWorks}
How search organizes information.
\newblock
  \url{https://www.google.com/search/howsearchworks/crawling-indexing/}.
\newblock online accessed: 2020-11-17.

\bibitem{Ananthakrishnan1995GMRT}
S.~{Ananthakrishnan}.
\newblock {The Giant Meterwave Radio Telescope / GMRT}.
\newblock {\em Journal of Astrophysics and Astronomy Supplement}, 16:427, 1995.

\bibitem{Arras2019UnifiedRadioImaging}
P.~Arras, P.~Frank, R.~Leike, R.~Westermann, and T.~A. En\ss{}lin.
\newblock Unified radio interferometric calibration and imaging with joint
  uncertainty quantification.
\newblock {\em A\&A}, 627:A134, 2019.

\bibitem{Arras2020RESOLVE}
P.~{Arras}, R.~A. {Perley}, H.~L. {Bester}, R.~{Leike}, O.~{Smirnov},
  R.~{Westermann}, and T.~A. {En{\ss}lin}.
\newblock \textit{(Preprint) arXiv}:2008.11435, v1, submitted: Aug
  \txtbf{2020}.

\bibitem{Blei2017VariationalInference}
David~M. Blei, Alp Kucukelbir, and Jon~D. McAuliffe.
\newblock Variational inference: A review for statisticians.
\newblock {\em Journal of the American Statistical Association},
  112(518):859--877, 2017.

\bibitem{Cai2017OnlineImaging}
X.~Cai, L.~Pratley, and J.~McEwen.
\newblock Online radio interferometric imaging: Assimilating and discarding
  visibilities on arrival.
\newblock {\em Monthly Notices of the Royal Astronomical Society}, 485, 12
  2017.

\bibitem{CoverElementsOfInformationTheory}
T.~M. Cover and J.~A. Thomas.
\newblock {\em Elements of information theory}.
\newblock Wiley-Interscience publication, Hoboken, NJ, 2nd edition, 2006.

\bibitem{Geiger2012SignalEnhancement}
B.~C. {Geiger} and G.~{Kubin}.
\newblock \textit{(Preprint) arXiv}:1205.6935, v2, submitted: Jan \txtbf{2013}.

\bibitem{Giraldi2018OptimalProjection}
L.~Giraldi, O.~P. Le~Maître, I.~Hoteit, and O.~M. Knio.
\newblock {Optimal projection of observations in a Bayesian setting}.
\newblock {\em Computational Statistics \& Data Analysis}, 124:252 -- 276,
  2018.

\bibitem{Grainge2017SKA}
K.~Grainge, B.~Alachkar, Shaun Amy, D.~Barbosa, M.~Bommineni, P.~Boven,
  R.~Braddock, J.~Davis, P.~Diwakar, V.~Francis, R.~Gabrielczyk, R.~Gamatham,
  S.~Garrington, T.~Gibbon, D.~Gozzard, S.~Gregory, Y.~Guo, Y.~Gupta,
  J.~Hammond, D.~Hindley, U.~Horn, R.~Hughes-Jones, M.~Hussey, S.~Lloyd,
  S.~Mammen, S.~Miteff, V.~Mohile, J.~Muller, S.~Natarajan, J.~Nicholls,
  R.~Oberland, M.~Pearson, T.~Rayner, S.~Schediwy, R.~Schilizzi, S.~Sharma,
  S.~Stobie, M.~Tearle, B.~Wang, B.~Wallace, L.~Wang, R.~Warange, R.~Whitaker,
  A.~Wilkinson, and N.~Wingfield.
\newblock Square kilometre array: The radio telescope of the xxi century.
\newblock {\em Astronomy Reports}, 61:288--296, 2017.

\bibitem{Jolliffe2010PCA}
I.~T. Jolliffe.
\newblock {\em Principal component analysis}.
\newblock Springer series in statistics. Springer, New York i.a., 2010.

\bibitem{Jones2001SciPy}
E.~Jones, T.~Oliphant, P.~Peterson, et~al.
\newblock {SciPy}: Open source scientific tools for {Python}, 2001.
\newblock {a}ccessed Oct 2020.

\bibitem{Junklewitz2016Resolve}
H.~Junklewitz, M.~R. Bell, M.~Selig, and T.~A. En\ss{}lin.
\newblock Resolve: A new algorithm for aperture synthesis imaging of extended
  emission in radio astronomy.
\newblock {\em A\&A}, 586:A76, 2016.

\bibitem{karhunen1947lineare}
K.~Karhunen.
\newblock {\em {\"U}ber lineare Methoden in der Wahrscheinlichkeitsrechnung}.
\newblock Annales Academiae Scientiarum Fennicae: Ser. A 1. Sana, 1947.

\bibitem{Knollmueller2018EncodingPriorKnowledge}
J.~Knollm{\"u}ller and T.~A. En{\ss}lin.
\newblock \textit{(Preprint) arXiv}:1812.04403, v1, submitted: Dec
  \txtbf{2013}.

\bibitem{Knollmueller2019MGVI}
J.~Knollm{\"u}ller and T.~A. En{\ss}lin.
\newblock \textit{(Preprint) arXiv}:1901.11033, v3, submitted: Jan
  \txtbf{2020}.

\bibitem{Kong2020UESS}
Lingqiang Kong, Zhifeng Liu, and Jianguo Wu.
\newblock A systematic review of big data-based urban sustainability research:
  State-of-the-science and future directions.
\newblock {\em Journal of Cleaner Production}, 273:123142, 2020.

\bibitem{kosambi1943statistics}
D.~D. Kosambi.
\newblock Statistics in function space.
\newblock {\em Journal of the Indian Mathematical Society}, 7:76--88, 1943.

\bibitem{ARPACK}
R.~Lehoucq, D.~Sorensen, and C.~Yang.
\newblock {\em ARPACK Users' Guide}.
\newblock Software, environments, tools. Society for Industrial and Applied
  Mathematics, Philadelphia, 1998.

\bibitem{LeikeOBA}
R.~Leike and T.~En{\ss}lin.
\newblock {Optimal Belief Approximation}.
\newblock {\em Entropy}, 19(8):402, 2017.

\bibitem{loeve1948functions}
M.~Lo{\`e}ve.
\newblock Functions aleatoires du second ordre.
\newblock {\em Processus stochastique et mouvement Brownien}, page 366, 1948.

\bibitem{Marx2013BigData}
Vivien Marx.
\newblock The big challenges of big data.
\newblock {\em Nature}, 498(7453):255--260, 2013.

\bibitem{Pearson1901LinesAndPlanes}
K.~Pearson.
\newblock Liii. on lines and planes of closest fit to systems of points in
  space.
\newblock {\em The London, Edinburgh, and Dublin Philosophical Magazine and
  Journal of Science}, 2(11):559--572, 1901.

\bibitem{Raja2014FaradaySlicing}
W.~Raja.
\newblock {\em Faraday Slicing Polarized Radio Sources}.
\newblock PhD thesis, Jawaharlal Nehru University New Delhi, 09 2014.

\bibitem{Vergassola2007Infotaxis}
M.~Vergassola, E.~Villermaux, and B.~I. Shraiman.
\newblock `infotaxis'as a strategy for searching without gradients.
\newblock {\em Nature}, 445(7126):406--409, 2007.

\bibitem{Wiener1949Extrapolation}
N.~Wiener.
\newblock {\em Extrapolation, interpolation and smoothing of stationary time
  series}.
\newblock Technology Press of the Mass. Inst. of Technology [u.a.], Cambridge,
  Mass. [u.a.], 1949.

\end{thebibliography}
